\documentclass[twocolumn,twocolappendix]{aastex631}

\usepackage{xspace}
\usepackage{amsmath}
\usepackage{url}
\usepackage{lipsum, babel, ulem, soul, cancel}
\usepackage[shortlabels]{enumitem}

\newcommand{\galah}{{\small GALAH}\xspace}
\newcommand{\esp}{{\small ESPRESSO}\xspace}
\newcommand{\apogee}{{\small APOGEE}\xspace}

\newcommand{\rave}{{\small RAVE}\xspace}
\newcommand{\kepler}{\textit{Kepler}\xspace}
\newcommand{\tess}{\textit{TESS}\xspace}
\newcommand{\gaia}{\textit{Gaia}\xspace}

\newcommand{\cosmic}{{\small COSMIC}\xspace}
\newcommand{\joker}{\textit{The Joker}\xspace}

\newcommand{\lk}{\texttt{lightkurve}\xspace}
\newcommand{\topcat}{\texttt{TOPCAT}\xspace}

\newcommand{\app}{$\sim$}
\newcommand{\logg}{$\log g$\xspace}
\newcommand{\teff}{T$_\textrm{eff}$\xspace}
\newcommand{\feh}{[Fe/H]\xspace}

\newcommand{\ali}{$A\textrm{(Li)}$\xspace}

\newcommand{\vbroad}{$v_\textrm{broad}$\xspace}

\newcommand{\kms}{km s$^{-1}$\xspace}

\newcommand{\ang}{$\textrm{\AA}$\xspace}

\newcommand{\msini}{$M\sin{i}$\xspace}

\newcommand{\lirich}{Li-rich\xspace}
\newcommand{\lin}{Li-normal\xspace}

\newcommand{\dg}{doppelg\"anger\xspace}

\newcommand{\sprocess}{$s$-process\xspace}

\newcommand{\total}{ten\xspace}
\newcommand{\target}{58\xspace}
\newcommand{\notbin}{48\xspace}

\newcommand{\columbia}{Department of Astronomy, Columbia University, 550 West 120th Street, New York, NY, USA}
\newcommand{\barnard}{Department of Physics and Astronomy, Barnard College, Columbia University, 3009 Broadway, New York, NY 10027, USA}
\newcommand{\anu}{Research School of Astronomy \& Astrophysics, Australian National
University, Canberra, ACT 2611, Australia}
\newcommand{\cca}{Center for Computational Astrophysics, Flatiron Institute, 162 5th Avenue, Manhattan, NY, USA}
\newcommand{\monash}{School of Physics \& Astronomy, Monash University, Clayton, Victoria 3800, Australia}
\newcommand{\uoft}{David A. Dunlap Department of Astronomy \& Astrophysics, University of Toronto, Toronto, ON M5S 3H4, Canada}

\newcommand{\new}[1]{\textcolor{black}{#1}}

\shorttitle{\esp and \lirich giants}
\shortauthors{Sayeed et al.}

\begin{document}

\title{Looking for Companionship: Radial Velocity Follow-Up of Lithium-Rich Giants with ESPRESSO}

\correspondingauthor{Maryum Sayeed}
\email{maryum.sayeed@columbia.edu}

\author[0000-0001-6180-8482]{Maryum Sayeed}
\affiliation{\uoft}
\affiliation{\columbia}

\author[0000-0003-0174-0564]{Andrew R. Casey}
\affiliation{\monash}
\affiliation{\cca}

\author[0000-0001-7516-8308]{Benjamin T. Montet}
\affiliation{School of Physics, University of New South Wales, Kensington, NSW 2052, Australia}

\author[0000-0001-5082-6693]{Melissa K. Ness}
\affiliation{\anu}

\author[0000-0003-0872-7098]{Adrian~M.~Price--Whelan}
\affiliation{\cca}

\author[0000-0001-8832-4488]{Daniel Huber}
\affiliation{Institute for Astronomy, University of Hawai`i, 2680 Woodlawn Drive, Honolulu, HI 96822, USA}

\author[0009-0009-8668-3060]{Madeline J. Maldonado Gutierrez}
\affiliation{\barnard}


\begin{abstract}
Lithium-rich red giants have been a long-standing mystery in stellar astrophysics. A leading theory to explain these chemically peculiar and rare objects is interactions with a close companion. To investigate their companion fraction, we collected high-resolution spectra of \target \lirich red giants using {\small ESPRESSO}, and constrained their orbital parameters using a Monte Carlo sampler, \textit{The Joker}. We recover an overall companion rate of 17\% (10/58) \new{consistent with the 16\% expected for normal red giants observed with {\small ESPRESSO} cadence and precision in \apogee DR16}. All ten binary candidates have $\rm A(Li) = 1.5-2.7 \; dex$ and $\log g = 2-3 \rm \; dex$; we find no binary candidates with $\rm A(Li) > 2.7 \; dex$. Secondary masses reveal three planetary mass companions ($ M\sin i \approx 7-11 \; \rm M_{Jup}$), one brown dwarf ($ M\sin i\approx 19 \; \rm M_{Jup}$), and six stellar mass companions ($M\sin i= 0.08-0.98 \;\rm  M_\odot$). \new{The significant difference in companion fraction between low and high lithium at $\rm A(Li) \approx 2.7 \; dex$ suggests that there are at least two formation mechanisms for \lirich giants: (i) red giants with low to moderate lithium ($\rm A(Li) \approx 1.5-2.7 \; dex$) become lithium enhanced via planet engulfment, mass transfer, or the Cameron--Fowler mechanism via tidal spin-up; (ii) the dearth of companions around higher lithium giants ($\rm A(Li) \gtrsim 2.7 \; dex$) could be due to stellar mergers or prior engulfment of companions.} Extended baselines and dedicated follow-up are required to confirm the orbital parameters of these systems and distinguish between mechanisms.
\end{abstract}


\section{Introduction}

Surface lithium is diluted when a star goes through the first dredge-up phase (FDU) on the lower red giant branch \citep[][]{Iben1967}. However, $\approx1\%$ of red giants are found to be enhanced in lithium, \ali $\geq 1.5 \rm \; dex$. Since their discovery by \cite{WallersteinSneden1982}, recent large-scale spectroscopic surveys have revealed thousands of \lirich giants \citep[e.g.,][]{gao_2019, casey_2019, Martell2021, sayeed_2024}. Many theories have been proposed to explain their formation such as enhanced mixing induced by the core helium flash \citep[e.g.,][]{kumar_2020, Zhang2021, Mallick_2023}, thermohaline mixing \citep[e.g.,][]{Lattanzio2015}, or engulfment of planets or sub-stellar companions \citep[e.g.,][]{Alexander1967, King1997, SiessLivio1999a, SiessLivio1999b, VillaverLivio2009, adamow_2012, carlberg_2012, gomez_2016a}. However, despite many studies on the formation of \lirich giants in the field \citep[e.g.,][]{singh_2019, deepak_reddy_2019, Gao2020, kumar_2020, deepak_2020, melinda_2021, wheeler2021, yan_2021, Ming-hao_2021, yan_2022, zhou_2022, chaname_2022}, \new{and in clusters \citep[e.g.,][]{Anthony-Twarog2013, Dorazi2015b, Kirby2016, BocekTopcu2016, Smiljanic2018, Anthony-Twarog2018, Sanna2020, Anthony-Twarog2020, Magrini2021, Magrini2021_2, Anthony-Twarog2021, Zhang2022, Sun2022, Carlos2023, Sun2023, Nagarajan2023, Anthony-Twarog2024, Nine2024, Anthony-Twarog2026, Holanda2026, Flaulhabe2026}}, their formation mechanism(s) remain disputed. 
 
Multiple theories suggest the presence of a companion as the cause of Li-enrichment of red giants, such as mass transfer from an asymptotic giant branch (AGB) star \citep[e.g.,][]{SackmannBoothroyd1992} or tidal spin-up induced by a companion \citep[e.g.,][]{casey_2019}. For instance, an intermediate-mass AGB companion would enrich the red giant via mass transfer, while a close companion, regardless of mass, would initiate enhanced mixing on the stellar surface retaining surface lithium; this enhanced mixing inside the red giant is required to initiate the Cameron--Fowler (CF) mechanism \citep[][]{cameron_1971}. Briefly, the CF mechanism occurs when beryllium ($^7$Be) is produced near the hydrogen-burning shell by $^3$He($\alpha,\gamma$)$^7$Be; the product of this reaction is then transported to cooler regions to create $^7$Li via electron capture. Without enhanced mixing, such as mixing induced via tidal interactions with a companion, $^7$Be stays in hotter regions of the stellar interiors where the subsequent lithium is destroyed by proton capture. Regardless, evidence of a companion around \lirich red giants would be strong support for stellar binarity as a formation mechanism for these objects.

Multiple studies have already tested the companion fraction of \lirich giants. \cite{Adamow2018} (hereon A18) conducted a radial velocity (RV) survey of 15 \lirich giants to search for low--mass substellar companions using data collected with the High Resolution Spectrograph (HRS) on \textit{Hobby--Eberly} Telescope (HET) \citep[][]{Ramsey1998, Tull1998}, and High Accuracy Radial Velocity Planet Searcher in the North Hemisphere \citep[HARPS--N, ][]{Cosentino2012}. Of the 15 red giants in their sample, they find a stellar binary frequency of 27\% (4/15), planetary mass companion frequency of $20-33\%$ ($3-5/15$), and an overall companion frequency of $47-60\%$; the latter two frequencies vary given that A18 argue for a planetary mass companion for two systems, but do not have sufficient data to confirm. However, they concluded that the binary fraction of their \lirich sample is similar to their \lin sample, but the planet fraction for the \lirich sample is twice that of the \lin sample. \new{Furthermore, although their sample consisted of both \lirich and \lin stars, eight targets have $\rm A(Li) = 1.4-1.6$ dex, near the edge of the \lirich threshold, and only two have a broadening velocity above 3 \kms.} Similarly, \cite{Castro-Tapia2024} combined radial velocity data from \gaia, \galah, and RAVE to test the binary fraction of \lirich giants on the first ascent red giant branch (RGB) as compared to those on the red clump (RC). They concluded red clump stars do not have a preference for being in binary systems, but they emphasize that the accuracy of their analysis is highly dependent on the precision of their radial velocity measurements. In a related study of \lirich giants in 32 open clusters, \cite{Tsantaki2023} find a correlation between red giants with higher lithium abundance and faster rotation. \new{However, their sample consists of only 14 \lirich stars of their total sample of 247 stars; furthermore, all stars are in young clusters and only two of the 14 \lirich stars have $\rm A(Li) > 1.6$ dex.} 

\new{Cluster studies are valuable in that they allow comparison of \lirich giants to stars of similar mass, age, composition, and evolutionary state, with the caveat that \lirich giants in clusters are rare and limited to small samples \citep[e.g.,][]{Anthony-Twarog2018, Anthony-Twarog2026}. However, dedicated effort has been made to study these objects in clusters with the WIYN Open Cluster Study (WOCS) which has revealed valuable information with respect to their age and evolutionary state. For instance, at an age of 1.5 Gyr, stars in NGC 7789 leaving the main sequence have a wide range in rotational speed, but spin down and achieve $\rm A(Li) \lesssim 1.4 \; dex$ once they reach the RC luminosity \citep{Anthony-Twarog2026}. This pattern is found for stars in clusters between 1.5 Gyr to $2 - 2.5 \rm \; Gyr$, and can be used to map the relationship between lithium and rotational speed from the subgiant phase to the base of the RGB \citep{Anthony-Twarog2018}. After its ascent up the RGB, lithium remains constant until the RGB bump then declines sharply along the upper RGB to the tip \citep{Charbonnel2010}. Beyond $2 - 2.5$ Gyr, stars at the turnoff spin down and reduce their surface lithium before entering the subgiant phase; in NGC 6819 with an age of 2.25 Gyr, red giants from the base to the tip of the RGB exhibited lithium non-detection with upper limits of  $\rm A(Li) = 0.7 \; dex$ at the base to $\rm A(Li)= -1 \; dex$ at the tip \citep{Deliyannis2019}. Furthermore, multiple studies reveal rapidly rotating \lirich giants \citep[e.g.,][]{Tayar2023}. However, despite mass and age constraints, some \lirich giants in clusters do not provide a measurable lithium abundance or constraint on evolutionary phase due to broadened lines and photometric variability \citep[e.g,][]{Anthony-Twarog2013, Anthony-Twarog2018, Anthony-Twarog2020, Anthony-Twarog2021}.}

Recently, \cite{sayeed_2024} investigated the formation mechanisms of \lirich giants by comparing stellar parameters and astrometric properties of \lirich giants relative to a \dg sample; they compared binary proxies between \lirich giants and \lin samples using RUWE (re-normalized unit weight error) values from \gaia's second data release (DR2) \citep[][]{gaia_collab, gaia_2018} and radial velocity errors from \gaia DR3 \citep[][]{paired, gaiadr3}, but found no significant difference in binary indicators between the two samples. They also found a subset of their sample with higher than average broadening velocities -- a possible indicator of enhanced rotation -- as well as enhanced \sprocess elements for giants at the \new{base} of the RGB, suggesting possible mass transfer from an intermediate-mass AGB companion. In addition, \cite{sayeed2025} used binary population synthesis code  \citep[\cosmic,][]{Breivik2020} to investigate resulting binary architectures of \lirich giants based on \galah observations. Under the assumption of a binary scenario where the red giant is enriched by an intermediate-mass AGB, they concluded that these systems are found at an average separation of $3.3\pm0.5$ AU. \new{Despite many studies aimed at uncovering the formation of \lirich giants by studying these objects in clusters or otherwise, their origins still remain unclear. Furthermore, in the context of theories relating to binaries, there does not exist a large, homogeneous sample of red giants that can be used to test for the companion fraction of \lirich giants in relation to a reference set. Until recently, direct evidence of companions around \lirich red giants has been limited due to small samples and insufficient RV precision and baselines.} 

Motivated by results from \cite{sayeed2025}, we use radial velocity data from \esp on the Very Large Telescope, and archival spectroscopy, to test for the companion fraction of \lirich giants. We use a custom rejection sampler \joker \citep{thejoker} to sample orbital parameters consistent with the data and derive companion masses. We make predictions of possible formation mechanisms based on our results.

\begin{figure}[t!]
    \centering
    \includegraphics[width=\linewidth]{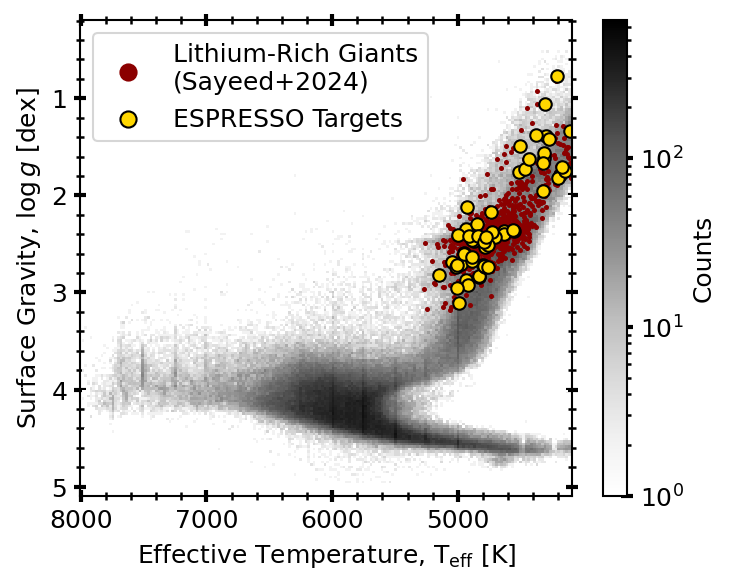}
    \caption{\new{Kiel} diagram showing the full sample of \lirich red giants from \cite{sayeed_2024} in red and \esp target sample from this work in gold overlaid on the full \galah sample (with good quality flags) for reference. Our \esp target sample covers the overall distribution of the reference \lirich sample along the red giant branch.}
    \label{fig:hrd_galah}
\end{figure}

\section{Data \& Methods}
\subsection{\lirich Sample}
\label{sec:lirich_def}

Our \esp targets were chosen from the parent sample of \lirich giants in \cite{sayeed_2024} (hereon S24). We briefly describe their sample here and refer the reader to the original paper for more details. S24 selected red giants from the third data release of the Galactic Archaeology with HERMES survey \citep[\galah DR3,][]{galah_survey, galah_sven_2021}. \galah is a high-resolution ($R\approx28,000$) spectroscopic survey with four bands in the optical ($ 4713-7887 \; \AA$). In addition to stellar parameters such as effective temperature, surface gravity, and metallicity, \galah measures 30 individual elemental abundances for 588,571 nearby stars. S24 define \lirich giants as objects with effective temperature $\rm T_{eff}=[3000,5730] \; K$, surface gravity $\log g = [-1,3.2] \rm \; dex$, and $\rm A(Li) \geq 1.5 \; dex$, where $\rm A(Li) = [Li/Fe] + [Fe/H] + 1.05$. 

\new{We note that the definition of ``Li-rich" remains disputed, and therefore will affect the final sample based on definition of choice. Traditionally, the threshold of $\rm A(Li) =1.5 \; dex$ is chosen given that a Population I star with similar metallicity to the Sun will have $\rm A(Li) = 3.3\; dex$ initially, similar to that of the interstellar medium. Post dredge-up, we expect the lithium in Population I stars to fall to $1.5 - 1.8$ dex \citep[][]{Lambert1980,Brown1989,Mallik1999,CharbonnelBalachandran2000}. In fact, \cite{karakas2014dawes} use the mass and composition of NGC 6397, combined with data from \cite{Lind2009} and models from \cite{Angelou2012} to conclude that the expected depletion of surface lithium before and after the FDU phase is $\sim 1 \; \rm dex$. Studies of \lirich red giants in clusters reveal that surface lithium post-FDU is a function of stellar mass and metallicity \citep[e.g.,][]{Sun2022, Anthony-Twarog2021, Tsantaki2023}. In addition, \cite{kumar_2020} argue that all low-mass stars undergo a lithium production phase between the tip of the RGB and the red clump, and therefore propose a new threshold of \ali $\geq -0.9 \rm \; dex$ to define \lirich red clump stars given that the traditional definition only accounts for 3\% of \lirich red clump stars. We chose the traditional definition of \lirich to define our sample consistent with S24. Unlike in clusters, our sample of field stars does not contain information on the turnoff mass, and so it is possible that our sample is biased towards higher mass stars (see Section \ref{sec:caveats} for a full discussion).}

\begin{figure}[t!]
    \centering
    \includegraphics[width=\linewidth]{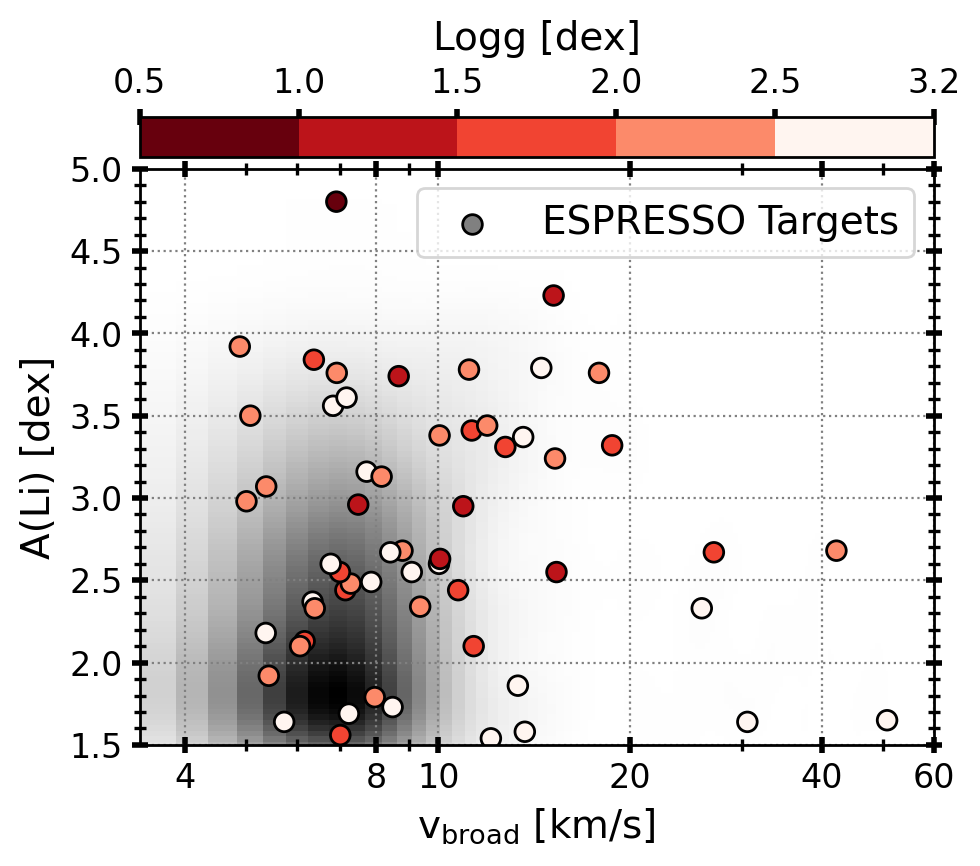}
    \caption{Lithium abundance (\ali) of our \esp targets as a function of broadening velocity (\vbroad) colored by their surface gravity (\logg). The complete sample of \lirich sample in \galah from S24 is shown as a density distribution for reference. Our \esp sample is diverse, spanning evolutionary state, lithium abundance, and broadening velocity.}
    \label{fig:hrd}
\end{figure}

Figure \ref{fig:hrd_galah} shows the \lirich sample from S24 (red points) and our \esp targets (gold points) on a Kiel diagram with the full \galah sample (grey points) shown for reference. Our \esp targets were chosen to span the range of \lirich red giants in \galah as well as surface gravities. Given that S24's \lirich sample includes stars with high broadening velocities (\vbroad), we chose our \esp targets to be representative of the \lirich sample. Figure \ref{fig:hrd} shows the targets on a Li--\vbroad plane, colored by \logg. Our target sample consists of \lirich giants with $\rm A(Li)=1.5-4.8 \; dex$, $ \log g=0.8-3.1\;\rm dex$, and $ v_{\rm broad}=4.9-50.7 \;\rm km \; s^{-1}$.

\begin{deluxetable*}{cccccccc}[t!]
\tabletypesize{\scriptsize}
    \tablecolumns{8}
    \tablewidth{0pt}
    \tablecaption{Properties of \target red giants in three \esp programs where stellar properties are from \galah. \label{tb:sample} }
    \tablehead{
    Red Giant ID & GALAH ID & \gaia DR3 ID & Object Name & \ali & [Fe/H] & \logg &  \vbroad
    \\
    & & & &  [dex] & [dex] &[dex] & [\kms] 
    }
    \startdata
    \multicolumn{8}{c}{\textbf{P112}} \\\hline
    1 & 150107004201104 & 5632958769593943424 & UCAC4 297-057956 & 3.41 & -0.28 & 1.79 & 11.29 \\
    2 & 151230003202196 & 5483084721864794752 & UCAC4 159-007674 & 2.10 & -0.28 & 1.96 & 11.37 \\
    4 & 161013005401317 & 3142806554658429952 & TYC 763-2824-1 & 2.68 & -0.47 & 2.18 & 8.79 \\
    ... & ...  & ...  & ...  &...  & ... & ... & ... \\\hline
    \multicolumn{8}{c}{\textbf{P113}} \\\hline
    1 & 140806001701013 & 5909827209206002688 & UCAC4 139-176959 & 3.32 & -0.43 & 1.82 & 18.76 \\
    2 & 170802003201248 & 5911546192556778624 & UCAC4 143-204046 & 1.56 & -0.08 & 1.75 & 7.02 \\
    3 & 150429004102159 & 6043491023860004096 & 2MASS J16094423-2557437 & 2.55 & -0.79 & 1.33 & 15.34 \\
    ... & ...  & ...  &...  &...  & ... & ... & ... \\\hline
    \multicolumn{8}{c}{\textbf{P116}} \\\hline
    1 & 170218001601395 & 2884674278524180608 & UCAC4 263-008944 & 2.44 & -0.61 & 1.76 & 7.15 \\
    2 & 160330002101014 & 5445499875375929728 & UCAC4 268-056168 & 2.13 & -0.56 & 1.73 & 6.17 \\
    3 & 150209002201366 & 2939800046333110272 & UCAC4 349-013340 & 3.13 & -0.22 & 2.47 & 8.15 \\
    ... & ...  & ...  &...  &...  & ... & ... & ... \\
    \enddata
    \tablecomments{Full table in machine-readable format is available online.}
\end{deluxetable*}

\begin{deluxetable*}{ccccc}[t!]
    \tablecolumns{5}
    \tablecaption{Radial velocity data from \esp, \galah, \apogee, and RAVE for \target targets in our sample. \label{tb:rv_data}}
    \tablehead{
    GALAH ID & Time & Radial Velocity & Radial Velocity Error & Source
    \\
    & [JD] & [\kms] & [$\rm m \; s^{-1}$] &
    }
    \startdata
    140311007101261 & 2460388.50 & 23.50 & 49 & ESPRESSO \\
    140311007101261 & 2460303.80 & 23.30 & 18 & ESPRESSO \\
    140311007101261 & 2460344.90 & 23.20 & 51 & ESPRESSO \\
    140311007101261 & 2456728.10 & 23.70 & 82 & GALAH \\
    140707003101315 & 2458732.80 & $-$44.80 & 55 & APOGEE \\
    140707003101315 & 2458731.80 & $-$44.60 & 58 & APOGEE \\
    140707003101315 & 2460473.90 & $-$44.40 & 50 & ESPRESSO \\
    140707003101315 & 2460458.90 & $-$44.50 & 50 & ESPRESSO \\
    140707003101315 & 2460534.70 & $-$44.10 & 62 & ESPRESSO \\
    140707003101315 & 2460493.90 & $-$44.40 & 42 & ESPRESSO \\
    140707003101315 & 2460517.60 & $-$45.00 & 69 & ESPRESSO \\
    140707003101315 & 2456846.20 & $-$44.80 & 112 & GALAH \\
    ... & ... & ... & ... & ...
    \enddata
    \tablecomments{Full table in machine-readable format is available online.}
\end{deluxetable*}

\subsection{\esp Data}
We obtain spectra with the \'Echelle Spectrograph for Rocky Planets and Stable Spectroscopic Observations \citep[\esp;][]{pepe2021}, a high-resolution spectrograph with $\mathrm{R \sim 140,000-190,000}$ on the Very Large Telescope (VLT). \esp covers the wavelength range $\mathrm{378.2 - 788.7} \rm \; nm$, and is designed to reach the ultimate precision of 10 cm s$^{-1}$ over 10 years \citep{pepe2021}. Our three programs, 112.25UC.001 (hereon P112), 113.26Q8.001 (hereon P113), and 116.290U.001 (hereon P116) ran over October 2023--March 2024, April 2024--September 2024, and November 2025--April 2026, respectively, collecting data for 25 red giants in each period in 1-UT mode. Throughout this work, we use the shorthand P112-X, P113-X or P116-X to refer to objects where P112/P113/P116 indicates the program followed by the red giant ID, X.

For P112, we obtained only 54 of the requested 125 observations due to lower rank of the proposal resulting in only ten stars with sufficient data; however, one of the targets was removed from our sample after follow-up analysis revealed it was a young stellar object. Furthermore, we removed another target from our P112 sample as it was a repeat target also found in P116; therefore, we combined the P112 and P116 RV measurements for this target and included it in P116 sample. All requested observations were obtained for P113 and P116. In P112 and P113, we requested five spectra over six months--but six spectra over six months in P116--with a spacing of $4-6$ weeks, and 300 seconds exposure to achieve a signal-to-noise ratio (SNR) of at least 25 per pixel at 550 nm. For our analysis, we excluded targets with fewer than four observations or if a given observation received a C grade with a final sample of \target total giants: eight in P112, and 25 each in P113 and P116. \new{Unlike P112 and P113 which contained targets with lithium values spanning the full range of \lirich giants in S24, P116 specifically targeted red giants with high lithium abundance, \ali $> 2.0$ dex, to further probe the correlation between binarity and lithium abundance.} Table \ref{tb:sample} includes object names, $\rm A(Li)$, \vbroad, \logg, and metallicity from \galah for all \target targets.   

Radial velocity measurements were obtained from \esp's own Data Reduction Software\footnote{\url{https://www.eso.org/sci/software/pipelines/espresso/espresso-pipe-recipes.html}} (DRS) versions 3.0.0 (P112), 3.1.2 (P113), and 3.3.13 (P116). The DRS cross-correlates the observed spectrum with a weighted binary mask based on the spectral type of the target \citep[e.g.,][]{Pepe2002, pepe2021}; the radial velocity is then measured by fitting a Gaussian to the cross-correlation function. The average SNR of the spectrum at 550 nm ranged from 12--94 per star for P112 ($\mu=31$), 5--106 per star for P113 ($\mu=25$), and 4--76 for P116 ($\mu=20$). The average uncertainty in RV per star ranged from $\mathrm{1-66 \;m \;s^{-1}}$ for targets in P112, $\mathrm{2-112 \;m \;s^{-1}}$ for targets in P113, and $\mathrm{3-83 \;m \;s^{-1}}$ for targets in P116. 

We supplement the \esp RV data with RV measurements from three additional surveys: \galah DR3 \citep[][]{galah_survey}, \apogee DR17 \citep[][]{apogee, apogee_dr17}, and RAVE DR6 \citep[][]{rave_dr6a, rave_dr6b}. We use the best-method radial velocity\footnote{\texttt{rv\_galah}} from \galah from \cite{Zwitter2021}
which are corrected for gravitational redshift (\texttt{rv\_obst}). \cite{Zwitter2021} provide RV measurements for \galah stars by creating noiseless spectra on a grid belonging to the same \teff, \logg, \feh bin with a width of 50 K in effective temperature, 0.2 dex in gravity, and 0.1 dex in metallicity. A given spectrum is then cross-correlated with these noiseless spectra to measure the radial velocity. \new{All of our \target targets have \esp and \galah data, nine targets also have RAVE data while four targets have \esp, \galah, and \apogee data, and only one target has data from all four sources.}

\esp, \galah, and \apogee provide barycentric RVs while RAVE provides heliocentric RVs. However, the mean error in RAVE RVs across the nine targets is $\rm 1.15 \; km \; s^{-1}$. Given that RAVE uncertainties are on the order of or larger than the expected RV variation, we ignore barycentric correction to RAVE RV data. Furthermore, we do not perform zero-point correction after combining data across RV surveys given that zero-point corrections are an order of magnitude smaller than expected RV variation for objects in our sample, and on the same order as the typical uncertainties in the respective survey. For instance, the \galah zero-point offset from \gaia is order of $40 \; \rm m \;  s^{-1}$ \citep[e.g.,][]{Tsantaki2022}, while between HARPS and \esp is $130 \; \rm m \;  s^{-1}$ \citep{Faria2020}. Furthermore, \cite{Li2023} find a zero-point offset of $149 \; \rm m \; s^{-1}$ between APOGEE DR17 and RAVE, and $31 \; \rm m \; s^{-1}$ between APOGEE DR17 and \galah, while typical uncertainties in \galah radial velocities are $100 \rm \; m \; s^{-1}$. Usually, zero-point offsets are incorporated in the Keplerian model as a free parameter; however, we are unable to fit for them as we have only one RV measurement from \galah and RAVE. 
Table \ref{tb:rv_data} provides radial velocity measurements for all targets across four surveys. 

\subsection{\joker}


To sample possible orbital solutions, we used \joker\footnote{Named after Johannes Kepler. Pronounced \textit{YO-ker}.}, a custom Monte Carlo (rejection) sampler specifically designed for sparse radial velocity measurements of a binary or exoplanet system \citep{thejoker}. A major advantage of \joker is that it produces independent samples even in cases with a multi-modal likelihood; if the probability density function is multi-modal, \joker produces all relevant solutions. We briefly describe \joker below, but refer the reader to the original paper for a more detailed discussion \citep[e.g.,][]{thejoker, apw_2020}. 

\begin{figure*}[t!]
    \centering
    \includegraphics[width=\linewidth]{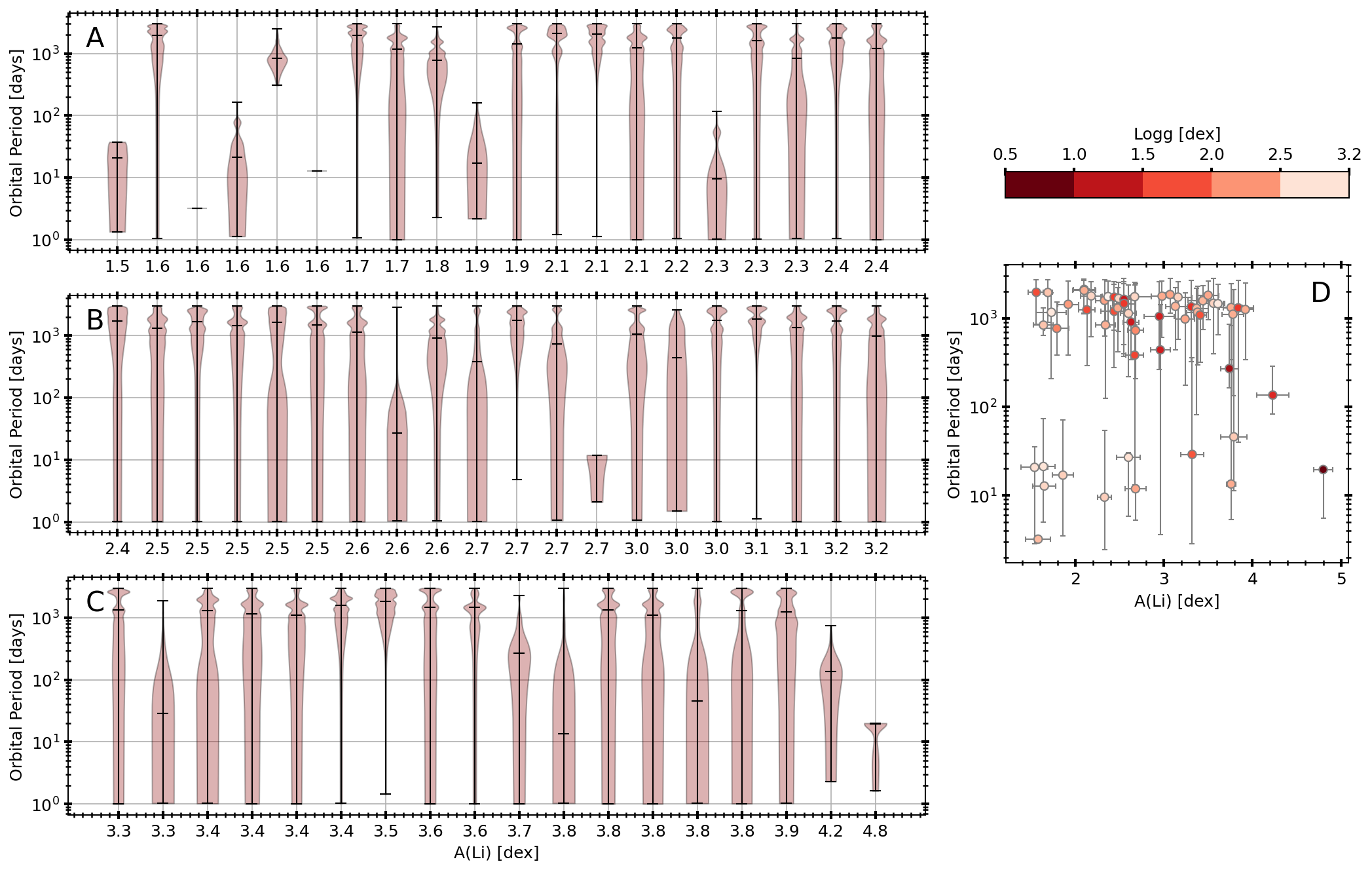}
    \caption{\textit{Panels A -- C:} Orbital periods returned from \joker's posterior samples for our \target targets as a function of lithium; the sample is shown on three subplots for clarity. The two uni-modal targets in Panel A are included for completeness, but shown as flat lines. \textit{Panel D:} The median period for each target, colored by its surface gravity, as a function of lithium abundance where the error bars represent the 16th and 84th percentiles. Since our period prior is log-uniform, any period distributions inconsistent with our choice of prior reveal orbital features of our systems. Panel D reveals no correlation between orbital period and lithium abundance.} 
    \label{fig:violin_plot}
\end{figure*}

\joker describes a radial velocity curve by six parameters: the period ($P$), eccentricity ($e$), pericenter phase ($\phi_0$) and argument ($\omega$), velocity semi-amplitude ($K$), and barycenter velocity ($v_0$). Analytically, the radial velocity $v$ at time $t$ is given by, 
\begin{align}
    v(t, \theta) &= v_0 + K [\cos(\omega + f) + e \cos\omega] \\ 
    \cos f &= \frac{\cos E - e}{1 - e\cos E} \\
    M &= \frac{2\pi t}{P} - \phi_0,\\
    M &= E - e\sin E
\end{align}
where $f$ is the true anomaly, $E$ is the eccentric anomaly, and $M$ is the mean anomaly. Four of the parameters -- $P$, $e$, $\omega$, $\phi_0$ -- are non-linear, and two -- $K$, $v_0$ -- are linearly related to the radial velocity, $v(t)$.
When the priors on $K$ and $v_0$ are Gaussian, \joker analytically marginalizes out the linear parameters during rejection sampling, thus reducing the computational cost of the rejection sampling procedure. The linear parameters are then sampled directly using the analytic posterior probability density function (pdf) over these parameters. \joker also incorporates and infers an unknown radial velocity scatter as an overall ``jitter'' parameter, $s^2$, which we also vary as a free parameter.

We use \textit{The Joker's} default prior pdf for all six parameters as detailed below: 
\begin{align}
    p(\ln P) &= \mathcal{U}(\ln P_{\rm min}, \ln P_{\rm max}),\\
    p(e) &= \frac{\Gamma(a+b)}{\Gamma(a)\Gamma(b)} e^{a-1}[1-e]^{b-1},\\
    p(\omega) &= \mathcal{U}(0, 2\pi) \; [\rm rad] \\ 
    p(\phi_0) &= \mathcal{U}(0, 2\pi) \; [\rm rad] \\
    p(K) &= \mathcal{N}(0, \sigma_K^2) \\
    p(v_0) &= \mathcal{N}(0, \sigma_{v_0}^2)\\
    p(\ln s) &= \mathcal{N}(\mu_s, \sigma_s^2)
    ,
\end{align}
where $\mathcal{U}(x_1, x_2)$ is the uniform distribution between $x_1$ and $x_2$, and $\mathcal{N}(\mu, \sigma^2)$ is the normal distribution with mean $\mu$ and variance $\sigma^2$. The prior for $e$ is described as a Beta distribution with $a=0.867$ and $b=3.03$ \citep[][]{Kipping2013, thejoker}. For linear parameters $K$ and $v_0$, the prior is a broad Gaussian centered on zero with a large input variance. We used a prior on $P$ from 1--3000 days, $\sigma_K$ of $\rm 20 \; km \; s^{-1}$, and $\sigma_{v_0}$ of $\rm 100 \; km \; s^{-1}$, and set $s$ as a free parameter with a log-Normal prior ($\mu$, $\sigma_s$) = (0, 0.5 $\rm km \; s^{-1}$). The prior and posterior size was set to 10 million and 10,000 samples, respectively; however, \joker can return fewer samples. For systems with a uni-modal solution, we chose two Monte Carlo chains. Finally, we also set a minimum RV error on all measurements to $\rm 100 \; m \; s^{-1}$ for any RV uncertainties less than this threshold.

We experimented with different priors for $P$ and $e$, namely uniform distribution for both as compared to the default choice of uniform in logarithmic space and Beta distribution, respectively. We found that the choice of prior did not significantly impact the final $P$ and $e$, and used the default priors for both of these parameters. It is important to note that the binary status of our systems is undetermined, and so it is possible \joker is being used to constrain binary orbital parameters for a single star system. 

\begin{figure*}
    \centering
    \includegraphics[width=0.8\linewidth]{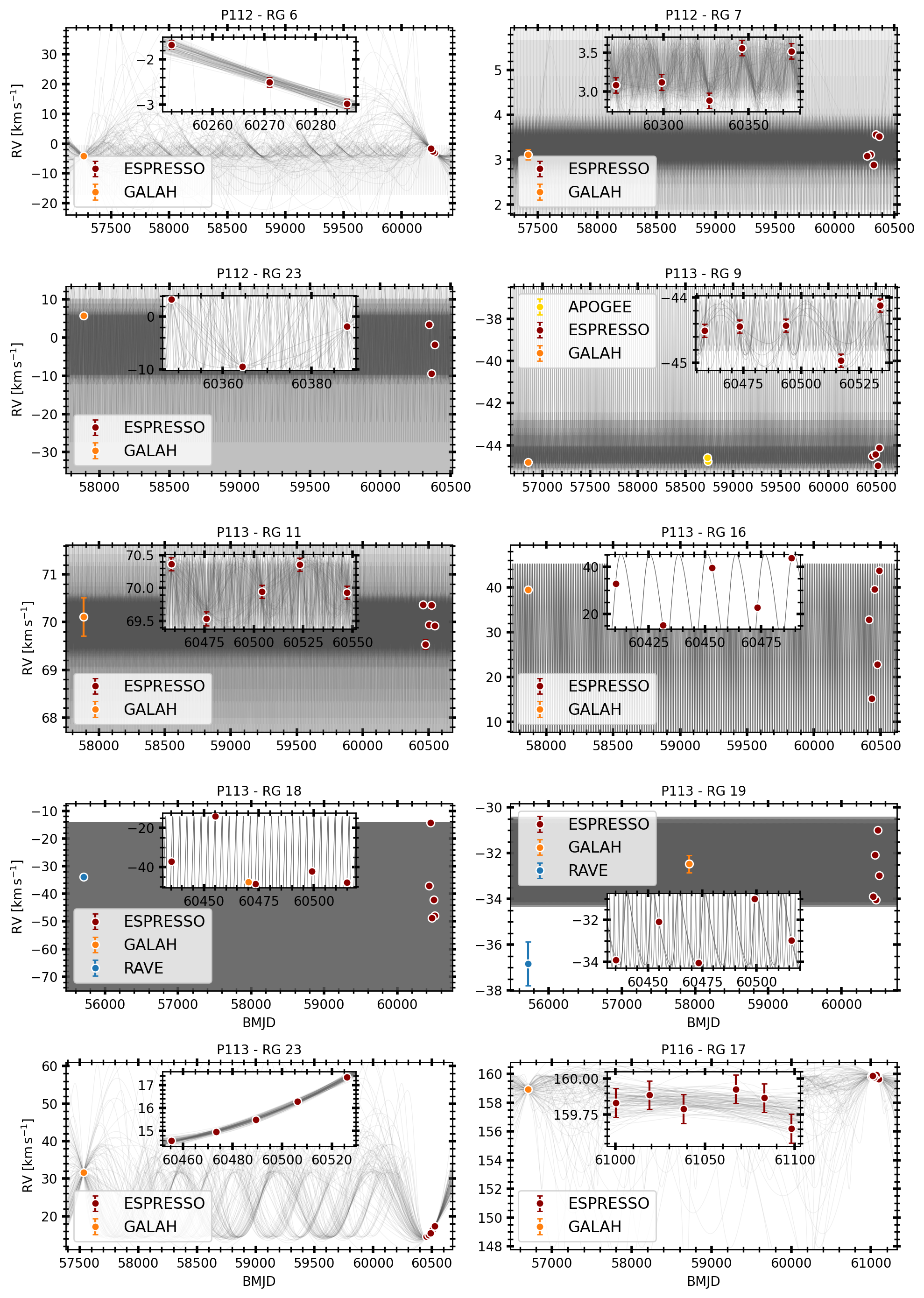}
    \caption{Samples from \joker after rejection sampling for our \total binary candidates. The curves show \joker's samples, and the points show the respective radial velocity data. The inset plot shows the \esp data points zoomed in. For the majority of our binary candidates, many diverse solutions are consistent with  current data motivating the need for extended baselines to constrain orbital parameters of our systems.} 
    \label{fig:joker} 
\end{figure*}

\section{Results}
Figure \ref{fig:violin_plot} shows the samples of orbital periods returned by \joker (after rejection sampling) for each system as a function of lithium abundance. Of the \target targets, two were uni-modal while the rest were multi-modal ranging from $3-3144$ samples. Figure \ref{fig:violin_plot}D shows no correlation between lithium abundance and the median orbital period of each system; to find potential binary candidates, we implemented a selection criterion using the framework developed by \cite{Hekker2008} (hereon H08).

H08 conducted a radial velocity survey of 179 $K$ giants using the Coude Auxiliary Telescope at UCO/Lick Observatory to investigate the source of radial velocity variations in giants. They found that single red giants show intrinsic radial velocity variations of $\sim100 \rm \; m \;s^{-1}$ which encompasses signal from stellar oscillations, granulation, super-granulation, and short-term and long-term variability such as rotation and magnetic cycles \citep[][]{Yu2018}. H08 measured non-periodic RV variation of single stars and established a lower limit of expected RV variation for single giants. Similar to previous findings \citep[e.g.,][]{Hatzes1998}, they found that the non-periodic radial velocity variation of giants decreases with increasing surface gravity described by the following relation,

\begin{equation}
    \log(v_r \; [\rm m/s]) = a \cdot \log(g) + b
    \label{eq:K}
\end{equation}
where $a=-0.60\pm0.04 \rm \; dex^{-1}$ and $b=3.31\pm0.10$. For instance, RV semi-amplitudes below $\sim \rm 550 \; m \; s^{-1}$ for red giants at $\log g = 2 \rm \; dex$ are likely to be intrinsic due to convective motion, and not bona fide binaries. For each of our targets with a given surface gravity from \galah, we calculated the expected semi-amplitude. In this work, we classify a red giant as a binary candidate if its 5th percentile semi-amplitude is above a factor of 3.5$\sigma$ of the expected semi-amplitude, where $\sigma$ is the $\sigma$ on the coefficients in Equation \ref{eq:K}. We note that this choice of selection criterion is conservative and chosen specifically to reduce false positives, but will likely produce false negatives as well. Based on this criterion, we find \total  binary candidates, including the two uni-modal targets, in our sample of \target targets with a recovery rate of 17\%. 

Figure \ref{fig:joker} shows the orbital solutions from \joker for our \total binary candidates, where the first 100 samples are shown for the eight multi-modal systems. The inset plot is zoomed into \esp data for clarity. The two targets with a uni-modal posterior have an orbital period of $\approx 3 \rm \; days$ and $\approx 13 \rm \; days$  (discussed in detail in Section \ref{sec:uni-modal}). Figure \ref{fig:joker} demonstrates the variety and number of possible orbital solutions given our current data set, and the need for extended baseline with \esp. 

\begin{figure}[t!]
    \centering
    \includegraphics[width=\linewidth]{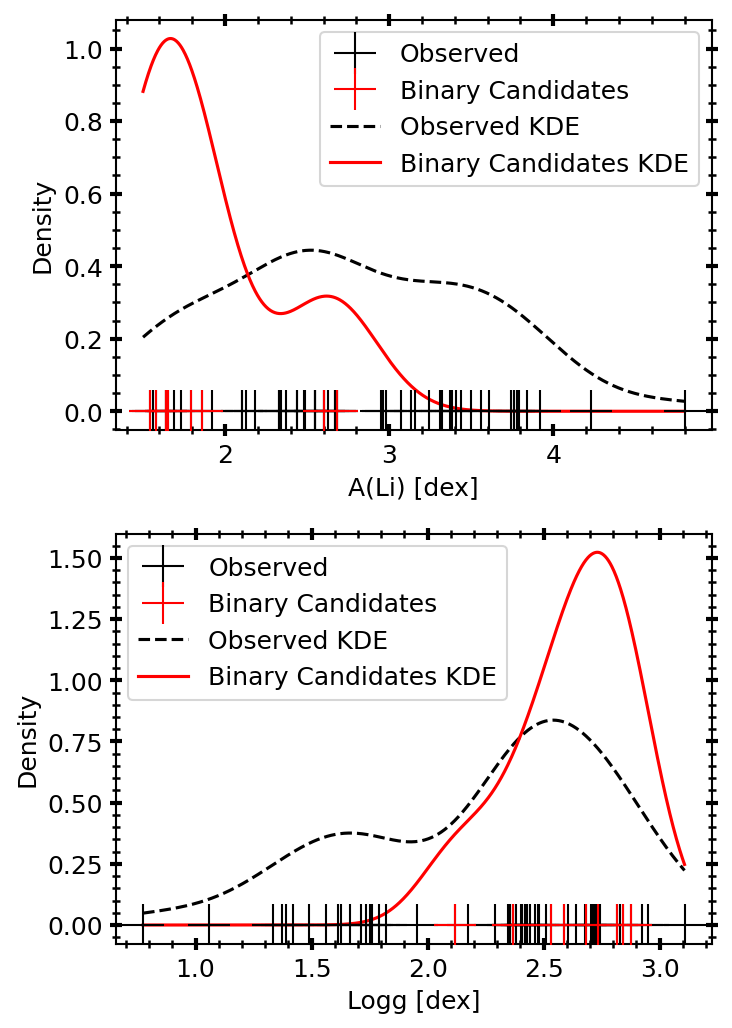}
    \caption{Kernel Density Estimate of the lithium abundance (top) and surface gravity (bottom) of \esp targets observed in black, dashed line and the \total targets found to be binary candidates with \joker in red, solid line. The markers at the bottom of each panel represent the distribution of lithium and \logg for observed (black) and binary candidates (red). Targets with lower lithium and higher \logg are over-represented in our binary candidate sample as compared to the target sample.}
    \label{fig:bin_frac}
\end{figure}


\begin{figure*}[t!]
    \centering
    \includegraphics[width=\linewidth]{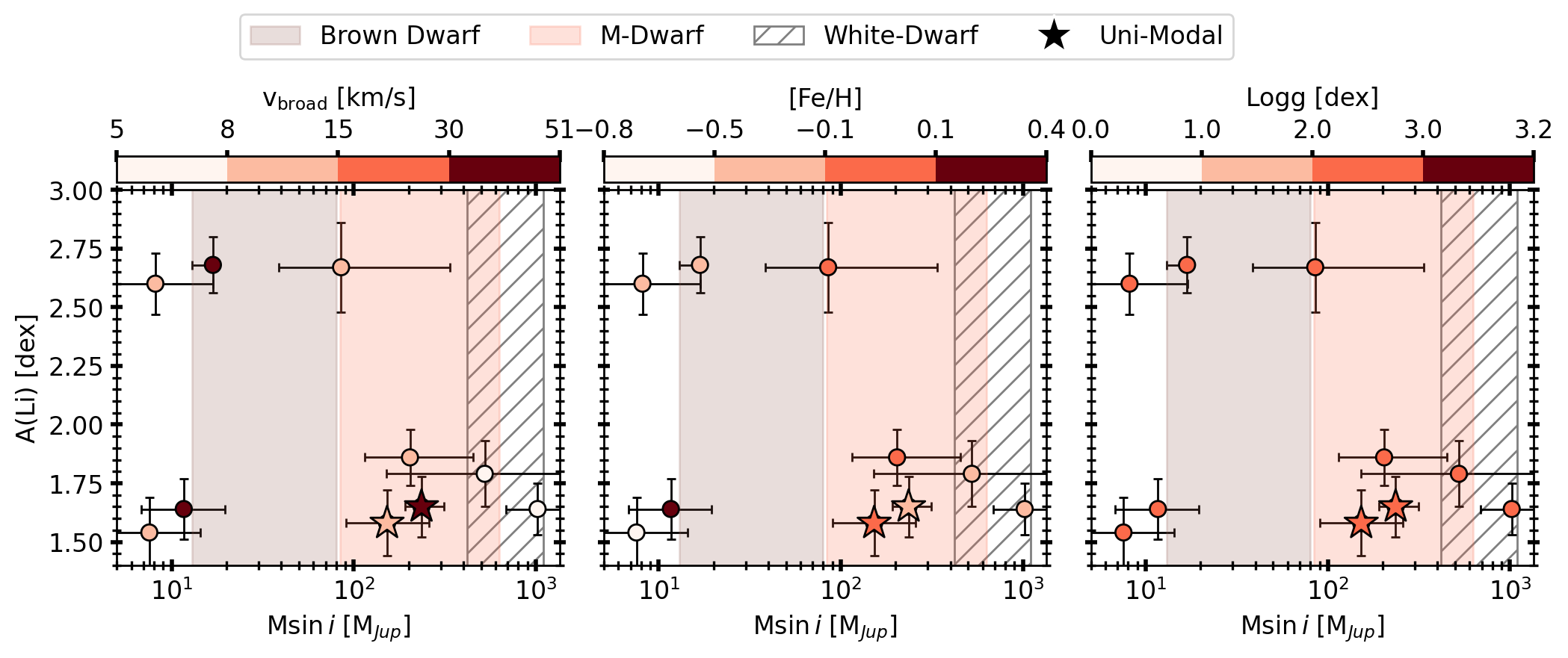}
    \caption{Lithium abundance as a function  of \msini (in $\rm M_{Jup}$) colored by \vbroad (left), metallicity (middle), and surface gravity (right). The brown, orange, and hatched areas indicate the brown dwarf, $M-$dwarf, and white dwarf regions, respectively, for reference. The companion masses of our \total binary candidates show no commonality nor any correlation with stellar parameters suggesting multiple formation pathways for Li-enrichment.}
    \label{fig:msini_mix}
\end{figure*}

In Figure \ref{fig:bin_frac}, we compare the distribution of lithium abundance (top) and surface gravity (bottom) as a Kernel Density Estimate between our binary and non-binary candidates. We find that giants with lower lithium abundance and higher \logg are over-represented in our binary candidate sample, despite our parent target sample spanning a wide range of \logg, \vbroad, and lithium abundance.

For instance, seven of our \total binary targets fall in the lithium range of $\rm A(Li) = 1.5 - 2.0 \; dex$. Of the eleven targets observed in this lithium bin, seven are binary candidates (or a rate of 64\%), and \new{their \vbroad ranges from $5.7-50.7 \rm \; km \; s^{-1}$, with a median of $8.5  \rm \; km \; s^{-1}$.} To calculate the significance of our results within $\rm A(Li) = 1.5 - 2.0 \; dex$, we can describe the detection of a binary candidate as a Bernoulli trial with probability of 10/58 ($\sim17\%$). Therefore, the probability of obtaining seven or more binary candidates out of eleven within this lithium bin is 0.08\%. The next highest number of giants targeted was thirteen with $\rm A(Li) = 2.5 - 3.0 \; dex$; however, we find only three binary candidates in this lithium range, or a binary rate of 23\%. One star has a $v_{\rm broad} = 42 \rm \; km \; s^{-1}$, second highest in our original target sample, and the rest have $v_{\rm broad} = 8- 10 \rm \; km \; s^{-1}$, similar to the average \vbroad of a typical red giant \citep[e.g.,][]{Carney2008, Patton_2023}.


If astrophysical, the higher binary fraction for giants with $\rm A(Li) = 1.5-2.0\; dex$ could be strong support for the formation of \lirich giants as a consequence of binarity. Interestingly, \cite{sayeed2025} argue that giants with $\rm A(Li) = 1.5-2.2\; dex$ could be lithium enhanced by inheriting lithium from their main--sequence phase without a need for external mechanisms (ie. mass transfer in a binary system) while those above this threshold ($\rm A(Li) \gtrsim 2.2\; dex$) became lithium enhanced due to binarity. However, we find only three possible binary candidates of the 38 stars above this threshold suggesting that another mechanism is responsible for lithium enhancement of the 38 red giants. \new{Interestingly, the lithium abundance of our \total binary candidates can be grouped in two groups: seven of them fall within $\rm A(Li) = 1.5-1.9\; dex$, and the remaining three stars have almost identical lithium abundance: $\rm A(Li) = 2.60  \; dex$, $\rm 2.67  \; dex$, and $\rm 2.68  \; dex$.}

\new{Furthermore, among the three \esp programs, P116 has the lowest binarity rate (1/25) as compared to P112 (3/8) and P113 (6/25). The baseline and observing cadence are similar between the three programs as well as the uncertainty in radial velocity measurements: $0.017 \pm 0.002 \; \rm km \; s^{-1}$, $0.033 \pm 0.003 \; \rm km \; s^{-1}$, and $0.026 \pm 0.002 \; \rm km \; s^{-1}$ for P112, P113, and P116, respectively. However, P116 was specifically targeted towards red giants with higher lithium abundance, $\rm A(Li) \gtrsim 2.0 \; dex$, as compared to the other two programs where the lower lithium limit was 1.5 dex. Given the low companion fraction of P116 with \esp, we argue for a possible anti-correlation between lithium abundance and binarity.}

In addition, all our \total binary candidates have $\log g = 2.0-3.0 \rm \; dex$, with a binary rate of 10/39 or 26\% in this \logg bin. The KDE in Figure \ref{fig:bin_frac} suggests that our binary candidates have higher surface gravity as compared to the overall target sample. Using the Binomial distribution, where $p=10/58$ and $n=39$, the probability of obtaining ten or more binary systems within $\log g = 2.0-3.0 \rm \; dex$ is 12\%.

Finally, we also compared the 30 elemental abundances of the binary and non-binary candidates against the \lirich giants sample from \cite{sayeed_2024} (see Figure \ref{fig:abundance} in Appendix). However, given the small number of binary candidates, we are unable to make population level conclusions in regards to chemical peculiarities.

\begin{figure*}[t!]
    \centering
    \includegraphics[width=\textwidth]{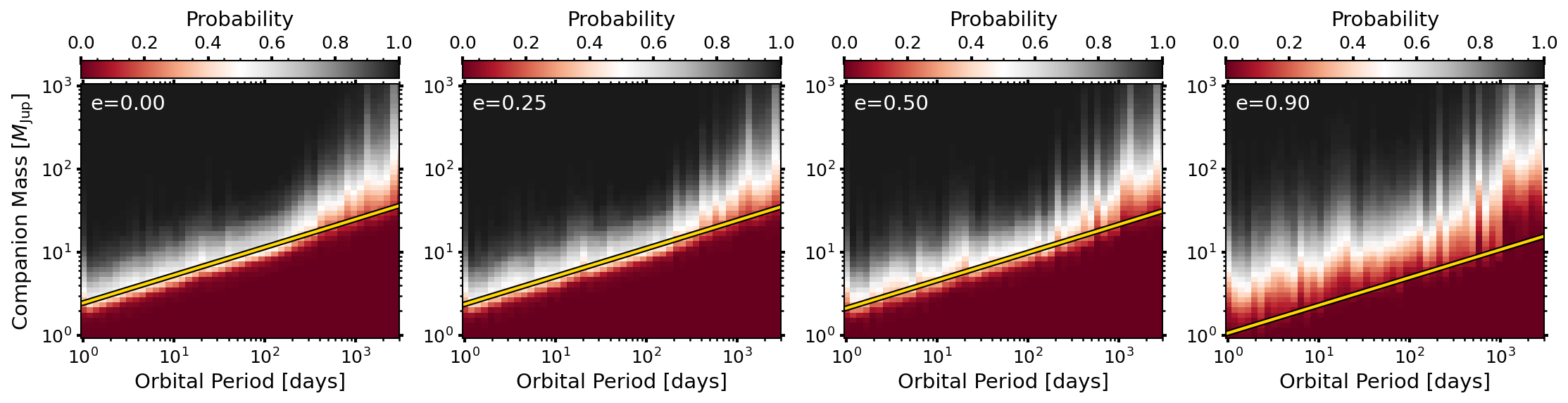}
    \caption{Average probability of observing a system with a given companion mass and orbital period using epochs from \notbin red giants with no indication of binarity. Panels indicate increasing eccentricity from 0 to 0.9, and the color bar indicates the probability. The yellow line shows a constant semi-amplitude of $0.5 \rm \; km \; s^{-1}$ for reference. The vertical structure present in non-zero eccentricities are due to time sampling where some orbital solutions are better for detecting systems with specific periods. } 
    \label{fig:non_binary}
\end{figure*}

\subsection{Companion Masses} \label{sec:companion_masses}

While \joker does not directly measure the companion mass, we can use posterior samples to put lower limits on the companion mass, \msini, using the following relation,

\begin{equation}
    M\sin{i} = (M_\star+M\sin{i})^{2/3}K \Bigg( \frac{P}{2\pi G}\Bigg)^{1/3}(1-e^2)^{1/2}
    \label{eq:msini}
\end{equation}

where $K$ is the semi-amplitude, $M\sin i$ is the companion mass, $M_\star$ is the stellar mass from \galah, $P$ is the period, and $e$ is the eccentricity, where $K$, $P$, and $e$ are measured with \joker. Given the uncertainty in stellar mass from \galah, we draw samples of $M_\star$ where the mean and standard deviation are the stellar mass and respective error from \galah DR3 \citep{sharma_bstep}. We then use samples of $K$, $P$, and $e$, and report the median \msini as the companion mass, where the 16th and 84th percentile values represent uncertainties on \msini.

In Figure \ref{fig:msini_mix}, we plot the lithium abundance of the red giant as a function of the companion mass, colored by \vbroad, stellar metallicity, and surface gravity from \galah. The colour bands indicate the brown dwarf range ($\rm 13-80 \; M_{Jup}$) and the M--dwarf range ($\rm 0.08-0.6 \; M_{\odot}$). Three systems fall within the planet companion region, one system falls within the brown dwarf region, and six systems fall within the stellar mass companion region ($M \sin i>0.08 \; M_\odot$). We see no correlation between companion mass and lithium abundance, nor a correlation with companion mass and other stellar parameters such as \vbroad, stellar metallicity, and surface gravity.  

\new{A possible formation mechanism for \lirich giants is mass transfer from an intermediate-mass AGB companion ($4-8 \; \rm M_\odot$) that produced lithium in its interior via Hot Bottom Burning (HBB). We calculated the progenitor mass of each companion using the initial-final mass relation (IFMR) \citep[][]{ElBadry2018, Cummings2018, Cunningham2024} to verify if these objects were intermediate-mass AGB stars. Based on our companion masses ranging from $M = 0.08-1.0 \; \rm M_\odot$, and using the following IFMR from \cite{Cummings2018},}
\new{
\begin{equation}
    \begin{split}
        M_{\rm f} \; = \; &(0.107 \pm 0.016) \times M_{\rm i}+(0.471\pm0.077) \, M_\odot \\
        & (3.60 \, M_\odot < M_{\rm i} < 7.20\, M_\odot)
    \end{split}
    \label{eq:imfr}
\end{equation}
}

\new{only one target would be eligible for this scenario with a progenitor mass of $\rm 4.8 \; M_\odot$, sufficiently high enough for an intermediate-mass AGB. \cite{Shahaf2025} studied spectroscopic binaries in an open cluster with a red giant and white dwarf companion with orbital periods $0.5-20$ years. They find that white dwarfs in binary systems are $\sim 20\%$ less massive than those in single systems. Correcting for this mass loss by inflating our final white dwarf companion masses by 20\%, we derive a progenitor mass of $M = 6.7 \; \rm M_\odot$, which again suggests an intermediate-mass AGB companion. }

\new{Furthermore, AGB mass transfer is also thought to be the dominant formation channel of blue straggler stars in open clusters older than $\sim 1 \; \rm Gyr.$ \cite{Nine2024} find that $43\pm11\%$ of blue stragglers in NGC 7789, NGC 6819, and M67 are barium-enriched, and that blue stragglers in older clusters are predominantly in long-period binaries. The accretors in these systems evolve onto the giant branch with rapid rotation and chemical peculiarities. However, AGB donors near the turnoff mass of these clusters $\leq 3 \; \rm M_\odot$ do not undergo HBB, suggesting that these stars are enhanced in $s-$process elements without lithium production. If our binary candidates were enriched by an AGB companion, the lack of [Ba/Fe] enhancement in Figure \ref{fig:abundance} could be due to a high-mass donor; however, the sample in Figure \ref{fig:abundance} is too small to make population level conclusions.
}


\subsection{Detection Sensitivity}

To better understand the detectability of systems with \esp, we performed an injection-recovery analysis using the observed times of the \notbin systems classified as non-binaries as input. We simulate RV curves for a range of 25 $\omega$ and 25 $\phi_0$ values between $0-2\pi$, and four eccentricities ($e\in \{ 0, 0.25, 0.50, 0.90\}$) over a grid of orbital periods and companion masses. For each target, we calculate the observed semi-amplitude as well as the expected semi-amplitude from Equation \ref{eq:K}. We label targets as observable if the observed semi-amplitude exceeds the threshold expected semi-amplitude. Figure \ref{fig:non_binary} shows the average probability of detecting a system as a binary with a given companion mass and orbital period, where the probability is defined as the number of systems that would be observed out of the 625 simulated solutions, marginalized over $\omega$, $\phi_0$, and the \notbin systems. Based on Figure \ref{fig:non_binary}, targets with periods and companion masses in the top left would have been detected by our pipeline given the data, while those in the red region in the bottom right would have not been detected given their low RV signal. Objects in the black and white region could have been detected based on the object's orbit. In summary, our pipeline cannot detect any companions below $\sim 40 \rm \; M_{Jup}$ on circular orbits above $\sim3000$ days.

\new{Using binary population synthesis \citep[\cosmic,][]{Breivik2020}, \cite{sayeed2025} predicted that binary systems with a white dwarf that was previously an intermediate-mass AGB companion would have separation of $\approx 3.3 \pm 0.5 \rm \; AU$. According to Figure \ref{fig:non_binary}, white-dwarf companions of $\approx0.6 \; \rm M_\odot$ ($\approx630 \; \rm M_{Jup}$) would be found at $P\approx1600 \; \rm days$ and $K\approx 7 \rm \; km \; s^{-1}$ with detection probability above 50\%. The small number of binary candidates in our sample with $\rm A(Li) \gtrsim 2.2 \; dex$ -- 3/38 -- disfavors the scenario of mass transfer from an intermediate-mass AGB companion for most of our sample. }

\begin{figure*}[t!]
    \centering
    \includegraphics[width=\linewidth]{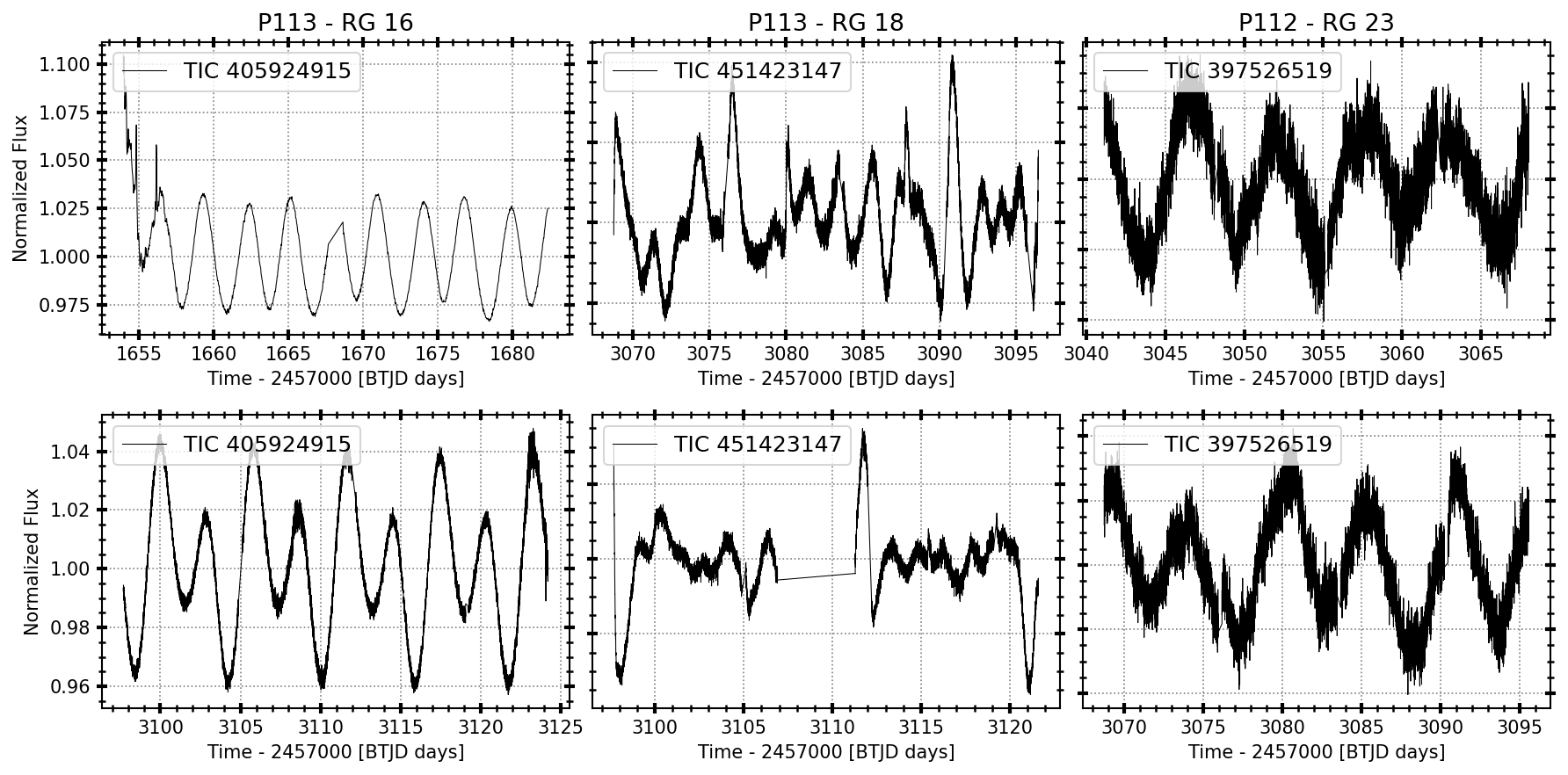}
    \caption{\tess lightcurves for three targets that showed clear variability in the photometry: TIC 405924915 (left), TIC 451423147 (middle), and TIC 397526519 (right). These targets show easily distinguishable, but unexpectedly fast, rotation rates for red giants which could be caused by binary interactions or contamination by nearby objects.}
    \label{fig:lc}
\end{figure*}

\subsection{Companions in \apogee DR16}

To set an expectation of the binary fraction in a background population of red giants, we test the binary fraction in \apogee DR16 data with \joker. We select red giants with similar stellar effective temperature and surface gravity as the parent sample in \cite{sayeed_2024} ($\rm T_{eff}=[3000,5730] \; K$ and $\log g = [-1,3.2] \rm \; dex$), and $V$ magnitude between 7--12. We limit to targets with at least four visits, and with a red giant classification (\texttt{WASH\_DDO51\_GIANT\_FLAG==1}). We also enforce quality flags following the methods by \cite{apw_2020} listed below,

\begin{enumerate}
    \item \texttt{STARFLAG} in allStar must not contain \texttt{VERY\_BRIGHT\_NEIGHBOR} or \texttt{SUSPECT\_RV\_COMBINATION}
    \item \texttt{ASPCAPFLAG} in allStar must not contain \texttt{TEFF\_BAD}, \texttt{LOGG\_BAD}, \texttt{VMICRO\_BAD}, \texttt{ROTATION\_BAD}, or \texttt{VSINI\_BAD}.
    \item STARFLAG in allVisit must not contain \texttt{VERY\_BRIGHT\_NEIGHBOR}, \texttt{SUSPECT\_RV\_COMBINATION}, \texttt{LOW\_SNR}, \texttt{PERSIST\_HIGH}, \texttt{PERSIST\_JUMP\_POS}, or \texttt{PERSIST\_JUMP\_NEG}.
\end{enumerate}

The final \apogee DR16 red giant sample contains 3159 stars after enforcing the above cuts. After running \joker, we recover 62 uni-modal systems. For the remaining multi-modal systems, we use Equation \ref{eq:K} to derive the expected semi-amplitude using the target's \logg (from \apogee DR16) and classify targets as binaries if the 5th percentile semi-amplitude is larger than $3.5\sigma$ of the expected semi-amplitude; 528 stars satisfy this condition which combined with the 62 uni-modal systems, yields a background companion fraction of $\sim19\%$. However, \apogee baseline of these $\sim19\%$ of binary candidates is significantly different (up to 2500 days) than \esp baseline. To remain consistent in our analysis, it is important to account for the RV precision, cadence, and baseline of \apogee data in comparison to \esp data. Therefore, we use our injection-recovery analysis of detectable systems in \esp data to weight each potential binary system in  \apogee by its probability of being detected (see Figure \ref{fig:non_binary}). First, we limit the sample to those with 4--5 visits, resulting in 435 multi-modal and 24 uni-modal systems. Then, we derive companion masses using median orbital parameter values for multi-modal systems and sampling over the reported primary mass and uncertainty from \cite{Queiroz2020}. We re-calculate the probabilities for circular orbits on a finer grid than in Figure \ref{fig:non_binary}. Summing over the probability of detection for 459 systems, we expect to recover 390 binary systems, or 85\% of known \apogee binaries. 

\new{To determine if \esp would have detected the binary systems in \apogee DR16, we calculate the probability below,
\begin{equation}
     P({\rm flagged}) = P({\rm binary}) \times P({\rm flagged | binary })
\end{equation}}

\new{where $P({\rm binary})$ is the probability of binaries in \apogee (19\%) and $P({\rm flagged | binary })$ (85\%) is the probability of \esp detecting a binary from a parent sample of giants in binary systems; as a result, our pipeline would have detected 16\% of all giants in binary systems in \apogee data. Therefore, our measured companion fraction of 17\% (10/58) is consistent with a background population of giants.} This inference assumes that the underlying distribution of masses and semi-major axes of binary companions is the same in our \lirich sample and in the red giants observed by \apogee DR16 as a whole. Further observations to extend our sensitivity to wider, lower-mass companions will enable us to separate between formation mechanisms.

\subsection{Variability}
We also checked \tess lightcurves of our \total potential binaries to ensure the orbital period fit by \joker was dissimilar to other variability at the same scale, such as a rotation period. We cross-matched \gaia RA and Dec with \tess Input Catalog v8.2 using \topcat with a matching radius of 2$\arcsec$. We then used \lk to download and process the data. All except for one target (TIC 264866723, P116-17) had lightcurves available. Three targets showed an easily distinguishable signal shown in Figure \ref{fig:lc}: TIC 405924915, P113--16 (left column), TIC 451423147, P113--18 (middle column), and TIC 397526519, P112--23 (right column). We used data from the Quick--Look Pipeline for TIC 405924915, but TESS--SPOC data for the other two targets \citep[][]{tess-spoc}. Note that targets TIC 405924915 and TIC 451423147 were uni-modal in \joker analysis. To measure their rotation period, we first removed outliers and NaNs, and calculated a periodogram with a maximum period of 100 days. The Lomb--Scargle model in \lk \citep[][]{Lomb1976, Scargle1982, VanderPlas2018} indicates a rotation period of 5.52 days for TIC 397526519 and 5.75 or 5.85 days for TIC 405924915; for the latter target, we chose the second highest peak in the periodogram as its rotation period since visual inspection shows a difference in flux amplitude that changes the flux amplitude every \app2.8 days. The rotation period for the third target, TIC 451423147, was inconclusive. 

Since evolved stars rotate slowly and very few have shown rotation periods less than \app 50 days \citep[e.g.,][]{Ceillier2017}; we checked the Full Frame Images (FFIs) to rule out contamination in the lightcurve. However, FFIs reveal a crowded field for both objects with other \gaia sources within the same pixel. Therefore, we cannot rule out contamination in the data due to the proximity of our targets to others in \tess FFIs.

To check the location of the rotation signal for TIC 397526519 and TIC 405924915, we use \texttt{TESS-localize} \citep{Higgins2023} to pinpoint the source of the variability on a \tess pixel. For TIC 405924915, we use an input frequency of 2.01 and 1.98 $\mu \rm Hz$ (or 5.75 and 5.85 days), and an input frequency of 2.10 $\mu \rm Hz$ (or 5.5 days) for TIC 397526519, where all frequencies were measured by \lk. For TIC 405924915, \texttt{TESS-localize} suggests that both the 5.75 day and 5.85 day signals originate from an adjacent pixel. For TIC 397526519, the \texttt{TESS-localize} analysis indicates the signal is originating from two pixels away. Further photometry is required to resolve the relevant targets and distinguish the source of variability.

\new{In addition to \tess, we also checked for photometric variability in ASAS-SN database \citep[][]{Shappee2014,Hart2023} using \texttt{pyasassn}. To account for the calibration offsets from each camera, we corrected for the offset between the primary camera--defined as the camera with the largest number of measurements--and the remaining cameras. We also ensured that at least 50 measurements were available from any given camera. Finally, we selected measurements with good quality flags and used only g band data. ASAS-SN data revealed a rotation period for three potential binary targets: TIC 405924915 (\galah 170418005701333, P113-16), TIC 451423147 (\galah 170801002501304, P113-18), and TIC 205846638 (\galah ID 140707003101315, P113-9), of which the first two targets also have \tess data. The rotation period, as measured with a Lomb--Scargle periodogram, is $\approx2.9 \; \rm days$ for TIC 405924915 (half of the period measured in 
\tess lightcurves), $\approx52.5 \; \rm days$ for TIC 451423147, and $\approx21.5 \; \rm days$ for TIC 205846638.}


\new{In addition to our \total binary candidates, we also investigated variability for the \notbin non-binary candidates in \tess. Of these, 47 targets were found in the \tess input catalog with QLP data and six targets also had SPOC data, one target had neither QLP and SPOC data, and 42 of the \notbin non-binary candidates were found in the ASAS-SN database \citep[][]{Shappee2014, Hart2023}. Within the \tess sample, only one star shows variability, TIC 24785283 (\galah 150107004201104, P112-1), which has a \vbroad in \galah of $\sim11 \; \rm km \, s^{-1}$. We also checked ASAS-SN data base where we measured a rotation period for only one star of the 42 with available data: P113-5 or GALAH 170416004301008 with a period of 267.2 days. While its \vbroad is only $6.9 \; \rm km \, s^{-1}$, it has the highest lithium abundance of 4.8 dex, as well as RV half-amplitude of $\approx 1.3 \; \rm km \; s^{-1}.$ Therefore, it is possible that P113-5 is in a wide-binary system.}



\section{Discussion}
\label{sec:discussion}



Based on Figure \ref{fig:bin_frac}, our binary candidates have higher \logg and lower lithium abundance relative to the target sample. \cite{sayeed_2024} found that red clump stars tend to have higher lithium abundance than \lirich stars on the red giant branch or at the base of the RGB. If most of our \total binary candidates with low lithium are indeed on the red clump, then our findings contradict those from \cite{sayeed_2024}. While we could argue for a mass transfer event from a lithium-producing intermediate-mass AGB star for our binary candidates, derived companion masses suggest that only one system would be eligible for this scenario (Section \ref{sec:companion_masses}).

\new{All our binary candidates span $\log(g) = 2-3 \; \rm dex$, but we cannot differentiate if these giants are first ascent or core helium-burning giants. To understand the distribution of red clump and red giant stars in our sample, we simulate the number of red clump stars in a synthetic population weighted by initial mass function and star formation rate, using MIST tracks and the target sample as input; the method is described in detail in Appendix \ref{sec:rc_contamination} and shown in Figure \ref{fig:rc_fraction}. We find that our sample contains a lower fraction of core helium-burning stars as compared to first ascent giants, up to at most 24\% for stars within $\log(g) = 2-3 \; \rm dex$. However, spectroscopic parameters alone cannot separate RC and RGB stars in this range, and therefore asteroseismology is required for definitive classification \citep[e.g.,][]{Bedding2011}. We also note that the lithium abundance measured is the red giant's current lithium and not its maximum. If lithium enrichment occurs near the helium flash \citep[e.g.,][]{kumar_2020}, a core helium-burning star may have since depleted its lithium to normal levels; first-ascent RGB stars have not yet undergone the flash.}



 \begin{figure}[t!]
    \centering
    \includegraphics[width=\linewidth]{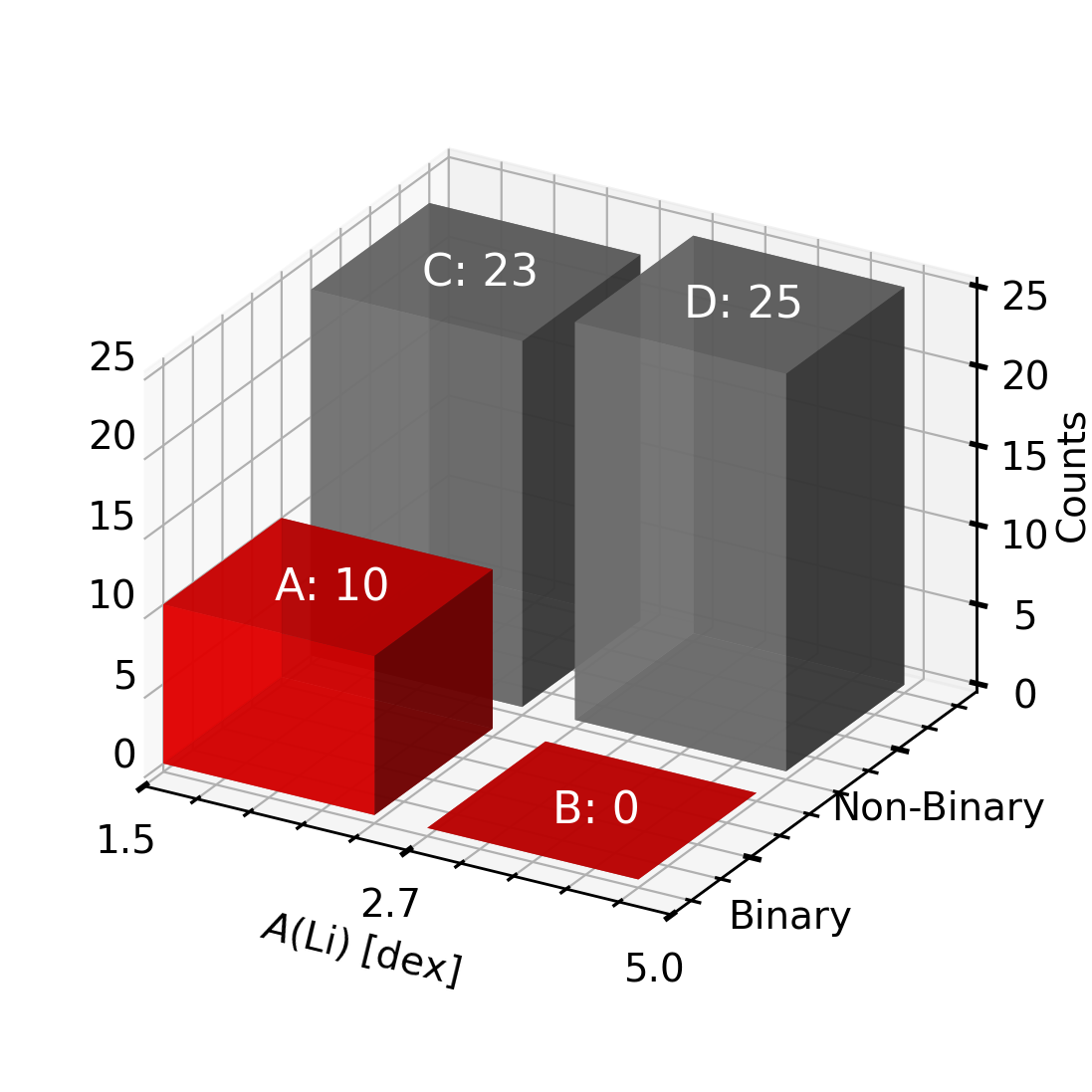}
    \caption{Distribution of giants with and without evidence of binarity in bins of lithium $\rm A(Li) = 1.5-2.7 \; dex$ and $\rm A(Li) =2.7-5.0 \; dex$. There are no binary candidates at $\rm A(Li) \geq 2.7 \; dex$ suggesting that different mechanisms are responsible for the production of \lirich red giants. }
    \label{fig:3d}
\end{figure}

\new{In Figure \ref{fig:3d}, we summarize the binary fraction as a function of lithium in two bins: $\rm A(Li)=1.5-2.7 \; dex$ and $\rm A(Li) =2.7-5.0 \; dex$. The low binary fraction for giants with higher lithium abundance suggests that another mechanism is responsible for Li-enrichment in evolved stars across the giant branch. If higher lithium stars have systematically lower \logg, then they would be classified as non-binaries with our pipeline. To determine the likelihood of this trend being astrophysical, we perform a two-sample Kolmogorov-Smirnov (KS) to determine if the low lithium and high lithium samples are drawn from the same underlying distributions. With a $p-$value of 0.078 (and $D=0.325)$, we cannot detect a difference between the two samples; however, given the similarity of our calculated $p-$value to the threshold of 0.05, it is possible the small sample size is limiting the effectiveness of the test. We also performed a two-sided Fisher exact test restricted to $\log g = 2-3\rm \; dex$ to determine whether there is a correlation between lithium abundance and binary classification; by restricting to a \logg range, we can test the result of binary status of low lithium stars indifferent of \logg. The $p-$value from the Fisher exact test is 0.002,  suggesting that if binarity were independent of lithium, a split at least this extreme would occur with probability 0.2\%. Therefore,  there is a statistical difference in the distribution shown in Figure \ref{fig:3d}: low-lithium giants are more likely to be in binary systems than high-lithium giants.}

\new{Below we summarize the possible formation mechanisms that could produce lithium-rich red giants based on our results. The two lithium bins at a threshold of $\rm A(Li)=2.7 \; dex$ are referred to as low- and high-lithium.}

\begin{enumerate}[i)]
    
    \item \new{\textbf{Low-lithium, binary status (Figure \ref{fig:3d}A):} 10 stars; there are multiple formation mechanisms for these objects such as planet engulfment in a multi-planet system, or tidal spin-up by a companion that induces the CF mechanism. The lithium abundance of red giants that has accreted a substellar object, including planets and brown dwarfs between $\rm 1 \; M_\oplus - 0.065 \; M_\odot $, is consistent with \lirich giants in this range up to a maximum of $\sim 2.2 \; \rm dex$; \cite{gomez_2016a} suggest that other mechanisms are responsible for lithium enrichment above this threshold. }
    
    \new{Alternatively, tidal-spin up by a close companion could result in lithium production via CF mechanism  regardless of companion mass \citep[][]{casey_2019}. Our analysis finds a range of planets--a brown dwarf, substellar, and stellar mass companions--, but the lithium quantity achieved is dependent on envelope mass and rotation induced mixing. Nevertheless, we explore Li-enrichment via CF mechanism given each type of companion in more detail below.}
    
    \new{Firstly, the occurrence rate of brown dwarfs around solar-type stars is $0.8\% ^{+0.3}_{-0.2}$ \citep{Unger2023}, as compared to the occurrence rate of stellar mass companions around solar-type stars \citep[e.g., 50\%,][]{Mo2017}. Studies indicate that brown dwarfs close to the host star $(<5 \rm \; AU)$ are rarer than giant planets ($<1\%$ vs. $5-10\%$) \citep{MaGe2014}. If our red giants are indeed hosting brown dwarfs, it is unlikely that these brown dwarfs formed in situ at these separations. It is more probable that the red giants in our sample hosted at least another planet that was engulfed by the red giant -- through evolution of the red giant or planet-planet scattering -- resulting in an eccentric system with a brown dwarf. Follow-up observations would provide companion mass as well as eccentricity to test this scenario.}

    \new{Another possible mechanism is mass transfer from an intermediate-mass AGB companion. Our analysis in Section \ref{sec:companion_masses} suggests that only one of the six stellar mass companions could have been an intermediate-mass AGB star that produced lithium via HBB and transferred its mass to the red giant primary enriching it in lithium. However, we emphasize that the choice of IMFR could affect our derived initial masses \citep[e.g.,][]{Marigo2007, Meng2008, Tremblay2016, Tremblay2019, ElBadry2018}, and direct detection of the companion is crucial in constraining its mass.
    }


    \item \new{\textbf{High-lithium, binary status (Figure \ref{fig:3d}B):} we find no binary candidates with high lithium abundance. However, it is possible that some of these are in face-on systems thereby preventing RV detection. Similarly, the companions may be smaller and more distant from their host star; longer and more precise follow-up is needed to confirm these companions. It is also possible that the companions are giant planets outside the sensitivity of \esp. Although smaller, Earth-sized planets have been discovered by \esp \citep[e.g.,][]{proxima_espresso, LilloBox2021} the number of observations for these studies was an order of magnitude larger than our observations (e.g., 41 measurements over \app 900 days for \cite{LilloBox2021} and typical RV noise of $\rm 18 \; cm \; s^{-1}$ or 63 observations over 250 days with photon noise of $\rm 18 \; cm \; s^{-1}$ for \cite{proxima_espresso}). }

    \item \new{\textbf{Low-lithium, non-binary status (Figure \ref{fig:3d}C):} for 23 stars, we found no evidence of a companion, but it is possible that these red giants become lithium enhanced via mechanisms that do not require the presence of a companion. Firstly, giants on the red clump could have become lithium enhanced via planet engulfment. Although we do not have constraints on the orbital periods of our targets, we would expect no companions within a few hundred days around red clump stars in our sample. Red clump stars have already undergone an envelope expansion phase when they ascend the giant branch. The red giant is expected to expand to $\sim \rm1 \; AU$ engulfing any close-in companions within this separation. \cite{apw_2020} found a decrease in binary fraction at the red clump that they attribute to companion engulfment. However, engulfment does not always leave a detectable chemical signature: even under optimistic detection thresholds (above 0.02 dex), \cite{Soares2025} find detectable signatures in at most 20\% of Sun-like stars, although the deep convective envelopes of giants would dilute such signatures further. Follow-up observations could test whether these stars lack close-in companions, as expected if they are red clump stars that engulfed them during the RGB phase.}
    
    \new{Another possible reason is that these are single stars that inherit lithium from their main sequence phase. \cite{sayeed2025} find that $\sim9\%$ of main sequence stars have $\rm A(Li) = 2.5-3.2\; dex$ which could produce lithium-rich red giants (ie. $\rm A(Li) \geq 1.5\; dex$) after the FDU phase. Therefore, the presence of a companion would not be required for red giants with current lithium abundance between $\sim 1.5-2.2 \rm \; dex$, consistent with some of our targets in this bin. This is further supported by the fact that the median \vbroad of these targets is $7.3 \rm \; km \; s^{-1}$, consistent with the average broadening velocity of a red giant without additional spinning \citep[e.g.,][]{Carney2008, Patton_2023}.} 

    \new{Engulfment of a brown dwarf could also produce these low-lithium, non-binary candidates. Low mass brown dwarfs (below $\sim63 \; \rm M_{\rm Jup}$) do not attain high enough temperatures to burn lithium ($\sim2.5 \times 10^{6} \; \rm K$); they retain their primordial lithium throughout their lifetime. For instance, Luhman 16 is the closest brown dwarf to the Sun at a distance of $2\pm0.15 \rm \; pc$, consisting of an L-dwarf primary and T-dwarf secondary \citep[e.g.,][]{Luhman2013,Burgasser2013, Faherty2014, Lodieu2015}. The optical spectra reveal strong lithium absorption line at $6708 $ \ang for both components \citep{Faherty2014}. Finally, single, lithium-rich red giants could also be a result of stellar mergers (discussed below).}
        
    \item \new{\textbf{High-lithium, non-binary status (Figure \ref{fig:3d}D):} we find no evidence of a companion around all 25 red giants with higher lithium abundance, $\rm A(Li)\gtrsim 2.7 \; dex$. } 
    
    \new{We argue that these single, higher lithium giants could be products of stellar mergers. We looked for possible support of mergers in the masses of these red giants, but predicted remnant masses differ between models; \cite{Rui2024} suggest merger products typically have $\rm M\gtrsim 1.5 \; M_\odot$, whereas \cite{Zhang2020} find that red clump remnants of helium white dwarf–red giant mergers span $0.8-1.8 \rm \; M_\odot$ with a peak at $\sim1.2 \; \rm M_\odot$. Our \galah DR3 masses ($M\sim0.9-2.0 \; M_\odot$, with only three targets above $\rm 1.5 \; M_\odot$) are consistent with the latter. The ten targets with \gaia DR3 {\small FLAME} masses\footnote{Found via \topcat with a radius of 2.0$\arcsec$.} have systematically higher masses, on average by $\sim \Delta= 2.0 \pm 0.5 \; \rm M_\odot$, but both \galah and {\small FLAME} rely on isochrone fitting, and {\small FLAME} for evolved stars carry large systematic uncertainties \citep{Creevey2023}. Current mass estimates therefore cannot confirm or rule out a merger scenario, and asteroseismic masses are needed to confirm. }

    \new{Alternatively, \cite{yan_2021} find that \lirich RGB stars show an upper limit of $\rm A(Li) = 2.6 \; dex$ while RC stars extend to higher values; our results, that higher lithium stars show no evidence of companions, could be because higher lithium stars are red clump stars which have already engulfed a companion. Further analysis into the stellar interior using asteroseismology would reveal if these objects are in fact products of a merger. }

    \new{Furthermore, using population synthesis to predict properties of stellar mergers between a giant primary and a main sequence or brown dwarf secondary, \cite{Politano2010} find that the median projected rotational velocity of merger remnants where the final product is a red giant is $v\sin i \approx 16 \rm \; km \; s^{-1}$, and ranges from $5.4 - 18.4 \rm \; km \; s^{-1}$ (dependent on input physics). The \vbroad of targets in this group ranges from $4.9-18.8\rm \; km \; s^{-1}$ within the range outlined by \cite{Politano2010}; therefore, it is possible that some of our red giants could be stellar mergers given their rotational velocity.} 
\end{enumerate}

\new{We highlight that stellar mergers are a possible scenario for \lirich red giants not only for objects without any evidence of a companion, but also for giants with a substellar companion. \cite{Hon2023} suggest that the planet 8 Ursae Minoris b should have been engulfed when its host -- a \lirich, core helium-burning giant with $\rm A(Li)=2.0 \pm 0.2 \; dex$ -- evolved off the main sequence. The planet avoided engulfment and remained in a stable orbit if its host merged with a white dwarf during the host's red giant phase. The merger ignited the host's core helium, and terminated its red giant expansion phase early thereby preventing planet engulfment despite the planet's close orbit ($\sim 93 \; \rm days$). Another possibility is that the planet was formed as a second-generation planet from the remnants of the stellar merger.}


\subsection{Caveats}
\label{sec:caveats}
\new{As mentioned in Section \ref{sec:lirich_def}, we chose the traditional threshold to define a star as \lirich, $\rm A(Li) \geq 1.5 \; dex$, based on expected dilution of lithium after FDU. However, cluster studies have found lithium after FDU phase varies as a function of mass and metallicity. \cite{Anthony-Twarog2026} found giants with lithium detection in NGC 7789 and NGC 2204 are on the RGB while those with upper limits are post RC stars. Without accurate masses, it is possible our chosen lithium threshold admits preferentially higher mass giants with lithium detections. Furthermore, standard evolutionary models suggest that lower mass giants experience higher lithium depletion than higher mass giants \citep{chaname_2022}; for instance, stars with $\rm 1 \; M_\odot$ and $\rm 2 \; M_\odot$ undergo depletion up to 7 dex and 2 dex, respectively. Therefore, the required enhancement needed to achieve \lirich status is higher for a $\rm 1 \; M_\odot$ star than a $\rm 2 \; M_\odot$ at similar conditions.} 

\new{Moreover, at higher masses the number of red clump stars is higher than the number of RGB and subgiant stars in a magnitude limited survey such as \galah \citep{Girardi2016}. In fact, the lifetime of RGB stars with $\rm M<3 \;M_\odot$ varies as $t_{\rm RGB}\sim M^{-5.5}$ while the lifetime of the RC phase is shorter (but increasing slightly) for $\rm M<2 \;M_\odot$ leading to a higher ratio of RC stars to RGB and subgiant stars with increasing mass \citep[e.g.,][]{Serenelli2007, Lagarde2012}. Therefore, our finding that all of our binary candidates are within $\log(g) = 2-3 \; \rm dex$ could be due to selection effects brought on by our lithium cut: including stars with $\rm A(Li) \geq 1.5 \; dex$ preferentially includes higher mass stars where the ratio of RC stars to RGB stars is higher. Because of this bias, many argue for a mass dependent or evolutionary dependent lithium cut \citep[e.g.,][]{gomez_2016a, gomez_2016b, kumar_2020, chaname_2022}, but unlike with stars in clusters, we do not have accurate masses for our field of \lirich giants without asteroseismology. Nevertheless, \cite{Sun2022} find four \lirich giants in cluster NGC 188, despite only two that satisfy traditional definition of \lirich, arguing that all four stars are enhanced in lithium relative to the cluster's own lithium non-detection baseline. Accurate masses would reveal if our results are biased based on stellar mass.}

\subsection{Tidal Interaction}

We also test tidal effects on the surface of the \lirich red giant. We calculate a dimensionless tidal parameter, $\epsilon$, which accounts for both stellar parameters and system configuration. We use the following equation \citep[e.g.,][]{Ogilvie2020}
\begin{equation}
    \epsilon = \frac{M_2}{M_1} \Bigg(\frac{R_1}{r_p} \Bigg)^3, \; r_p = a(1-e)
    \label{eq:eps}
\end{equation}
which takes into account the stellar and companion mass ($M_1$, $M_2$), the periastron radius ($r_p$), stellar radius ($R_1)$, system separation ($a$), and eccentricity ($e$). We use \joker's samples of semi-amplitude, orbital period, and eccentricity to calculate $\epsilon$ where we use the median for each target as the value of $\epsilon$, and the 16th and 84th percentile values as its uncertainties. The median $\epsilon$ in our sample of \target targets is $2.6\times10^{-6}$, with a range from $6\times10^{-8}-2$.

\begin{figure}[t!]
    \centering
    \includegraphics[width=\linewidth]{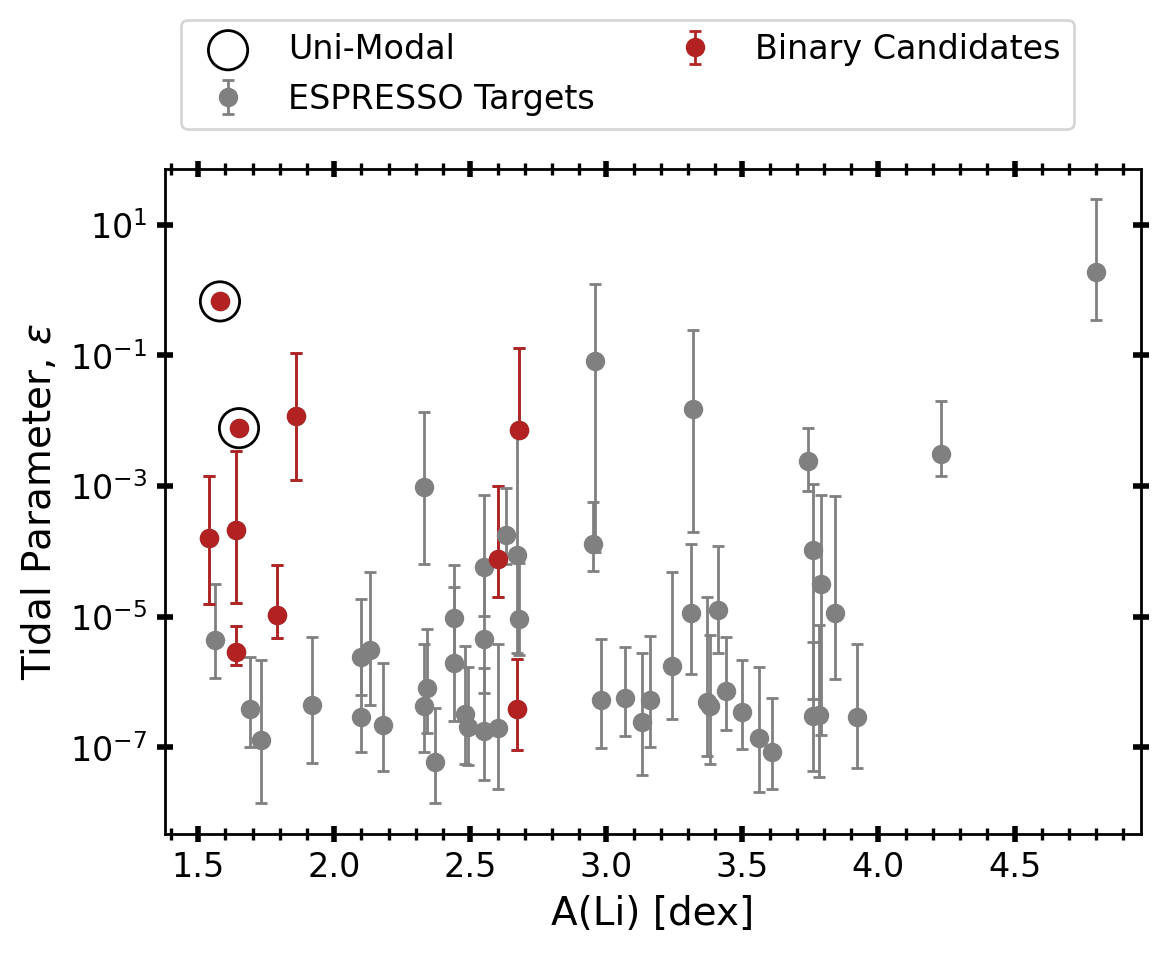}
    \caption{Dimensionless tidal parameter (Equation \ref{eq:eps}) for our full \esp sample (grey) and \total binary candidates (red). The two uni-modal systems (black circle) have a relatively high tidal parameter, orders of magnitude larger than the remaining sample suggesting strong evidence for tidal interactions.}
    \label{fig:eps}
\end{figure}

Figure \ref{fig:eps} shows the relationship between $\epsilon$ and lithium abundance. While we see no correlation between $\epsilon$ and our \total binary candidates, our two uni-modal systems have $\epsilon$ of 0.15 and 0.01. In comparison, the tidal parameter for Moon's orbit around the Earth is $\sim6\times 10^{-8}$. Similarly, WASP--12 a system known for its strong tidal interaction between the planet and its host star, has a tidal parameter of $\epsilon = 4\times 10^{-5}$ using planetary and stellar parameters from \cite{Akinsanmi2024}. Kepler--1658 (KOI--4.01), \textit{Kepler}'s first new planet candidate, hosts a giant planet around an evolved star with detected inspiral \citep[e.g.,][]{Vissapragada2022}; its tidal parameter is $\epsilon \sim 6\times 10^{-5}$. The tidal parameter for our uni-modal systems are therefore orders of magnitude larger than known tidally interacting star--planet systems, suggesting these systems might be undergoing strong tidal interactions.

\subsection{Comparison to Literature}
\new{While there does not yet exist a large and homogeneous radial velocity follow-up of red giants specifically targeting \lirich stars,  we compare our results to similar existing studies.} Our companion rate of lithium-rich red giants (17\%) is dissimilar to overall companion rate of A18 and H08 \citep[][]{Hekker2008, Adamow2018}, as well as the overall binary frequency of $\sim50$\% for solar-type stars \citep[][]{, Raghavan2010, Mo2017}. The overall companion rate as found by A18 and H08 is 47\% (confirmed companions only) and 37\%, while the stellar companion rate is 27\% and 13\% (23/179), respectively. However, our stellar companion rate (10\% or 6/58) is similar to the stellar companion fraction of 13\%  in H08 although their sample was not specifically targeted to \lirich red giants. Therefore, our companion fraction is more similar to a reference sample of red giants in H08 than the companion frequency of \lirich giants in A18. \new{Similarly, \cite{Castro-Tapia2024} find no preference for RC \lirich giants to be in binary systems, but their sample of RGB stars have a non-negligible preference for hosting close companions.} \new{Nevertheless, without a large, homogeneous sample of red giants and their companion fraction for reference, the significance of our results remains to be confirmed.}

Additionally, A18 recover a planetary companion rate of $20-33\%$ for their sample of \lirich giants as compared to our rate of 5\% (3/58). However, we are only sensitive to planets down to $10 \; \rm M_{Jup}$ and therefore 5\% is a lower limit on our planetary companion occurrence rate. The occurrence rate of red giants hosting giant planets is $10-20\%$ \citep[e.g.,][]{Johnson2007a, Johnson2010, Reffert2015, Ghezzi2018}. Close-in giant planets around red giants are rare, and are often destroyed due to engulfment or tidal decay \citep[e.g.,][]{Grunblatt2018}. For instance, \cite{Grunblatt2018} found three close-in giant planets around subgiants that they suggest are transient and will eventually be engulfed by the red giant. Similarly, our recovered fraction of brown dwarf in our sample of red giant binary candidates is dissimilar between this work (10\%), and those from H08 (30\%) and A18 (none).

Lastly, A18 also derive companion masses for their sample of planetary mass companions, as low as $1.3 \; \rm M_{Jup}$ (potential companion for TYC 3663-01966-1) and $1.7  \; \rm  M_{Jup}$ for $\rm BD+48 740$. Our planet companion masses are comparatively larger, $\rm M = 7.2 \;  M_{Jup}$, $8.0 \;  \rm M_{Jup}$, and $11.4 \;  \rm M_{Jup}$, due to our detection limits. Our injection-recovery tests show that we are not sensitive to companions below $\rm 10 \;  M_{Jup}$ orbits longer than $\sim 200$ days due to intrinsic RV jitter and our six month baseline (see Figure \ref{fig:non_binary}) as compared to A18's multi-year monitoring. Our detected planet masses therefore lie near our detection threshold.

\subsection{Interesting systems} \label{sec:uni-modal}

\subsubsection{P113--16, TIC 405924915, \gaia DR3 6708750736102048256} 
This uni-modal system in \joker has a period of $12.96^{+0.05}_{-0.02}$ days, or a separation of $0.123 \pm 0.005 \rm \; AU$, and an eccentricity of $0.11^{+0.21}_{-0.08}$. \tess lightcurves show a rotation period of 5.85 days as measured by \lk (see left column in Figure \ref{fig:lc}), but further investigation reveals the signal could originate from a nearby pixel. Fast rotating red giants are highly unusual; for instance, \cite{Ceillier2017} derived rotation periods for 361 red giants in \kepler of which only 1.1\% were below 10 days. \tess target pixel images show nearby objects within the same aperture as the target; given that this target is not completely resolved in \tess, the fast variability could be due to contamination. Furthermore, the red giant's broadening velocity of 51 \kms is larger than the expected circular velocity suggesting an additional cause for its faster than normal rotation period. Combined with its tidal parameter of 0.01, orders of magnitude larger than known tidally interacting planet-star systems, this system could be tidally interacting with a close companion, inducing the Cameron--Fowler mechanism, and causing Li-enrichment. Additional RV measurements are required to confirm the presence and orbital parameters of this potential binary system. 

Furthermore, fast rotating red giants have been hypothesized to be a member of a hierarchical three body systems. For instance, in a sample of 168 red giants, \cite{Colman2017} find 81 targets that show anomalous high-amplitude in the frequency domain which they attribute to common envelope or hierarchical triple systems (e.g., a red giant orbited by an inner contact binary). In their sample of 81 red giants hypothesized to be in this configuration, 12 (15\%) show a period between $\rm 1-5 \; days$. Additional RV measurements at high-resolution and longer baselines are needed to confirm if this system is in fact a hierarchical triple system.  

\subsubsection{P113--18, TIC 451423147, \gaia DR3 5819206285484386304} 
This uni-modal system in \joker has a period of $3.265^{+0.595}_{-0.047}$ days, or a separation of $0.054 \pm 0.006 \rm \; AU$, and eccentricity of $0.15\pm 0.06$. Its broadening velocity is \app 14 \kms. \tess data (middle column in Figure \ref{fig:lc}) reveal no periodic signal in its lightcurves. Its dimensionless tidal parameter is $\epsilon = 0.15$ a few orders of magnitude larger than \textit{Kepler}--1658 and WASP--12, and factor of $10^4$ larger than that of the Moon around the Earth. We note that at this separation the giant would be close to filling its Roche lobe, and a non-zero eccentricity ($e = 0.15$) would be unexpected given the short tidal circularization timescales of close giant binaries \citep{Verbunt1995}; together with the much longer photometric period ($\approx52.5 \rm \; days$), this suggests the $3.3-$day solution may be an alias, but additional RVs are needed to confirm it.

 
\subsubsection{P112--23, TIC 397526519, \gaia DR3 5232071641786703488} 
With only \esp and \galah data, this system has an observed semi-amplitude of 7.5 \kms (or 6.3 \kms excluding \galah data), and 22 RV solutions from \joker. Its \tess lightcurve from the TESS--SPOC pipeline \citep[][]{tess-spoc} shows a clear periodic signal with a rotation period of 5.5 days (see right column in Figure \ref{fig:lc}). However, target pixel files reveal nearby objects within the same aperture of the red giant in \tess data; therefore, the fast rotation could be from another object. Further RV measurements at high-resolution as well as ground-based photometry would help to extract the rotational signal and confirm the existence of a companion.

\section{Conclusion}
We tested the companion fraction for \target \lirich giants using high-resolution spectroscopy from \esp, supplemented with archival radial velocity data from \galah DR3, \apogee DR17, and \rave DR6. We recovered \total potential binary candidates, with a final companion rate of 17\%, using a custom Monte-Carlo sampler (\joker) to sample orbital solutions. To establish an expectation for a background population of red giants, we performed the same analysis with red giants in \apogee DR16 data weighted by probability of detection for \esp baseline. For a background population of red giants of similar magnitude and stellar parameters in known binary systems, our pipeline recovers a companion fraction of 16\% in \apogee data similar to our inferred companion rate of 17\% for \lirich stars in \esp data. \new{The similarity in companion fraction between a sub-sample of \lirich giants and a background population of binaries suggests that \lirich giants have a binary fraction similar to that of the background population, unless their companions are preferentially in orbits and mass ratios different to \apogee binaries.} 


We \new{further} caution that our recovered companion fraction of 17\% is preliminary and should not be generalized to the full population of \lirich giants. We utilized \esp data, with a baseline of six months, to detect a binary based on the cadence of observations and period distribution in \joker. While we selected our targets to span relevant parameter spaces (e.g., lithium abundance, surface gravity, broadening velocity), we were still only limited to \target targets. Extended baselines and additional RV data for larger samples of \lirich giants are crucial in interpreting the significance of our companion binary fraction. Our main findings are as follows:
\begin{enumerate}
    \item We find \lirich giants with a lower lithium abundance ($\rm A(Li) = 1.5-2.0 \; dex$) have a higher binary rate (64\%) than \lirich red giants with $\rm A(Li) \geq 2.0 \; dex$, while the companion fraction is zero for $\rm A(Li) > 2.7 \; dex$. Our findings suggest that multiple mechanisms could be responsible for lithium enhancement of red giants. See Figure \ref{fig:bin_frac}.
    
    \item \new{Our findings suggest that binarity is not the only possible pathway for \lirich giants given we find a companion fraction of 17\%. Notably, the lack of binary candidates with $\rm A(Li)\geq2.7\; \rm dex$ suggests two possibilities. See Figure \ref{fig:3d}.} 
    
    \new{First, the lack of binarity for single red giants with higher lithium ($\rm A(Li)\gtrsim2.7\; \rm dex$) and low to moderate broadening velocity suggest that stellar mergers are a possibility for lithium enhancement of red giants. Stellar mergers have been hypothesized to create chemically peculiar objects \citep[e.g.,][]{Zhang2013, Zhang2020, Rui2024} that could be investigated by detailed, high-resolution asteroseismology. While asteroseismology would reveal various internal properties of our targets, including mass, rotation, and age, it is beyond the scope of this work.}

    \new{Second, \lirich red giants below this threshold can still be explained by planet engulfment (up to $\rm A(Li)\sim2.2\; \rm dex$), CF mechanism via tidal spin-up, and mass transfer from a lithium-producing, intermediate-mass AGB star. Extended baselines are required to constrain orbital properties and differentiate between scenarios for our objects.}
    
    \item \new{All our binary candidates have $\log g = 2-3 \rm \; dex$ which encompasses the typical \logg of core helium-burning red giants. Derivation of contamination rate using a synthetic population of stars shows that core helium-burning red giants comprise at most 24\% of this sample. This result suggests that (i) binarity is not the only possible mechanism for Li-enrichment of red clump stars, and (ii) red giants on the tip of the giant branch or those at the base of the giant branch become lithium enhanced via other mechanisms than binarity, although our sensitivity to companions is lower for giants with $\log g < 2 \rm \; dex$. We further note that the current lithium abundance may not be the giant's maximum lithium abundance due to lithium depletion post helium flash. Given that most of our binary candidates have $\rm A(Li)<2.0\; \rm dex$, it is therefore expected that the lithium of red clump stars would be lower than first ascent giants which have not yet undergone helium flash. See Figure \ref{fig:bin_frac} and Figure \ref{fig:rc_fraction}.} 

    \new{In addition, it is possible our lithium threshold cutoff ($\rm A(Li) \geq 1.5 \; dex$) preferentially admits higher mass stars where the ratio of red clump stars to red giant branch stars is higher than those at lower masses leading to higher prevalence of potential red clump stars in our binary candidate sample. Accurate masses from asteroseismology would reveal if our initial sample of field \lirich giants is biased towards higher masses.}

    \item We derived lower limits on companion masses, and recovered three potential planetary companions, one brown dwarf companion, and six stellar mass companions. Two of our \total systems are uni-modal with orbital periods of 3.3 days and 13.0 days, and eccentricities of 0.15 and 0.11. See Figure \ref{fig:msini_mix}.

    
\end{enumerate}

Our analysis used precise radial velocity measurements to investigate the companion fraction of \lirich giants. Despite our \esp programs spanning a wide parameter space in lithium, \logg, and \vbroad, we find that our \total binary candidates share a common \logg range and similar lithium abundances. However, our program was restricted to a baseline of six months for a limited number of stars; longer-baseline measurements are therefore crucial to confirm binarity of \lirich giants, and differentiate between multiple mechanisms. The upcoming \gaia DR4 release will be pivotal in uncovering hidden companions around \lirich giants. In addition, ground-based follow-up over long baselines with high-resolution spectrographs will be useful as \gaia DR4 alone will not be sufficient to uncover binarity due to factors that affect the detectability of a system, such as sky location and apparent magnitude.


%
\new{Finally, we emphasize that the focus of this work was to look for companions around \lirich giants, but other mechanisms could also cause Li-enrichment of red giants such as evolutionary based mechanisms (e.g., helium flash), novae \citep[e.g.,][]{Borisov2024}, or others. If multiple mechanisms are responsible, differentiating between them would require dedicated, large, and homogeneous radial velocity programs with extended baselines, multi-band high-resolution spectroscopy to investigate abundance gradients, and asteroseismology to probe stellar interiors.}

\section{Acknowledgments}
We are thankful to our anonymous referee for helpful comments that improved this manuscript. Based on observations collected at the European Organisation for Astronomical Research in the Southern Hemisphere under ESO programmes 112.25UC.001, 113.26Q8.001, and  116.290U.001. We acknowledge the use of TESS High Level Science Products (HLSP) produced by the Quick--Look Pipeline (QLP) at the TESS Science Office at MIT, which are publicly available from the Mikulski Archive for Space Telescopes (MAST). Funding for the TESS mission is provided by NASA's Science Mission directorate.

\software{astropy \citep{2013A&A...558A..33A,2018AJ....156..123A}, lightkurve \citep{lkurve}, Matplotlib \citep{matplotlib}, MIST \citep{choi_2016}, NumPy \citep{harris2020array}, 
Pandas \citep{mckinney-proc-scipy-2010}, pyASAS-SN \citep{Hart2023}, SciPy \citep{2020SciPy-NMeth}, TESS-localize \citep{Higgins2023}, \joker \citep{thejoker}, TOPCAT \citep{topcat} }

\facilities{\apogee, \esp, \galah, \rave, \tess}

\appendix
\label{figuresappendix} 
\renewcommand{\thefigure}{A\arabic{figure}} 
\setcounter{figure}{0}

\section{GALAH Abundances}
\begin{figure*}
    \centering
    \includegraphics[width=\linewidth]{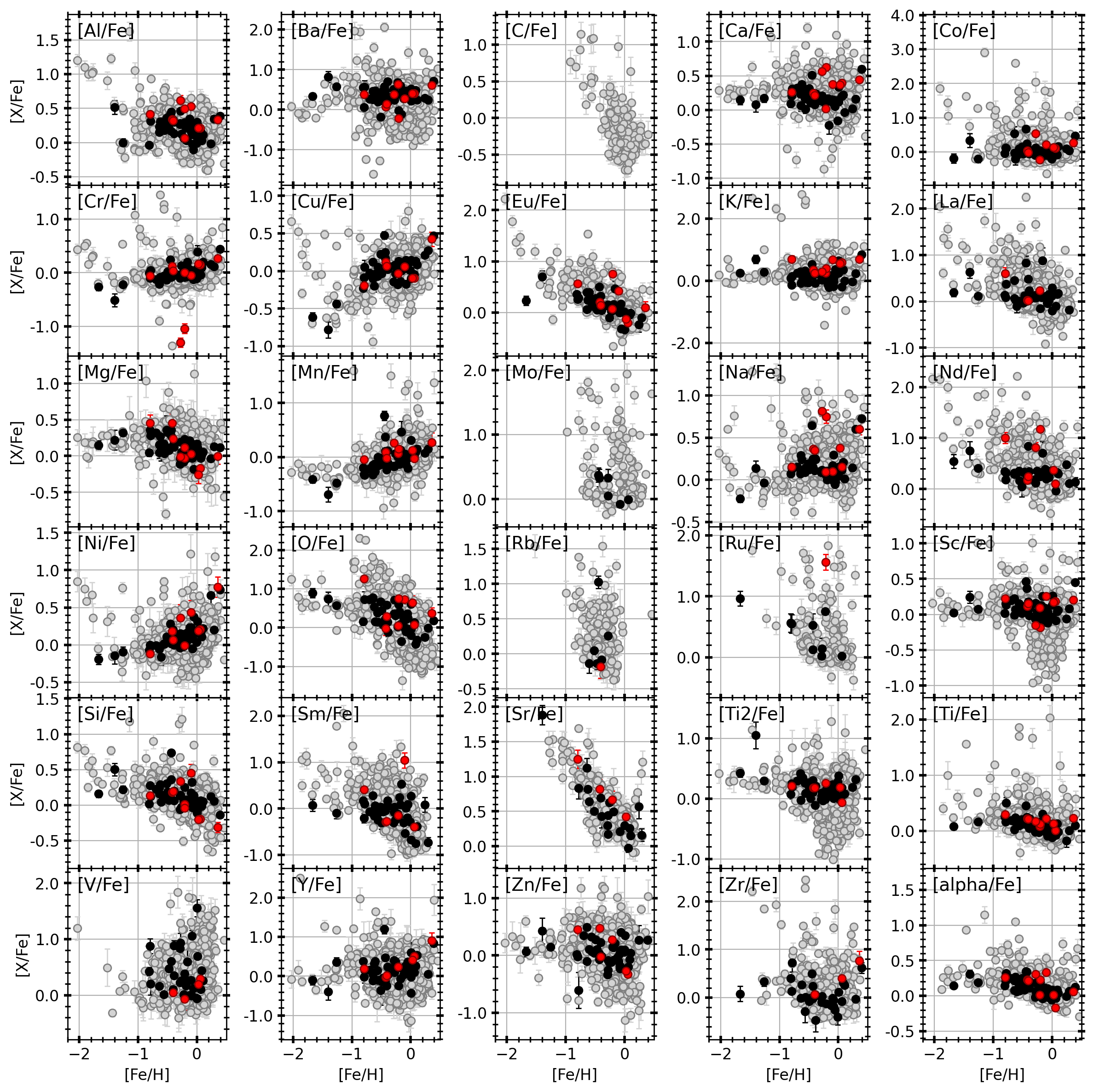}
    \caption{Comparing the chemical abundances from \galah of \total binary candidates (red) and \notbin non-binary stars in the \esp target sample against the \lirich giants sample from \cite{sayeed_2024} (grey) as a reference. Given the limited number of binary candidates, we do not see significant trends in elemental abundances between the binary and non-binary samples.}
    \label{fig:abundance}
\end{figure*}

Figure \ref{fig:abundance} compares 30 abundances from \galah DR3 for our target sample to the \total binary candidates against the overall distribution of \lirich giants in S24. Measurements are not available for all stars for each element (ie., see C/Fe in first row). The limited sample size makes it difficult to distinguish the significance of element abundances in our binary candidate sample.

\section{Red Clump Contamination}
\label{sec:rc_contamination}

\new{To understand the distribution of core helium-burning and first-ascent giants in our sample, we use equivalent evolutionary points (EEP) from MIST\footnote{MIST v2.5} to simulate the relative contribution from each phase \citep{choi_2016, Dotter2016, Dotter2026, Bauer2026}. The MIST tracks are valuable in that they contain the ground truth stage of a star, where every model point is labeled with its evolutionary phase e.g., red giant branch, core helium-burning, and early AGB phase. We use the MIST grids nearest our sample of 39 stars, $\rm [Fe/H] \in [-1.75, 0.50]$ with 0.25 dex spacing, $\rm [\alpha/Fe] = 0$, and no rotation ($v/v_{\rm crit}=0$) with a total of 1196 tracks. We limit our synthesis sample to stars in the red giant branch, core helium burning, or early-AGB phase, $\log g \in [0.0, 3.9] \rm \; dex$, and $T_{\rm eff} \in [3800, \; 5450] \; \rm K$. Every row of every MIST track is treated as one synthetic star carrying a weight calculated as the number of stars within a given mass range, and the time spent in that phase described by,}

\begin{equation}
    w_{j,k} = \xi(M_j)\Delta M_j \times \psi(\tau_k) \Delta\tau_k
\end{equation}

\new{where for row $k$ of track $j$, $\Delta M_j$ is the mass-grid spacing, $\xi$ is the initial mass function (IMF), $\psi$ is the star formation history (SFH) $\psi(\tau) \propto \exp[-(t_0-\tau)/\tau_{\rm SFH}]$ with $t_0=12 \; \rm Gyr$, and $\Delta\tau_k = (\tau_{k+1} - \tau_k)/2$ is the time difference between adjacent EEPs. We experiment with three IMF choices, Salpeter \citep{salpeter1955}, Kroupa \citep{Kroupa2001}, and Chabrier \citep{Chabrier2003}, and three SFH including a constant SFH, and two exponentially declining models with $\tau_{\rm SFH} \in \{6, 12\} \rm \; Gyr$. The population size ranges from $5\times10^4-1.1\times10^5$ synthetic stars based on model choice, and the given age range of $0.2-13.5 \rm \; Gyr$, $1.0-13.5 \rm \; Gyr$, and $4.0-13.5 \rm \; Gyr$. The $\Delta \tau$ factor is key in deciding the ratio of RC to RGB stars; a low mass star crosses $\log(g) \sim 2.4 \;\rm dex$ in $\sim 20 \; \rm Myr$, but spends $\sim 100 \; \rm Myr$ burning helium.} 

 \begin{figure*}[t!]
    \centering
    \includegraphics[width=\linewidth]{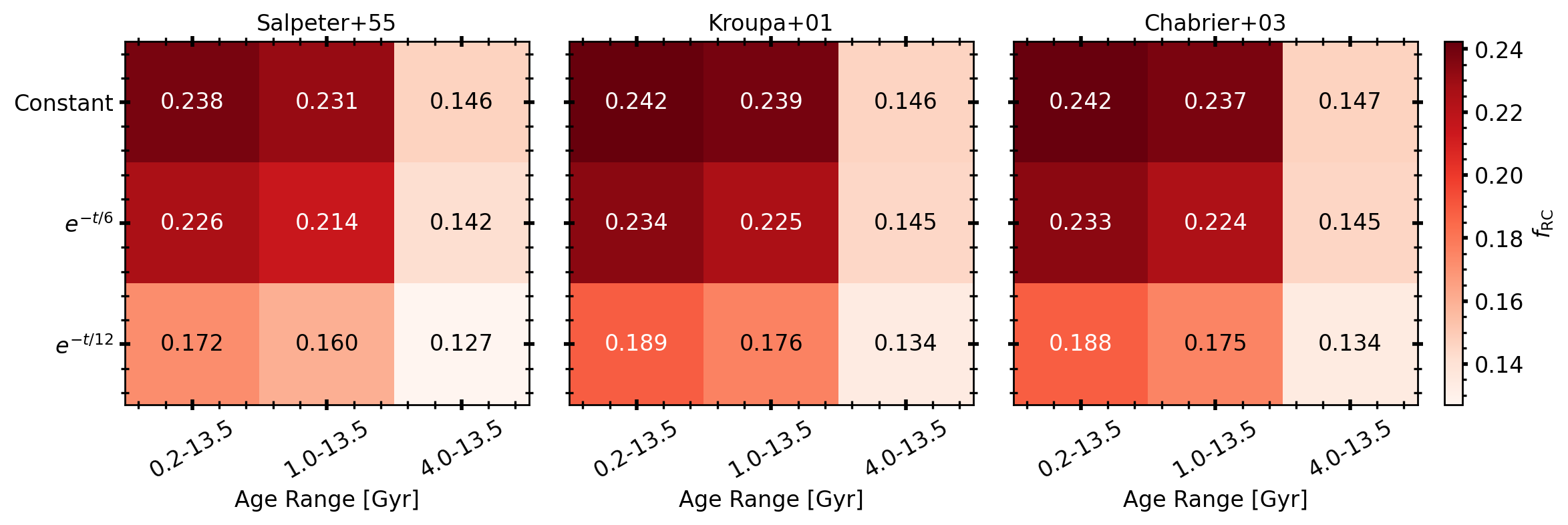}
    \caption{Estimated fraction of red clump stars in a sample of 39 red giants using a synthetic population where darker color corresponds to a higher fraction. The three panels show the results for three chosen IMFs, the y-axis lists the SFH models, and x-axis shows the age range of the population. The estimated red clump contamination in our sample is at most $\sim24\%$.}
    \label{fig:rc_fraction}
\end{figure*}
\new{For each observed star, we select all synthetic points within $\pm1\sigma$ of its measured $T_{\rm eff}$ and $\log g$, and within $\pm(\sigma_{\rm [Fe/H]}^2 + \Delta^2/12)^{1/2}$ in [Fe/H], where $\Delta = 0.25 \rm \; dex$ is the MIST grid spacing. We do not select on stellar mass as the \galah stellar masses come from isochrone fitting, and are dependent on the model prior. $P\rm(RC)$ is the summed weight of core helium-burning points in a $\pm 1\sigma$ box divided by the total weight in the box. Therefore, the expected number of core helium-burning stars in the sample is $N_{\rm CHeB} = \sum_i P_i$, with uncertainty $\sigma_N = [\sum_i P_i(1-P_i)]^{1/2}$. Across the 27 models, the contamination fraction ranges from 13\% to 24\%, each with a statistical uncertainty of about $\pm3\%$. The RC fraction in our sample is shown in Figure \ref{fig:rc_fraction} with SFH on the vertical axis and age range on the horizontal axis, and the different panels indicate the IMF choice.}

\bibliography{references}{}

@ARTICLE{pepe2021,
       author = {{Pepe}, F. and {Cristiani}, S. and {Rebolo}, R. and {Santos}, N.~C. and {Dekker}, H. and {Cabral}, A. and {Di Marcantonio}, P. and {Figueira}, P. and {Lo Curto}, G. and {Lovis}, C. and {Mayor}, M. and {M{\'e}gevand}, D. and {Molaro}, P. and {Riva}, M. and {Zapatero Osorio}, M.~R. and {Amate}, M. and {Manescau}, A. and {Pasquini}, L. and {Zerbi}, F.~M. and {Adibekyan}, V. and {Abreu}, M. and {Affolter}, M. and {Alibert}, Y. and {Aliverti}, M. and {Allart}, R. and {Allende Prieto}, C. and {{\'A}lvarez}, D. and {Alves}, D. and {Avila}, G. and {Baldini}, V. and {Bandy}, T. and {Barros}, S.~C.~C. and {Benz}, W. and {Bianco}, A. and {Borsa}, F. and {Bourrier}, V. and {Bouchy}, F. and {Broeg}, C. and {Calderone}, G. and {Cirami}, R. and {Coelho}, J. and {Conconi}, P. and {Coretti}, I. and {Cumani}, C. and {Cupani}, G. and {D'Odorico}, V. and {Damasso}, M. and {Deiries}, S. and {Delabre}, B. and {Demangeon}, O.~D.~S. and {Dumusque}, X. and {Ehrenreich}, D. and {Faria}, J.~P. and {Fragoso}, A. and {Genolet}, L. and {Genoni}, M. and {G{\'e}nova Santos}, R. and {Gonz{\'a}lez Hern{\'a}ndez}, J.~I. and {Hughes}, I. and {Iwert}, O. and {Kerber}, F. and {Knudstrup}, J. and {Landoni}, M. and {Lavie}, B. and {Lillo-Box}, J. and {Lizon}, J. -L. and {Maire}, C. and {Martins}, C.~J.~A.~P. and {Mehner}, A. and {Micela}, G. and {Modigliani}, A. and {Monteiro}, M.~A. and {Monteiro}, M.~J.~P.~F.~G. and {Moschetti}, M. and {Murphy}, M.~T. and {Nunes}, N. and {Oggioni}, L. and {Oliveira}, A. and {Oshagh}, M. and {Pall{\'e}}, E. and {Pariani}, G. and {Poretti}, E. and {Rasilla}, J.~L. and {Rebord{\~a}o}, J. and {Redaelli}, E.~M. and {Santana Tschudi}, S. and {Santin}, P. and {Santos}, P. and {S{\'e}gransan}, D. and {Schmidt}, T.~M. and {Segovia}, A. and {Sosnowska}, D. and {Sozzetti}, A. and {Sousa}, S.~G. and {Span{\`o}}, P. and {Su{\'a}rez Mascare{\~n}o}, A. and {Tabernero}, H. and {Tenegi}, F. and {Udry}, S. and {Zanutta}, A.},
        title = "{ESPRESSO at VLT. On-sky performance and first results}",
      journal = {\aap},
         year = 2021,
        month = jan,
       volume = {645},
          eid = {A96},
        pages = {A96},
          doi = {10.1051/0004-6361/202038306},
archivePrefix = {arXiv},
       eprint = {2010.00316},
 primaryClass = {astro-ph.IM},
       adsurl = {https://ui.adsabs.harvard.edu/abs/2021A&A...645A..96P}
}

@ARTICLE{Yu2018,
       author = {{Yu}, Jie and {Huber}, Daniel and {Bedding}, Timothy R. and {Stello}, Dennis},
        title = "{Predicting radial-velocity jitter induced by stellar oscillations based on Kepler data}",
      journal = {\mnras},
         year = 2018,
        month = oct,
       volume = {480},
       number = {1},
        pages = {L48-L53},
          doi = {10.1093/mnrasl/sly123},
archivePrefix = {arXiv},
       eprint = {1807.00096},
 primaryClass = {astro-ph.EP},
       adsurl = {https://ui.adsabs.harvard.edu/abs/2018MNRAS.480L..48Y}
}

@ARTICLE{Hekker2008,
       author = {{Hekker}, S. and {Snellen}, I.~A.~G. and {Aerts}, C. and {Quirrenbach}, A. and {Reffert}, S. and {Mitchell}, D.~S.},
        title = "{Precise radial velocities of giant stars. IV. A correlation between surface gravity and radial velocity variation and a statistical investigation of companion properties}",
      journal = {\aap},
         year = 2008,
        month = mar,
       volume = {480},
       number = {1},
        pages = {215-222},
          doi = {10.1051/0004-6361:20078321},
archivePrefix = {arXiv},
       eprint = {0801.0741},
 primaryClass = {astro-ph},
       adsurl = {https://ui.adsabs.harvard.edu/abs/2008A&A...480..215H}
}

@ARTICLE{Breivik2020,
       author = {{Breivik}, Katelyn and {Coughlin}, Scott and {Zevin}, Michael and {Rodriguez}, Carl L. and {Kremer}, Kyle and {Ye}, Claire S. and {Andrews}, Jeff J. and {Kurkowski}, Michael and {Digman}, Matthew C. and {Larson}, Shane L. and {Rasio}, Frederic A.},
        title = "{COSMIC Variance in Binary Population Synthesis}",
      journal = {\apj},
         year = 2020,
        month = jul,
       volume = {898},
       number = {1},
          eid = {71},
        pages = {71},
          doi = {10.3847/1538-4357/ab9d85},
archivePrefix = {arXiv},
       eprint = {1911.00903},
 primaryClass = {astro-ph.HE},
       adsurl = {https://ui.adsabs.harvard.edu/abs/2020ApJ...898...71B}
}

@ARTICLE{paired,
       author = {{Chance}, Quadry and {Foreman-Mackey}, Daniel and {Ballard}, Sarah and {Casey}, Andrew and {David}, Trevor and {Price-Whelan}, Adrian},
        title = "{paired: A Statistical Framework for Determining Stellar Binarity with Gaia RVs. I. Planet Hosting Binaries}",
      journal = {arXiv e-prints},
         year = 2022,
        month = jun,
          eid = {arXiv:2206.11275},
        pages = {arXiv:2206.11275},
          doi = {10.48550/arXiv.2206.11275},
archivePrefix = {arXiv},
       eprint = {2206.11275},
 primaryClass = {astro-ph.EP},
       adsurl = {https://ui.adsabs.harvard.edu/abs/2022arXiv220611275C}
}

@ARTICLE{thejoker,
       author = {{Price-Whelan}, Adrian M. and {Hogg}, David W. and {Foreman-Mackey}, Daniel and {Rix}, Hans-Walter},
        title = "{The Joker: A Custom Monte Carlo Sampler for Binary-star and Exoplanet Radial Velocity Data}",
      journal = {\apj},
         year = 2017,
        month = mar,
       volume = {837},
       number = {1},
          eid = {20},
        pages = {20},
          doi = {10.3847/1538-4357/aa5e50},
archivePrefix = {arXiv},
       eprint = {1610.07602},
 primaryClass = {astro-ph.SR},
       adsurl = {https://ui.adsabs.harvard.edu/abs/2017ApJ...837...20P}
}

@ARTICLE{apw_2020,
       author = {{Price-Whelan}, Adrian M. and {Hogg}, David W. and {Rix}, Hans-Walter and {Beaton}, Rachael L. and {Lewis}, Hannah M. and {Nidever}, David L. and {Almeida}, Andr{\'e}s and {Badenes}, Carles and {Barba}, Rodolfo and {Beers}, Timothy C. and {Carlberg}, Joleen K. and {De Lee}, Nathan and {Fern{\'a}ndez-Trincado}, Jos{\'e} G. and {Frinchaboy}, Peter M. and {Garc{\'\i}a-Hern{\'a}ndez}, D.~A. and {Green}, Paul J. and {Hasselquist}, Sten and {Longa-Pe{\~n}a}, Pen{\'e}lope and {Majewski}, Steven R. and {Nitschelm}, Christian and {Sobeck}, Jennifer and {Stassun}, Keivan G. and {Stringfellow}, Guy S. and {Troup}, Nicholas W.},
        title = "{Close Binary Companions to APOGEE DR16 Stars: 20,000 Binary-star Systems Across the Color-Magnitude Diagram}",
      journal = {\apj},
         year = 2020,
        month = may,
       volume = {895},
       number = {1},
          eid = {2},
        pages = {2},
          doi = {10.3847/1538-4357/ab8acc},
archivePrefix = {arXiv},
       eprint = {2002.00014},
 primaryClass = {astro-ph.SR},
       adsurl = {https://ui.adsabs.harvard.edu/abs/2020ApJ...895....2P}
}

@ARTICLE{Martell2021,
       author = {{Martell}, Sarah L. and {Simpson}, Jeffrey D. and {Balasubramaniam}, Adithya G. and {Buder}, Sven and {Sharma}, Sanjib and {Hon}, Marc and {Stello}, Dennis and {Ting}, Yuan-Sen and {Asplund}, Martin and {Bland-Hawthorn}, Joss and {De Silva}, Gayandhi M. and {Freeman}, Ken C. and {Hayden}, Michael and {Kos}, Janez and {Lewis}, Geraint F. and {Lind}, Karin and {Zucker}, Daniel B. and {Zwitter}, Toma{\v{z}} and {Campbell}, Simon W. and {{\v{C}}otar}, Klemen and {Horner}, Jonathan and {Montet}, Benjamin and {Wittenmyer}, Rob},
        title = "{The GALAH survey: a census of lithium-rich giant stars}",
      journal = {\mnras},
         year = 2021,
        month = aug,
       volume = {505},
       number = {4},
        pages = {5340-5355},
          doi = {10.1093/mnras/stab1356},
archivePrefix = {arXiv},
       eprint = {2006.02106},
 primaryClass = {astro-ph.SR},
       adsurl = {https://ui.adsabs.harvard.edu/abs/2021MNRAS.505.5340M}
}

@ARTICLE{galah_sven_2021,
       author = {{Buder}, Sven and {Sharma}, Sanjib and {Kos}, Janez and {Amarsi}, Anish M. and {Nordlander}, Thomas and {Lind}, Karin and {Martell}, Sarah L. and {Asplund}, Martin and {Bland-Hawthorn}, Joss and {Casey}, Andrew R. and {de Silva}, Gayandhi M. and {D'Orazi}, Valentina and {Freeman}, Ken C. and {Hayden}, Michael R. and {Lewis}, Geraint F. and {Lin}, Jane and {Schlesinger}, Katharine J. and {Simpson}, Jeffrey D. and {Stello}, Dennis and {Zucker}, Daniel B. and {Zwitter}, Toma{\v{z}} and {Beeson}, Kevin L. and {Buck}, Tobias and {Casagrande}, Luca and {Clark}, Jake T. and {{\v{C}}otar}, Klemen and {da Costa}, Gary S. and {de Grijs}, Richard and {Feuillet}, Diane and {Horner}, Jonathan and {Kafle}, Prajwal R. and {Khanna}, Shourya and {Kobayashi}, Chiaki and {Liu}, Fan and {Montet}, Benjamin T. and {Nandakumar}, Govind and {Nataf}, David M. and {Ness}, Melissa K. and {Spina}, Lorenzo and {Tepper-Garc{\'\i}a}, Thor and {Ting}, Yuan-Sen and {Traven}, Gregor and {Vogrin{\v{c}}i{\v{c}}}, Rok and {Wittenmyer}, Robert A. and {Wyse}, Rosemary F.~G. and {{\v{Z}}erjal}, Maru{\v{s}}a and {GALAH Collaboration}},
        title = "{The GALAH+ survey: Third data release}",
      journal = {\mnras},
         year = 2021,
        month = sep,
       volume = {506},
       number = {1},
        pages = {150-201},
          doi = {10.1093/mnras/stab1242},
archivePrefix = {arXiv},
       eprint = {2011.02505},
 primaryClass = {astro-ph.GA},
       adsurl = {https://ui.adsabs.harvard.edu/abs/2021MNRAS.506..150B}
}

@ARTICLE{Castro-Tapia2024,
       author = {{Castro-Tapia}, Matias and {Aguilera-G{\'o}mez}, Claudia and {Chanam{\'e}}, Julio},
        title = "{Are lithium-rich giants binaries? A radial velocity variability analysis of 1400 giants}",
      journal = {\aap},
         year = 2024,
        month = oct,
       volume = {690},
          eid = {A367},
        pages = {A367},
          doi = {10.1051/0004-6361/202349106},
archivePrefix = {arXiv},
       eprint = {2401.00049},
 primaryClass = {astro-ph.SR},
       adsurl = {https://ui.adsabs.harvard.edu/abs/2024A&A...690A.367C}
}

@ARTICLE{Pepe2002,
       author = {{Pepe}, F. and {Mayor}, M. and {Galland}, F. and {Naef}, D. and {Queloz}, D. and {Santos}, N.~C. and {Udry}, S. and {Burnet}, M.},
        title = "{The CORALIE survey for southern extra-solar planets VII. Two short-period Saturnian companions to <ASTROBJ>HD 108147</ASTROBJ> and <ASTROBJ>HD 168746</ASTROBJ>}",
      journal = {\aap},
         year = 2002,
        month = jun,
       volume = {388},
        pages = {632-638},
          doi = {10.1051/0004-6361:20020433},
archivePrefix = {arXiv},
       eprint = {astro-ph/0202457},
 primaryClass = {astro-ph},
       adsurl = {https://ui.adsabs.harvard.edu/abs/2002A&A...388..632P}
}

@ARTICLE{sayeed_2024,
       author = {{Sayeed}, Maryum and {Ness}, Melissa K. and {Montet}, Benjamin T. and {Cantiello}, Matteo and {Casey}, Andrew R. and {Buder}, Sven and {Bedell}, Megan and {Breivik}, Katelyn and {Metzger}, Brian D. and {Martell}, Sarah L. and {McGee-Gold}, Leah},
        title = "{Many Roads Lead to Lithium: Formation Pathways For Lithium-rich Red Giants}",
      journal = {\apj},
         year = 2024,
        month = mar,
       volume = {964},
       number = {1},
          eid = {42},
        pages = {42},
          doi = {10.3847/1538-4357/ad1936},
archivePrefix = {arXiv},
       eprint = {2306.03323},
 primaryClass = {astro-ph.SR},
       adsurl = {https://ui.adsabs.harvard.edu/abs/2024ApJ...964...42S}
}

@ARTICLE{salpeter1955,
       author = {{Salpeter}, Edwin E.},
        title = "{The Luminosity Function and Stellar Evolution.}",
      journal = {\apj},
         year = 1955,
        month = jan,
       volume = {121},
        pages = {161},
          doi = {10.1086/145971},
       adsurl = {https://ui.adsabs.harvard.edu/abs/1955ApJ...121..161S}
}

@ARTICLE{Lind2009,
       author = {{Lind}, K. and {Primas}, F. and {Charbonnel}, C. and {Grundahl}, F. and {Asplund}, M.},
        title = "{Signatures of intrinsic Li depletion and Li-Na anti-correlation in the metal-poor globular cluster NGC 6397}",
      journal = {\aap},
         year = 2009,
        month = aug,
       volume = {503},
       number = {2},
        pages = {545-557},
          doi = {10.1051/0004-6361/200912524},
archivePrefix = {arXiv},
       eprint = {0906.2876},
 primaryClass = {astro-ph.SR},
       adsurl = {https://ui.adsabs.harvard.edu/abs/2009A&A...503..545L}
}

@ARTICLE{2018AJ....156..123A,
       author = {{Astropy Collaboration} and {Price-Whelan}, A.~M. and {Sip{\H{o}}cz}, B.~M. and {G{\"u}nther}, H.~M. and {Lim}, P.~L. and {Crawford}, S.~M. and {Conseil}, S. and {Shupe}, D.~L. and {Craig}, M.~W. and {Dencheva}, N. and {Ginsburg}, A. and {VanderPlas}, J.~T. and {Bradley}, L.~D. and {P{\'e}rez-Su{\'a}rez}, D. and {de Val-Borro}, M. and {Aldcroft}, T.~L. and {Cruz}, K.~L. and {Robitaille}, T.~P. and {Tollerud}, E.~J. and {Ardelean}, C. and {Babej}, T. and {Bach}, Y.~P. and {Bachetti}, M. and {Bakanov}, A.~V. and {Bamford}, S.~P. and {Barentsen}, G. and {Barmby}, P. and {Baumbach}, A. and {Berry}, K.~L. and {Biscani}, F. and {Boquien}, M. and {Bostroem}, K.~A. and {Bouma}, L.~G. and {Brammer}, G.~B. and {Bray}, E.~M. and {Breytenbach}, H. and {Buddelmeijer}, H. and {Burke}, D.~J. and {Calderone}, G. and {Cano Rodr{\'\i}guez}, J.~L. and {Cara}, M. and {Cardoso}, J.~V.~M. and {Cheedella}, S. and {Copin}, Y. and {Corrales}, L. and {Crichton}, D. and {D'Avella}, D. and {Deil}, C. and {Depagne}, {\'E}. and {Dietrich}, J.~P. and {Donath}, A. and {Droettboom}, M. and {Earl}, N. and {Erben}, T. and {Fabbro}, S. and {Ferreira}, L.~A. and {Finethy}, T. and {Fox}, R.~T. and {Garrison}, L.~H. and {Gibbons}, S.~L.~J. and {Goldstein}, D.~A. and {Gommers}, R. and {Greco}, J.~P. and {Greenfield}, P. and {Groener}, A.~M. and {Grollier}, F. and {Hagen}, A. and {Hirst}, P. and {Homeier}, D. and {Horton}, A.~J. and {Hosseinzadeh}, G. and {Hu}, L. and {Hunkeler}, J.~S. and {Ivezi{\'c}}, {\v{Z}}. and {Jain}, A. and {Jenness}, T. and {Kanarek}, G. and {Kendrew}, S. and {Kern}, N.~S. and {Kerzendorf}, W.~E. and {Khvalko}, A. and {King}, J. and {Kirkby}, D. and {Kulkarni}, A.~M. and {Kumar}, A. and {Lee}, A. and {Lenz}, D. and {Littlefair}, S.~P. and {Ma}, Z. and {Macleod}, D.~M. and {Mastropietro}, M. and {McCully}, C. and {Montagnac}, S. and {Morris}, B.~M. and {Mueller}, M. and {Mumford}, S.~J. and {Muna}, D. and {Murphy}, N.~A. and {Nelson}, S. and {Nguyen}, G.~H. and {Ninan}, J.~P. and {N{\"o}the}, M. and {Ogaz}, S. and {Oh}, S. and {Parejko}, J.~K. and {Parley}, N. and {Pascual}, S. and {Patil}, R. and {Patil}, A.~A. and {Plunkett}, A.~L. and {Prochaska}, J.~X. and {Rastogi}, T. and {Reddy Janga}, V. and {Sabater}, J. and {Sakurikar}, P. and {Seifert}, M. and {Sherbert}, L.~E. and {Sherwood-Taylor}, H. and {Shih}, A.~Y. and {Sick}, J. and {Silbiger}, M.~T. and {Singanamalla}, S. and {Singer}, L.~P. and {Sladen}, P.~H. and {Sooley}, K.~A. and {Sornarajah}, S. and {Streicher}, O. and {Teuben}, P. and {Thomas}, S.~W. and {Tremblay}, G.~R. and {Turner}, J.~E.~H. and {Terr{\'o}n}, V. and {van Kerkwijk}, M.~H. and {de la Vega}, A. and {Watkins}, L.~L. and {Weaver}, B.~A. and {Whitmore}, J.~B. and {Woillez}, J. and {Zabalza}, V. and {Astropy Contributors}},
        title = "{The Astropy Project: Building an Open-science Project and Status of the v2.0 Core Package}",
      journal = {\aj},
         year = 2018,
        month = sep,
       volume = {156},
       number = {3},
          eid = {123},
        pages = {123},
          doi = {10.3847/1538-3881/aabc4f},
archivePrefix = {arXiv},
       eprint = {1801.02634},
 primaryClass = {astro-ph.IM},
       adsurl = {https://ui.adsabs.harvard.edu/abs/2018AJ....156..123A}
}

@ARTICLE{2013A&A...558A..33A,
       author = {{Astropy Collaboration} and {Robitaille}, Thomas P. and
         {Tollerud}, Erik J. and {Greenfield}, Perry and {Droettboom}, Michael and
         {Bray}, Erik and {Aldcroft}, Tom and {Davis}, Matt and
         {Ginsburg}, Adam and {Price-Whelan}, Adrian M. and
         {Kerzendorf}, Wolfgang E. and {Conley}, Alexander and {Crighton}, Neil and
         {Barbary}, Kyle and {Muna}, Demitri and {Ferguson}, Henry and
         {Grollier}, Fr{\'e}d{\'e}ric and {Parikh}, Madhura M. and
         {Nair}, Prasanth H. and {Unther}, Hans M. and {Deil}, Christoph and
         {Woillez}, Julien and {Conseil}, Simon and {Kramer}, Roban and
         {Turner}, James E.~H. and {Singer}, Leo and {Fox}, Ryan and
         {Weaver}, Benjamin A. and {Zabalza}, Victor and {Edwards}, Zachary I. and
         {Azalee Bostroem}, K. and {Burke}, D.~J. and {Casey}, Andrew R. and
         {Crawford}, Steven M. and {Dencheva}, Nadia and {Ely}, Justin and
         {Jenness}, Tim and {Labrie}, Kathleen and {Lim}, Pey Lian and
         {Pierfederici}, Francesco and {Pontzen}, Andrew and {Ptak}, Andy and
         {Refsdal}, Brian and {Servillat}, Mathieu and {Streicher}, Ole},
        title = "{Astropy: A community Python package for astronomy}",
      journal = {\aap},
         year = "2013",
        month = "Oct",
       volume = {558},
          eid = {A33},
        pages = {A33},
          doi = {10.1051/0004-6361/201322068},
archivePrefix = {arXiv},
       eprint = {1307.6212},
 primaryClass = {astro-ph.IM},
       adsurl = {https://ui.adsabs.harvard.edu/abs/2013A&A...558A..33A}
}

@ARTICLE{Rui2024,
       author = {{Rui}, Nicholas Z. and {Fuller}, Jim},
        title = "{Finding the unusual red giant remnants of cataclysmic variable mergers}",
      journal = {The Open Journal of Astrophysics},
         year = 2024,
        month = sep,
       volume = {7},
          eid = {81},
        pages = {81},
          doi = {10.33232/001c.123878},
archivePrefix = {arXiv},
       eprint = {2404.14474},
 primaryClass = {astro-ph.SR},
       adsurl = {https://ui.adsabs.harvard.edu/abs/2024OJAp....7E..81R}
}

@ARTICLE{Zhang2013,
       author = {{Zhang}, Xianfei and {Jeffery}, C. Simon},
        title = "{White dwarf-red giant mergers, early-type R stars, J stars and lithium}",
      journal = {\mnras},
         year = 2013,
        month = apr,
       volume = {430},
       number = {3},
        pages = {2113-2120},
          doi = {10.1093/mnras/stt035},
archivePrefix = {arXiv},
       eprint = {1301.0766},
 primaryClass = {astro-ph.SR},
       adsurl = {https://ui.adsabs.harvard.edu/abs/2013MNRAS.430.2113Z}
}

@ARTICLE{Zhang2020,
       author = {{Zhang}, Xianfei and {Jeffery}, C. Simon and {Li}, Yaguang and {Bi}, Shaolan},
        title = "{Population Synthesis of Helium White Dwarf-Red Giant Star Mergers and the Formation of Lithium-rich Giants and Carbon Stars}",
      journal = {\apj},
         year = 2020,
        month = jan,
       volume = {889},
       number = {1},
          eid = {33},
        pages = {33},
          doi = {10.3847/1538-4357/ab5e89},
archivePrefix = {arXiv},
       eprint = {2001.05600},
 primaryClass = {astro-ph.SR},
       adsurl = {https://ui.adsabs.harvard.edu/abs/2020ApJ...889...33Z}
}

@ARTICLE{casey_2019,
       author = {{Casey}, Andrew R. and {Ho}, Anna Y.~Q. and {Ness}, Melissa and {Hogg}, David W. and {Rix}, Hans-Walter and {Angelou}, George C. and {Hekker}, Saskia and {Tout}, Christopher A. and {Lattanzio}, John C. and {Karakas}, Amanda I. and {Woods}, Tyrone E. and {Price-Whelan}, Adrian M. and {Schlaufman}, Kevin C.},
        title = "{Tidal Interactions between Binary Stars Can Drive Lithium Production in Low-mass Red Giants}",
      journal = {\apj},
         year = 2019,
        month = aug,
       volume = {880},
       number = {2},
          eid = {125},
        pages = {125},
          doi = {10.3847/1538-4357/ab27bf},
archivePrefix = {arXiv},
       eprint = {1902.04102},
 primaryClass = {astro-ph.SR},
       adsurl = {https://ui.adsabs.harvard.edu/abs/2019ApJ...880..125C}
}

@ARTICLE{Ming-hao_2021,
       author = {{Ming-hao}, DU and {Shao-lan}, BI and {Jian-rong}, SHI and {Hong-liang}, YAN},
        title = "{Surface Rotation of LAMOST-Kepler Li-rich Giant Stars}",
      journal = {\caa},
         year = 2021,
        month = jan,
       volume = {45},
       number = {1},
        pages = {45-57},
          doi = {10.1016/j.chinastron.2021.02.003},
       adsurl = {https://ui.adsabs.harvard.edu/abs/2021ChA&A..45...45M}
}

@ARTICLE{kumar_2020,
       author = {{Kumar}, Yerra Bharat and {Reddy}, Bacham E. and {Campbell}, Simon W. and {Maben}, Sunayana and {Zhao}, Gang and {Ting}, Yuan-Sen},
        title = "{Discovery of ubiquitous lithium production in low-mass stars}",
      journal = {Nature Astronomy},
         year = 2020,
        month = jul,
       volume = {4},
        pages = {1059-1063},
          doi = {10.1038/s41550-020-1139-7},
archivePrefix = {arXiv},
       eprint = {2007.07045},
 primaryClass = {astro-ph.SR},
       adsurl = {https://ui.adsabs.harvard.edu/abs/2020NatAs...4.1059K}
}

@ARTICLE{yan_2022,
       author = {{Yan}, T. -S. and {Shi}, J. -R. and {Wang}, L. and {Yan}, H. -L. and {Zhou}, Z. -M. and {Zhou}, Y. -T. and {Fang}, X. -S. and {Li}, C. -Q. and {Chen}, T. -Y. and {Xie}, X. -J.},
        title = "{Discovery of Nine Super Li-rich Unevolved Stars from the LAMOST Survey}",
      journal = {\apjl},
         year = 2022,
        month = apr,
       volume = {929},
       number = {1},
          eid = {L14},
        pages = {L14},
          doi = {10.3847/2041-8213/ac63a5},
archivePrefix = {arXiv},
       eprint = {2204.03180},
 primaryClass = {astro-ph.SR},
       adsurl = {https://ui.adsabs.harvard.edu/abs/2022ApJ...929L..14Y}
}

@ARTICLE{proxima_espresso,
       author = {{Su{\'a}rez Mascare{\~n}o}, A. and {Faria}, J.~P. and {Figueira}, P. and {Lovis}, C. and {Damasso}, M. and {Gonz{\'a}lez Hern{\'a}ndez}, J.~I. and {Rebolo}, R. and {Cristiani}, S. and {Pepe}, F. and {Santos}, N.~C. and {Zapatero Osorio}, M.~R. and {Adibekyan}, V. and {Hojjatpanah}, S. and {Sozzetti}, A. and {Murgas}, F. and {Abreu}, M. and {Affolter}, M. and {Alibert}, Y. and {Aliverti}, M. and {Allart}, R. and {Allende Prieto}, C. and {Alves}, D. and {Amate}, M. and {Avila}, G. and {Baldini}, V. and {Bandi}, T. and {Barros}, S.~C.~C. and {Bianco}, A. and {Benz}, W. and {Bouchy}, F. and {Broeng}, C. and {Cabral}, A. and {Calderone}, G. and {Cirami}, R. and {Coelho}, J. and {Conconi}, P. and {Coretti}, I. and {Cumani}, C. and {Cupani}, G. and {D'Odorico}, V. and {Deiries}, S. and {Delabre}, B. and {Di Marcantonio}, P. and {Dumusque}, X. and {Ehrenreich}, D. and {Fragoso}, A. and {Genolet}, L. and {Genoni}, M. and {G{\'e}nova Santos}, R. and {Hughes}, I. and {Iwert}, O. and {Kerber}, F. and {Knusdstrup}, J. and {Landoni}, M. and {Lavie}, B. and {Lillo-Box}, J. and {Lizon}, J. and {Lo Curto}, G. and {Maire}, C. and {Manescau}, A. and {Martins}, C.~J.~A.~P. and {M{\'e}gevand}, D. and {Mehner}, A. and {Micela}, G. and {Modigliani}, A. and {Molaro}, P. and {Monteiro}, M.~A. and {Monteiro}, M.~J.~P.~F.~G. and {Moschetti}, M. and {Mueller}, E. and {Nunes}, N.~J. and {Oggioni}, L. and {Oliveira}, A. and {Pall{\'e}}, E. and {Pariani}, G. and {Pasquini}, L. and {Poretti}, E. and {Rasilla}, J.~L. and {Redaelli}, E. and {Riva}, M. and {Santana Tschudi}, S. and {Santin}, P. and {Santos}, P. and {Segovia}, A. and {Sosnowska}, D. and {Sousa}, S. and {Span{\`o}}, P. and {Tenegi}, F. and {Udry}, S. and {Zanutta}, A. and {Zerbi}, F.},
        title = "{Revisiting Proxima with ESPRESSO}",
      journal = {\aap},
         year = 2020,
        month = jul,
       volume = {639},
          eid = {A77},
        pages = {A77},
          doi = {10.1051/0004-6361/202037745},
archivePrefix = {arXiv},
       eprint = {2005.12114},
 primaryClass = {astro-ph.EP},
       adsurl = {https://ui.adsabs.harvard.edu/abs/2020A&A...639A..77S}
}

@ARTICLE{LilloBox2021,
       author = {{Lillo-Box}, J. and {Faria}, J.~P. and {Su{\'a}rez Mascare{\~n}o}, A. and {Figueira}, P. and {Sousa}, S.~G. and {Tabernero}, H. and {Lovis}, C. and {Silva}, A.~M. and {Demangeon}, O.~D.~S. and {Benatti}, S. and {Santos}, N.~C. and {Mehner}, A. and {Pepe}, F.~A. and {Sozzetti}, A. and {Zapatero Osorio}, M.~R. and {Gonz{\'a}lez Hern{\'a}ndez}, J.~I. and {Micela}, G. and {Hojjatpanah}, S. and {Rebolo}, R. and {Cristiani}, S. and {Adibekyan}, V. and {Allart}, R. and {Allende Prieto}, C. and {Cabral}, A. and {Damasso}, M. and {Di Marcantonio}, P. and {Lo Curto}, G. and {Martins}, C.~J.~A.~P. and {Megevand}, D. and {Molaro}, P. and {Nunes}, N.~J. and {Pall{\'e}}, E. and {Pasquini}, L. and {Poretti}, E. and {Udry}, S.},
        title = "{HD 22496 b: The first ESPRESSO stand-alone planet discovery}",
      journal = {\aap},
         year = 2021,
        month = oct,
       volume = {654},
          eid = {A60},
        pages = {A60},
          doi = {10.1051/0004-6361/202141714},
archivePrefix = {arXiv},
       eprint = {2109.00226},
 primaryClass = {astro-ph.EP},
       adsurl = {https://ui.adsabs.harvard.edu/abs/2021A&A...654A..60L}
}

@ARTICLE{Tremblay2016,
       author = {{Tremblay}, P. -E. and {Cummings}, J. and {Kalirai}, J.~S. and {G{\"a}nsicke}, B.~T. and {Gentile-Fusillo}, N. and {Raddi}, R.},
        title = "{The field white dwarf mass distribution}",
      journal = {\mnras},
         year = 2016,
        month = sep,
       volume = {461},
       number = {2},
        pages = {2100-2114},
          doi = {10.1093/mnras/stw1447},
archivePrefix = {arXiv},
       eprint = {1606.05292},
 primaryClass = {astro-ph.SR},
       adsurl = {https://ui.adsabs.harvard.edu/abs/2016MNRAS.461.2100T}
}

@ARTICLE{Tremblay2019,
       author = {{Tremblay}, P. -E. and {Cukanovaite}, E. and {Gentile Fusillo}, N.~P. and {Cunningham}, T. and {Hollands}, M.~A.},
        title = "{Fundamental parameter accuracy of DA and DB white dwarfs in Gaia Data Release 2}",
      journal = {\mnras},
         year = 2019,
        month = feb,
       volume = {482},
       number = {4},
        pages = {5222-5232},
          doi = {10.1093/mnras/sty3067},
archivePrefix = {arXiv},
       eprint = {1811.03084},
 primaryClass = {astro-ph.SR},
       adsurl = {https://ui.adsabs.harvard.edu/abs/2019MNRAS.482.5222T}
}

@ARTICLE{Shahaf2025,
       author = {{Shahaf}, Sahar},
        title = "{The White Dwarf Pareto: Tracing Mass Loss in Binary Systems}",
      journal = {\apj},
         year = 2025,
        month = mar,
       volume = {981},
       number = {1},
          eid = {54},
        pages = {54},
          doi = {10.3847/1538-4357/adb156},
archivePrefix = {arXiv},
       eprint = {2501.09586},
 primaryClass = {astro-ph.SR},
       adsurl = {https://ui.adsabs.harvard.edu/abs/2025ApJ...981...54S}
}

@ARTICLE{Unger2023,
       author = {{Unger}, N. and {S{\'e}gransan}, D. and {Barbato}, D. and {Delisle}, J. -B. and {Sahlmann}, J. and {Holl}, B. and {Udry}, S.},
        title = "{Exploring the brown dwarf desert with precision radial velocities and Gaia DR3 astrometric orbits}",
      journal = {\aap},
         year = 2023,
        month = dec,
       volume = {680},
          eid = {A16},
        pages = {A16},
          doi = {10.1051/0004-6361/202347578},
archivePrefix = {arXiv},
       eprint = {2310.02758},
 primaryClass = {astro-ph.EP},
       adsurl = {https://ui.adsabs.harvard.edu/abs/2023A&A...680A..16U}
}

@ARTICLE{Verbunt1995,
       author = {{Verbunt}, F. and {Phinney}, E.~S.},
        title = "{Tidal circularization and the eccentricity of binaries containing giant stars.}",
      journal = {\aap},
         year = 1995,
        month = apr,
       volume = {296},
        pages = {709},
       adsurl = {https://ui.adsabs.harvard.edu/abs/1995A&A...296..709V}
}

@ARTICLE{Tsantaki2023,
       author = {{Tsantaki}, M. and {Delgado-Mena}, E. and {Bossini}, D. and {Sousa}, S.~G. and {Pancino}, E. and {Martins}, J.~H.~C.},
        title = "{Search for lithium-rich giants in 32 open clusters with high-resolution spectroscopy★}",
      journal = {\aap},
         year = 2023,
        month = jun,
       volume = {674},
          eid = {A157},
        pages = {A157},
          doi = {10.1051/0004-6361/202244374},
archivePrefix = {arXiv},
       eprint = {2303.16124},
 primaryClass = {astro-ph.SR},
       adsurl = {https://ui.adsabs.harvard.edu/abs/2023A&A...674A.157T}
}

@ARTICLE{zhou_2022,
       author = {{Zhou}, Yutao and {Wang}, Chun and {Yan}, Hongliang and {Huang}, Yang and {Zhang}, Bo and {Ting}, Yuan-Sen and {Zhang}, Huawei and {Shi}, Jianrong},
        title = "{Li-rich Giants in LAMOST Survey. III. The Statistical Analysis of Li-rich Giants}",
      journal = {\apj},
         year = 2022,
        month = jun,
       volume = {931},
       number = {2},
          eid = {136},
        pages = {136},
          doi = {10.3847/1538-4357/ac6b3a},
archivePrefix = {arXiv},
       eprint = {2204.12759},
 primaryClass = {astro-ph.SR},
       adsurl = {https://ui.adsabs.harvard.edu/abs/2022ApJ...931..136Z}
}

@ARTICLE{singh_2019,
       author = {{Singh}, Raghubar and {Reddy}, Bacham E. and {Bharat Kumar}, Yerra and {Antia}, H.~M.},
        title = "{Survey of Li-rich Giants among Kepler and LAMOST Fields: Determination of Li-rich Giants{\textquoteright} Evolutionary Phase}",
      journal = {\apjl},
         year = 2019,
        month = jun,
       volume = {878},
       number = {1},
          eid = {L21},
        pages = {L21},
          doi = {10.3847/2041-8213/ab2599},
archivePrefix = {arXiv},
       eprint = {1906.03198},
 primaryClass = {astro-ph.SR},
       adsurl = {https://ui.adsabs.harvard.edu/abs/2019ApJ...878L..21S}
}

@ARTICLE{gao_2019,
       author = {{Gao}, Qi and {Shi}, Jian-Rong and {Yan}, Hong-Liang and {Yan}, Tai-Sheng and {Xiang}, Mao-Sheng and {Zhou}, Yu-Tao and {Li}, Chun-Qian and {Zhao}, Gang},
        title = "{Lithium-rich Giants in LAMOST Survey. I. The Catalog}",
      journal = {\apjs},
         year = 2019,
        month = dec,
       volume = {245},
       number = {2},
          eid = {33},
        pages = {33},
          doi = {10.3847/1538-4365/ab505c},
archivePrefix = {arXiv},
       eprint = {1912.08431},
 primaryClass = {astro-ph.SR},
       adsurl = {https://ui.adsabs.harvard.edu/abs/2019ApJS..245...33G}
}

@ARTICLE{melinda_2021,
       author = {{Soares-Furtado}, Melinda and {Cantiello}, Matteo and {MacLeod}, Morgan and {Ness}, Melissa K.},
        title = "{Lithium Enrichment Signatures of Planetary Engulfment Events in Evolved Stars}",
      journal = {\aj},
         year = 2021,
        month = dec,
       volume = {162},
       number = {6},
          eid = {273},
        pages = {273},
          doi = {10.3847/1538-3881/ac273c},
archivePrefix = {arXiv},
       eprint = {2002.05275},
 primaryClass = {astro-ph.SR},
       adsurl = {https://ui.adsabs.harvard.edu/abs/2021AJ....162..273S}
}

@ARTICLE{cameron_1971,
       author = {{Cameron}, A.~G.~W. and {Fowler}, W.~A.},
        title = "{Lithium and the s-PROCESS in Red-Giant Stars}",
      journal = {\apj},
         year = 1971,
        month = feb,
       volume = {164},
        pages = {111},
          doi = {10.1086/150821},
       adsurl = {https://ui.adsabs.harvard.edu/abs/1971ApJ...164..111C}
}

@ARTICLE{gomez_2016b,
       author = {{Aguilera-G{\'o}mez}, Claudia and {Chanam{\'e}}, Julio and {Pinsonneault}, Marc H. and {Carlberg}, Joleen K.},
        title = "{On Lithium-rich Red Giants: Engulfment on the Giant Branch of Trumpler 20}",
      journal = {\apjl},
         year = 2016,
        month = dec,
       volume = {833},
       number = {2},
          eid = {L24},
        pages = {L24},
          doi = {10.3847/2041-8213/833/2/L24},
archivePrefix = {arXiv},
       eprint = {1609.07492},
 primaryClass = {astro-ph.SR},
       adsurl = {https://ui.adsabs.harvard.edu/abs/2016ApJ...833L..24A}
}

@ARTICLE{gomez_2016a,
       author = {{Aguilera-G{\'o}mez}, Claudia and {Chanam{\'e}}, Julio and {Pinsonneault}, Marc H. and {Carlberg}, Joleen K.},
        title = "{On Lithium-rich Red Giants. I. Engulfment of Substellar Companions}",
      journal = {\apj},
         year = 2016,
        month = oct,
       volume = {829},
       number = {2},
          eid = {127},
        pages = {127},
          doi = {10.3847/0004-637X/829/2/127},
archivePrefix = {arXiv},
       eprint = {1605.05332},
 primaryClass = {astro-ph.SR},
       adsurl = {https://ui.adsabs.harvard.edu/abs/2016ApJ...829..127A}
}

@ARTICLE{adamow_2012,
       author = {{Adam{\'o}w}, M. and {Niedzielski}, A. and {Villaver}, E. and {Nowak}, G. and {Wolszczan}, A.},
        title = "{BD+48 740{\textemdash}Li Overabundant Giant Star with a Planet: A Case of Recent Engulfment?}",
      journal = {\apjl},
         year = 2012,
        month = jul,
       volume = {754},
       number = {1},
          eid = {L15},
        pages = {L15},
          doi = {10.1088/2041-8205/754/1/L15},
archivePrefix = {arXiv},
       eprint = {1206.4938},
 primaryClass = {astro-ph.SR},
       adsurl = {https://ui.adsabs.harvard.edu/abs/2012ApJ...754L..15A}
}

@ARTICLE{galah_survey,
       author = {{De Silva}, G.~M. and {Freeman}, K.~C. and {Bland-Hawthorn}, J. and {Martell}, S. and {de Boer}, E. Wylie and {Asplund}, M. and {Keller}, S. and {Sharma}, S. and {Zucker}, D.~B. and {Zwitter}, T. and {Anguiano}, B. and {Bacigalupo}, C. and {Bayliss}, D. and {Beavis}, M.~A. and {Bergemann}, M. and {Campbell}, S. and {Cannon}, R. and {Carollo}, D. and {Casagrande}, L. and {Casey}, A.~R. and {Da Costa}, G. and {D'Orazi}, V. and {Dotter}, A. and {Duong}, L. and {Heger}, A. and {Ireland}, M.~J. and {Kafle}, P.~R. and {Kos}, J. and {Lattanzio}, J. and {Lewis}, G.~F. and {Lin}, J. and {Lind}, K. and {Munari}, U. and {Nataf}, D.~M. and {O'Toole}, S. and {Parker}, Q. and {Reid}, W. and {Schlesinger}, K.~J. and {Sheinis}, A. and {Simpson}, J.~D. and {Stello}, D. and {Ting}, Y. -S. and {Traven}, G. and {Watson}, F. and {Wittenmyer}, R. and {Yong}, D. and {{\v{Z}}erjal}, M.},
        title = "{The GALAH survey: scientific motivation}",
      journal = {\mnras},
         year = 2015,
        month = may,
       volume = {449},
       number = {3},
        pages = {2604-2617},
          doi = {10.1093/mnras/stv327},
archivePrefix = {arXiv},
       eprint = {1502.04767},
 primaryClass = {astro-ph.GA},
       adsurl = {https://ui.adsabs.harvard.edu/abs/2015MNRAS.449.2604D}
}

@ARTICLE{deepak_2020,
       author = {{Deepak} and {Lambert}, David L. and {Reddy}, Bacham E.},
        title = "{Abundance analyses of Li-enriched and normal giants in the GALAH survey}",
      journal = {\mnras},
         year = 2020,
        month = may,
       volume = {494},
       number = {1},
        pages = {1348-1365},
          doi = {10.1093/mnras/staa729},
archivePrefix = {arXiv},
       eprint = {2003.05381},
 primaryClass = {astro-ph.SR},
       adsurl = {https://ui.adsabs.harvard.edu/abs/2020MNRAS.494.1348D}
}

@ARTICLE{chaname_2022,
       author = {{Chanam{\'e}}, Julio and {Pinsonneault}, Marc H. and {Aguilera-G{\'o}mez}, Claudia and {Zinn}, Joel C.},
        title = "{Mass Matters: No Evidence for Ubiquitous Lithium Production in Low-mass Clump Giants}",
      journal = {\apj},
         year = 2022,
        month = jul,
       volume = {933},
       number = {1},
          eid = {58},
        pages = {58},
          doi = {10.3847/1538-4357/ac70c8},
archivePrefix = {arXiv},
       eprint = {2109.13955},
 primaryClass = {astro-ph.SR},
       adsurl = {https://ui.adsabs.harvard.edu/abs/2022ApJ...933...58C}
}

@ARTICLE{choi_2016,
       author = {{Choi}, Jieun and {Dotter}, Aaron and {Conroy}, Charlie and {Cantiello}, Matteo and {Paxton}, Bill and {Johnson}, Benjamin D.},
        title = "{Mesa Isochrones and Stellar Tracks (MIST). I. Solar-scaled Models}",
      journal = {\apj},
         year = 2016,
        month = jun,
       volume = {823},
       number = {2},
          eid = {102},
        pages = {102},
          doi = {10.3847/0004-637X/823/2/102},
archivePrefix = {arXiv},
       eprint = {1604.08592},
 primaryClass = {astro-ph.SR},
       adsurl = {https://ui.adsabs.harvard.edu/abs/2016ApJ...823..102C}
}

@ARTICLE{carlberg_2012,
       author = {{Carlberg}, Joleen K. and {Cunha}, Katia and {Smith}, Verne V. and {Majewski}, Steven R.},
        title = "{Observable Signatures of Planet Accretion in Red Giant Stars. I. Rapid Rotation and Light Element Replenishment}",
      journal = {\apj},
         year = 2012,
        month = oct,
       volume = {757},
       number = {2},
          eid = {109},
        pages = {109},
          doi = {10.1088/0004-637X/757/2/109},
archivePrefix = {arXiv},
       eprint = {1208.1775},
 primaryClass = {astro-ph.SR},
       adsurl = {https://ui.adsabs.harvard.edu/abs/2012ApJ...757..109C}
}

@ARTICLE{deepak_reddy_2019,
       author = {{Deepak} and {Reddy}, Bacham E.},
        title = "{Study of Lithium-rich giants with the GALAH spectroscopic survey}",
      journal = {\mnras},
         year = 2019,
        month = apr,
       volume = {484},
       number = {2},
        pages = {2000-2008},
          doi = {10.1093/mnras/stz128},
archivePrefix = {arXiv},
       eprint = {1901.03832},
 primaryClass = {astro-ph.SR},
       adsurl = {https://ui.adsabs.harvard.edu/abs/2019MNRAS.484.2000D}
}

@ARTICLE{yan_2021,
       author = {{Yan}, Hong-Liang and {Zhou}, Yu-Tao and {Zhang}, Xianfei and {Li}, Yaguang and {Gao}, Qi and {Shi}, Jian-Rong and {Zhao}, Gang and {Aoki}, Wako and {Matsuno}, Tadafumi and {Li}, Yan and {Xu}, Xiao-Dong and {Li}, Haining and {Wu}, Ya-Qian and {Jin}, Meng-Qi and {Mosser}, Benoit and {Bi}, Shao-Lan and {Fu}, Jian-Ning and {Pan}, Kaike and {Suda}, Takuma and {Liu}, Yu-Juan and {Zhao}, Jing-Kun and {Liang}, Xi-Long},
        title = "{Most lithium-rich low-mass evolved stars revealed as red clump stars by asteroseismology and spectroscopy}",
      journal = {Nature Astronomy},
         year = 2021,
        month = jan,
       volume = {5},
        pages = {86-93},
          doi = {10.1038/s41550-020-01217-8},
archivePrefix = {arXiv},
       eprint = {2010.02106},
 primaryClass = {astro-ph.SR},
       adsurl = {https://ui.adsabs.harvard.edu/abs/2021NatAs...5...86Y}
}

@ARTICLE{SackmannBoothroyd1992,
       author = {{Sackmann}, I. -J. and {Boothroyd}, Arnold I.},
        title = "{The Creation of Superrich Lithium Giants}",
      journal = {\apjl},
         year = 1992,
        month = jun,
       volume = {392},
        pages = {L71},
          doi = {10.1086/186428},
       adsurl = {https://ui.adsabs.harvard.edu/abs/1992ApJ...392L..71S}
}

@ARTICLE{Alexander1967,
       author = {{Alexander}, J.~B.},
        title = "{A possible source of lithium in the atmospheres of some red giants}",
      journal = {The Observatory},
         year = 1967,
        month = oct,
       volume = {87},
        pages = {238-240},
       adsurl = {https://ui.adsabs.harvard.edu/abs/1967Obs....87..238A}
}

@ARTICLE{SiessLivio1999b,
       author = {{Siess}, Lionel and {Livio}, Mario},
        title = "{The accretion of brown dwarfs and planets by giant stars - II. Solar-mass stars on the red giant branch}",
      journal = {\mnras},
         year = 1999,
        month = oct,
       volume = {308},
       number = {4},
        pages = {1133-1149},
          doi = {10.1046/j.1365-8711.1999.02784.x},
archivePrefix = {arXiv},
       eprint = {astro-ph/9905235},
 primaryClass = {astro-ph},
       adsurl = {https://ui.adsabs.harvard.edu/abs/1999MNRAS.308.1133S}
}

@ARTICLE{SiessLivio1999a,
       author = {{Siess}, Lionel and {Livio}, Mario},
        title = "{The accretion of brown dwarfs and planets by giant stars - I. Asymptotic giant branch stars}",
      journal = {\mnras},
         year = 1999,
        month = apr,
       volume = {304},
       number = {4},
        pages = {925-937},
          doi = {10.1046/j.1365-8711.1999.02376.x},
archivePrefix = {arXiv},
       eprint = {astro-ph/9812220},
 primaryClass = {astro-ph},
       adsurl = {https://ui.adsabs.harvard.edu/abs/1999MNRAS.304..925S}
}

@ARTICLE{VillaverLivio2009,
       author = {{Villaver}, Eva and {Livio}, Mario},
        title = "{The Orbital Evolution of Gas Giant Planets Around Giant Stars}",
      journal = {\apjl},
         year = 2009,
        month = nov,
       volume = {705},
       number = {1},
        pages = {L81-L85},
          doi = {10.1088/0004-637X/705/1/L81},
archivePrefix = {arXiv},
       eprint = {0910.2396},
 primaryClass = {astro-ph.EP},
       adsurl = {https://ui.adsabs.harvard.edu/abs/2009ApJ...705L..81V}
}

@ARTICLE{King1997,
       author = {{King}, Jeremy R. and {Deliyannis}, Constantine P. and {Hiltgen}, Daniel D. and {Stephens}, Alex and {Cunha}, Katia and {Boesgaard}, Ann Merchant},
        title = "{Lithium Abundances in the Solar Twins 16 CYG A and B and the Solar Analog alpha CEN A, Calibration of the 6707 Angstrom Li Region Linelist, and Implications}",
      journal = {\aj},
         year = 1997,
        month = may,
       volume = {113},
        pages = {1871},
          doi = {10.1086/118399},
       adsurl = {https://ui.adsabs.harvard.edu/abs/1997AJ....113.1871K}
}

@ARTICLE{Magrini2021,
       author = {{Magrini}, L. and {Lagarde}, N. and {Charbonnel}, C. and {Franciosini}, E. and {Randich}, S. and {Smiljanic}, R. and {Casali}, G. and {Viscasillas V{\'a}zquez}, C. and {Spina}, L. and {Biazzo}, K. and {Pasquini}, L. and {Bragaglia}, A. and {Van der Swaelmen}, M. and {Tautvai{\v{s}}ien{\.{e}}}, G. and {Inno}, L. and {Sanna}, N. and {Prisinzano}, L. and {Degl'Innocenti}, S. and {Prada Moroni}, P. and {Roccatagliata}, V. and {Tognelli}, E. and {Monaco}, L. and {de Laverny}, P. and {Delgado-Mena}, E. and {Baratella}, M. and {D'Orazi}, V. and {Vallenari}, A. and {Gonneau}, A. and {Worley}, C. and {Jim{\'e}nez-Esteban}, F. and {Jofre}, P. and {Bensby}, T. and {Fran{\c{c}}ois}, P. and {Guiglion}, G. and {Bayo}, A. and {Jeffries}, R.~D. and {Binks}, A.~S. and {Gilmore}, G. and {Damiani}, F. and {Korn}, A. and {Pancino}, E. and {Sacco}, G.~G. and {Hourihane}, A. and {Morbidelli}, L. and {Zaggia}, S.},
        title = "{The Gaia-ESO survey: Mixing processes in low-mass stars traced by lithium abundance in cluster and field stars}",
      journal = {\aap},
         year = 2021,
        month = jul,
       volume = {651},
          eid = {A84},
        pages = {A84},
          doi = {10.1051/0004-6361/202140935},
archivePrefix = {arXiv},
       eprint = {2105.04866},
 primaryClass = {astro-ph.SR},
       adsurl = {https://ui.adsabs.harvard.edu/abs/2021A&A...651A..84M}
}

@ARTICLE{wheeler2021,
       author = {{Wheeler}, Adam J. and {Hogg}, David W. and {Ness}, Melissa},
        title = "{An Unsupervised Method for Identifying X-enriched Stars Directly from Spectra: Li in LAMOST}",
      journal = {\apj},
         year = 2021,
        month = feb,
       volume = {908},
       number = {2},
          eid = {247},
        pages = {247},
          doi = {10.3847/1538-4357/abd544},
archivePrefix = {arXiv},
       eprint = {2009.03920},
 primaryClass = {astro-ph.SR},
       adsurl = {https://ui.adsabs.harvard.edu/abs/2021ApJ...908..247W}
}

@ARTICLE{Gao2020,
       author = {{Gao}, Xudong and {Lind}, Karin and {Amarsi}, Anish M. and {Buder}, Sven and {Bland-Hawthorn}, Joss and {Campbell}, Simon W. and {Asplund}, Martin and {Casey}, Andrew R. and {de Silva}, Gayandhi M. and {Freeman}, Ken C. and {Hayden}, Michael R. and {Lewis}, Geraint F. and {Martell}, Sarah L. and {Simpson}, Jeffrey D. and {Sharma}, Sanjib and {Zucker}, Daniel B. and {Zwitter}, Toma{\v{z}} and {Horner}, Jonathan and {Munari}, Ulisse and {Nordlander}, Thomas and {Stello}, Dennis and {Ting}, Yuan-Sen and {Traven}, Gregor and {Wittenmyer}, Robert A. and {GALAH Collaboration}},
        title = "{The GALAH survey: a new constraint on cosmological lithium and Galactic lithium evolution from warm dwarf stars}",
      journal = {\mnras},
         year = 2020,
        month = sep,
       volume = {497},
       number = {1},
        pages = {L30-L34},
          doi = {10.1093/mnrasl/slaa109},
archivePrefix = {arXiv},
       eprint = {2006.05173},
 primaryClass = {astro-ph.SR},
       adsurl = {https://ui.adsabs.harvard.edu/abs/2020MNRAS.497L..30G}
}

@ARTICLE{WallersteinSneden1982,
       author = {{Wallerstein}, G. and {Sneden}, C.},
        title = "{A K giant with an unusually high abundance of lithium : HD 112127.}",
      journal = {\apj},
         year = 1982,
        month = apr,
       volume = {255},
        pages = {577-584},
          doi = {10.1086/159859},
       adsurl = {https://ui.adsabs.harvard.edu/abs/1982ApJ...255..577W}
}

@ARTICLE{Brown1989,
       author = {{Brown}, Jeffery A. and {Sneden}, Christopher and {Lambert}, David L. and {Dutchover}, Edward, Jr.},
        title = "{A Search for Lithium-rich Giant Stars}",
      journal = {\apjs},
         year = 1989,
        month = oct,
       volume = {71},
        pages = {293},
          doi = {10.1086/191375},
       adsurl = {https://ui.adsabs.harvard.edu/abs/1989ApJS...71..293B}
}

@ARTICLE{Dorazi2015b,
       author = {{D'Orazi}, V. and {Gratton}, R.~G. and {Angelou}, G.~C. and {Bragaglia}, A. and {Carretta}, E. and {Lattanzio}, J.~C. and {Lucatello}, S. and {Momany}, Y. and {Sollima}, A. and {Beccari}, G.},
        title = "{Lithium abundances in globular cluster giants: NGC 1904, NGC 2808, and NGC 362}",
      journal = {\mnras},
         year = 2015,
        month = jun,
       volume = {449},
       number = {4},
        pages = {4038-4047},
          doi = {10.1093/mnras/stv612},
archivePrefix = {arXiv},
       eprint = {1503.05925},
 primaryClass = {astro-ph.SR},
       adsurl = {https://ui.adsabs.harvard.edu/abs/2015MNRAS.449.4038D}
}

@ARTICLE{CharbonnelBalachandran2000,
       author = {{Charbonnel}, C. and {Balachandran}, S.~C.},
        title = "{The Nature of the lithium rich giants. Mixing episodes on the RGB and early-AGB}",
      journal = {\aap},
         year = 2000,
        month = jul,
       volume = {359},
        pages = {563-572},
archivePrefix = {arXiv},
       eprint = {astro-ph/0005280},
 primaryClass = {astro-ph},
       adsurl = {https://ui.adsabs.harvard.edu/abs/2000A&A...359..563C}
}

@ARTICLE{Sanna2020,
       author = {{Sanna}, N. and {Franciosini}, E. and {Pancino}, E. and {Mucciarelli}, A. and {Tsantaki}, M. and {Charbonnel}, C. and {Smiljanic}, R. and {Fu}, X. and {Bragaglia}, A. and {Lagarde}, N. and {Tautvai{\v{s}}iene}, G. and {Magrini}, L. and {Randich}, S. and {Bensby}, T. and {Korn}, A.~J. and {Bayo}, A. and {Bergemann}, M. and {Carraro}, G. and {Morbidelli}, L.},
        title = "{The Gaia-ESO Survey: an extremely Li-rich giant in globular cluster NGC 1261}",
      journal = {\aap},
         year = 2020,
        month = jul,
       volume = {639},
          eid = {L2},
        pages = {L2},
          doi = {10.1051/0004-6361/202038435},
archivePrefix = {arXiv},
       eprint = {2006.04393},
 primaryClass = {astro-ph.SR},
       adsurl = {https://ui.adsabs.harvard.edu/abs/2020A&A...639L...2S}
}

@ARTICLE{Johnson2010,
       author = {{Johnson}, John Asher and {Aller}, Kimberly M. and {Howard}, Andrew W. and {Crepp}, Justin R.},
        title = "{Giant Planet Occurrence in the Stellar Mass-Metallicity Plane}",
      journal = {\pasp},
         year = 2010,
        month = aug,
       volume = {122},
       number = {894},
        pages = {905},
          doi = {10.1086/655775},
archivePrefix = {arXiv},
       eprint = {1005.3084},
 primaryClass = {astro-ph.EP},
       adsurl = {https://ui.adsabs.harvard.edu/abs/2010PASP..122..905J}
}

@ARTICLE{Carney2008,
       author = {{Carney}, Bruce W. and {Gray}, David F. and {Yong}, David and {Latham}, David W. and {Manset}, Nadine and {Zelman}, Rachel and {Laird}, John B.},
        title = "{Rotation and Macroturbulence in Metal-Poor Field Red Giant and Red Horizontal Branch Stars}",
      journal = {\aj},
         year = 2008,
        month = mar,
       volume = {135},
       number = {3},
        pages = {892-906},
          doi = {10.1088/0004-6256/135/3/892},
archivePrefix = {arXiv},
       eprint = {0711.4984},
 primaryClass = {astro-ph},
       adsurl = {https://ui.adsabs.harvard.edu/abs/2008AJ....135..892C}
}

@ARTICLE{gaia_2018,
       author = {{Gaia Collaboration} and {Brown}, A.~G.~A. and {Vallenari}, A. and {Prusti}, T. and {de Bruijne}, J.~H.~J. and {Babusiaux}, C. and {Bailer-Jones}, C.~A.~L. and {Biermann}, M. and {Evans}, D.~W. and {Eyer}, L. and {Jansen}, F. and {Jordi}, C. and {Klioner}, S.~A. and {Lammers}, U. and {Lindegren}, L. and {Luri}, X. and {Mignard}, F. and {Panem}, C. and {Pourbaix}, D. and {Randich}, S. and {Sartoretti}, P. and {Siddiqui}, H.~I. and {Soubiran}, C. and {van Leeuwen}, F. and {Walton}, N.~A. and {Arenou}, F. and {Bastian}, U. and {Cropper}, M. and {Drimmel}, R. and {Katz}, D. and {Lattanzi}, M.~G. and {Bakker}, J. and {Cacciari}, C. and {Casta{\~n}eda}, J. and {Chaoul}, L. and {Cheek}, N. and {De Angeli}, F. and {Fabricius}, C. and {Guerra}, R. and {Holl}, B. and {Masana}, E. and {Messineo}, R. and {Mowlavi}, N. and {Nienartowicz}, K. and {Panuzzo}, P. and {Portell}, J. and {Riello}, M. and {Seabroke}, G.~M. and {Tanga}, P. and {Th{\'e}venin}, F. and {Gracia-Abril}, G. and {Comoretto}, G. and {Garcia-Reinaldos}, M. and {Teyssier}, D. and {Altmann}, M. and {Andrae}, R. and {Audard}, M. and {Bellas-Velidis}, I. and {Benson}, K. and {Berthier}, J. and {Blomme}, R. and {Burgess}, P. and {Busso}, G. and {Carry}, B. and {Cellino}, A. and {Clementini}, G. and {Clotet}, M. and {Creevey}, O. and {Davidson}, M. and {De Ridder}, J. and {Delchambre}, L. and {Dell'Oro}, A. and {Ducourant}, C. and {Fern{\'a}ndez-Hern{\'a}ndez}, J. and {Fouesneau}, M. and {Fr{\'e}mat}, Y. and {Galluccio}, L. and {Garc{\'\i}a-Torres}, M. and {Gonz{\'a}lez-N{\'u}{\~n}ez}, J. and {Gonz{\'a}lez-Vidal}, J.~J. and {Gosset}, E. and {Guy}, L.~P. and {Halbwachs}, J. -L. and {Hambly}, N.~C. and {Harrison}, D.~L. and {Hern{\'a}ndez}, J. and {Hestroffer}, D. and {Hodgkin}, S.~T. and {Hutton}, A. and {Jasniewicz}, G. and {Jean-Antoine-Piccolo}, A. and {Jordan}, S. and {Korn}, A.~J. and {Krone-Martins}, A. and {Lanzafame}, A.~C. and {Lebzelter}, T. and {L{\"o}ffler}, W. and {Manteiga}, M. and {Marrese}, P.~M. and {Mart{\'\i}n-Fleitas}, J.~M. and {Moitinho}, A. and {Mora}, A. and {Muinonen}, K. and {Osinde}, J. and {Pancino}, E. and {Pauwels}, T. and {Petit}, J. -M. and {Recio-Blanco}, A. and {Richards}, P.~J. and {Rimoldini}, L. and {Robin}, A.~C. and {Sarro}, L.~M. and {Siopis}, C. and {Smith}, M. and {Sozzetti}, A. and {S{\"u}veges}, M. and {Torra}, J. and {van Reeven}, W. and {Abbas}, U. and {Abreu Aramburu}, A. and {Accart}, S. and {Aerts}, C. and {Altavilla}, G. and {{\'A}lvarez}, M.~A. and {Alvarez}, R. and {Alves}, J. and {Anderson}, R.~I. and {Andrei}, A.~H. and {Anglada Varela}, E. and {Antiche}, E. and {Antoja}, T. and {Arcay}, B. and {Astraatmadja}, T.~L. and {Bach}, N. and {Baker}, S.~G. and {Balaguer-N{\'u}{\~n}ez}, L. and {Balm}, P. and {Barache}, C. and {Barata}, C. and {Barbato}, D. and {Barblan}, F. and {Barklem}, P.~S. and {Barrado}, D. and {Barros}, M. and {Barstow}, M.~A. and {Bartholom{\'e} Mu{\~n}oz}, S. and {Bassilana}, J. -L. and {Becciani}, U. and {Bellazzini}, M. and {Berihuete}, A. and {Bertone}, S. and {Bianchi}, L. and {Bienaym{\'e}}, O. and {Blanco-Cuaresma}, S. and {Boch}, T. and {Boeche}, C. and {Bombrun}, A. and {Borrachero}, R. and {Bossini}, D. and {Bouquillon}, S. and {Bourda}, G. and {Bragaglia}, A. and {Bramante}, L. and {Breddels}, M.~A. and {Bressan}, A. and {Brouillet}, N. and {Br{\"u}semeister}, T. and {Brugaletta}, E. and {Bucciarelli}, B. and {Burlacu}, A. and {Busonero}, D. and {Butkevich}, A.~G. and {Buzzi}, R. and {Caffau}, E. and {Cancelliere}, R. and {Cannizzaro}, G. and {Cantat-Gaudin}, T. and {Carballo}, R. and {Carlucci}, T. and {Carrasco}, J.~M. and {Casamiquela}, L. and {Castellani}, M. and {Castro-Ginard}, A. and {Charlot}, P. and {Chemin}, L. and {Chiavassa}, A. and {Cocozza}, G. and {Costigan}, G. and {Cowell}, S. and {Crifo}, F. and {Crosta}, M. and {Crowley}, C. and {Cuypers}, J. and {Dafonte}, C. and {Damerdji}, Y. and {Dapergolas}, A. and {David}, P. and {David}, M. and {de Laverny}, P. and {De Luise}, F. and {De March}, R. and {de Martino}, D. and {de Souza}, R. and {de Torres}, A. and {Debosscher}, J. and {del Pozo}, E. and {Delbo}, M. and {Delgado}, A. and {Delgado}, H.~E. and {Di Matteo}, P. and {Diakite}, S. and {Diener}, C. and {Distefano}, E. and {Dolding}, C. and {Drazinos}, P. and {Dur{\'a}n}, J. and {Edvardsson}, B. and {Enke}, H. and {Eriksson}, K. and {Esquej}, P. and {Eynard Bontemps}, G. and {Fabre}, C. and {Fabrizio}, M. and {Faigler}, S. and {Falc{\~a}o}, A.~J. and {Farr{\`a}s Casas}, M. and {Federici}, L. and {Fedorets}, G. and {Fernique}, P. and {Figueras}, F. and {Filippi}, F. and {Findeisen}, K. and {Fonti}, A. and {Fraile}, E. and {Fraser}, M. and {Fr{\'e}zouls}, B. and {Gai}, M. and {Galleti}, S. and {Garabato}, D. and {Garc{\'\i}a-Sedano}, F. and {Garofalo}, A. and {Garralda}, N. and {Gavel}, A. and {Gavras}, P. and {Gerssen}, J. and {Geyer}, R. and {Giacobbe}, P. and {Gilmore}, G. and {Girona}, S. and {Giuffrida}, G. and {Glass}, F. and {Gomes}, M. and {Granvik}, M. and {Gueguen}, A. and {Guerrier}, A. and {Guiraud}, J. and {Guti{\'e}rrez-S{\'a}nchez}, R. and {Haigron}, R. and {Hatzidimitriou}, D. and {Hauser}, M. and {Haywood}, M. and {Heiter}, U. and {Helmi}, A. and {Heu}, J. and {Hilger}, T. and {Hobbs}, D. and {Hofmann}, W. and {Holland}, G. and {Huckle}, H.~E. and {Hypki}, A. and {Icardi}, V. and {Jan{\ss}en}, K. and {Jevardat de Fombelle}, G. and {Jonker}, P.~G. and {Juh{\'a}sz}, {\'A}. L. and {Julbe}, F. and {Karampelas}, A. and {Kewley}, A. and {Klar}, J. and {Kochoska}, A. and {Kohley}, R. and {Kolenberg}, K. and {Kontizas}, M. and {Kontizas}, E. and {Koposov}, S.~E. and {Kordopatis}, G. and {Kostrzewa-Rutkowska}, Z. and {Koubsky}, P. and {Lambert}, S. and {Lanza}, A.~F. and {Lasne}, Y. and {Lavigne}, J. -B. and {Le Fustec}, Y. and {Le Poncin-Lafitte}, C. and {Lebreton}, Y. and {Leccia}, S. and {Leclerc}, N. and {Lecoeur-Taibi}, I. and {Lenhardt}, H. and {Leroux}, F. and {Liao}, S. and {Licata}, E. and {Lindstr{\o}m}, H.~E.~P. and {Lister}, T.~A. and {Livanou}, E. and {Lobel}, A. and {L{\'o}pez}, M. and {Managau}, S. and {Mann}, R.~G. and {Mantelet}, G. and {Marchal}, O. and {Marchant}, J.~M. and {Marconi}, M. and {Marinoni}, S. and {Marschalk{\'o}}, G. and {Marshall}, D.~J. and {Martino}, M. and {Marton}, G. and {Mary}, N. and {Massari}, D. and {Matijevi{\v{c}}}, G. and {Mazeh}, T. and {McMillan}, P.~J. and {Messina}, S. and {Michalik}, D. and {Millar}, N.~R. and {Molina}, D. and {Molinaro}, R. and {Moln{\'a}r}, L. and {Montegriffo}, P. and {Mor}, R. and {Morbidelli}, R. and {Morel}, T. and {Morris}, D. and {Mulone}, A.~F. and {Muraveva}, T. and {Musella}, I. and {Nelemans}, G. and {Nicastro}, L. and {Noval}, L. and {O'Mullane}, W. and {Ord{\'e}novic}, C. and {Ord{\'o}{\~n}ez-Blanco}, D. and {Osborne}, P. and {Pagani}, C. and {Pagano}, I. and {Pailler}, F. and {Palacin}, H. and {Palaversa}, L. and {Panahi}, A. and {Pawlak}, M. and {Piersimoni}, A.~M. and {Pineau}, F. -X. and {Plachy}, E. and {Plum}, G. and {Poggio}, E. and {Poujoulet}, E. and {Pr{\v{s}}a}, A. and {Pulone}, L. and {Racero}, E. and {Ragaini}, S. and {Rambaux}, N. and {Ramos-Lerate}, M. and {Regibo}, S. and {Reyl{\'e}}, C. and {Riclet}, F. and {Ripepi}, V. and {Riva}, A. and {Rivard}, A. and {Rixon}, G. and {Roegiers}, T. and {Roelens}, M. and {Romero-G{\'o}mez}, M. and {Rowell}, N. and {Royer}, F. and {Ruiz-Dern}, L. and {Sadowski}, G. and {Sagrist{\`a} Sell{\'e}s}, T. and {Sahlmann}, J. and {Salgado}, J. and {Salguero}, E. and {Sanna}, N. and {Santana-Ros}, T. and {Sarasso}, M. and {Savietto}, H. and {Schultheis}, M. and {Sciacca}, E. and {Segol}, M. and {Segovia}, J.~C. and {S{\'e}gransan}, D. and {Shih}, I. -C. and {Siltala}, L. and {Silva}, A.~F. and {Smart}, R.~L. and {Smith}, K.~W. and {Solano}, E. and {Solitro}, F. and {Sordo}, R. and {Soria Nieto}, S. and {Souchay}, J. and {Spagna}, A. and {Spoto}, F. and {Stampa}, U. and {Steele}, I.~A. and {Steidelm{\"u}ller}, H. and {Stephenson}, C.~A. and {Stoev}, H. and {Suess}, F.~F. and {Surdej}, J. and {Szabados}, L. and {Szegedi-Elek}, E. and {Tapiador}, D. and {Taris}, F. and {Tauran}, G. and {Taylor}, M.~B. and {Teixeira}, R. and {Terrett}, D. and {Teyssandier}, P. and {Thuillot}, W. and {Titarenko}, A. and {Torra Clotet}, F. and {Turon}, C. and {Ulla}, A. and {Utrilla}, E. and {Uzzi}, S. and {Vaillant}, M. and {Valentini}, G. and {Valette}, V. and {van Elteren}, A. and {Van Hemelryck}, E. and {van Leeuwen}, M. and {Vaschetto}, M. and {Vecchiato}, A. and {Veljanoski}, J. and {Viala}, Y. and {Vicente}, D. and {Vogt}, S. and {von Essen}, C. and {Voss}, H. and {Votruba}, V. and {Voutsinas}, S. and {Walmsley}, G. and {Weiler}, M. and {Wertz}, O. and {Wevers}, T. and {Wyrzykowski}, {\L}. and {Yoldas}, A. and {{\v{Z}}erjal}, M. and {Ziaeepour}, H. and {Zorec}, J. and {Zschocke}, S. and {Zucker}, S. and {Zurbach}, C. and {Zwitter}, T.},
        title = "{Gaia Data Release 2. Summary of the contents and survey properties}",
      journal = {\aap},
         year = 2018,
        month = aug,
       volume = {616},
          eid = {A1},
        pages = {A1},
          doi = {10.1051/0004-6361/201833051},
archivePrefix = {arXiv},
       eprint = {1804.09365},
 primaryClass = {astro-ph.GA},
       adsurl = {https://ui.adsabs.harvard.edu/abs/2018A&A...616A...1G}
}

@ARTICLE{2020SciPy-NMeth,
  author  = {Virtanen, Pauli and Gommers, Ralf and Oliphant, Travis E. and
            Haberland, Matt and Reddy, Tyler and Cournapeau, David and
            Burovski, Evgeni and Peterson, Pearu and Weckesser, Warren and
            Bright, Jonathan and {van der Walt}, St{\'e}fan J. and
            Brett, Matthew and Wilson, Joshua and Millman, K. Jarrod and
            Mayorov, Nikolay and Nelson, Andrew R. J. and Jones, Eric and
            Kern, Robert and Larson, Eric and Carey, C J and
            Polat, {\.I}lhan and Feng, Yu and Moore, Eric W. and
            {VanderPlas}, Jake and Laxalde, Denis and Perktold, Josef and
            Cimrman, Robert and Henriksen, Ian and Quintero, E. A. and
            Harris, Charles R. and Archibald, Anne M. and
            Ribeiro, Ant{\^o}nio H. and Pedregosa, Fabian and
            {van Mulbregt}, Paul and {SciPy 1.0 Contributors}},
  title   = {{{SciPy} 1.0: Fundamental Algorithms for Scientific
            Computing in Python}},
  journal = {Nature Methods},
  year    = {2020},
  volume  = {17},
  pages   = {261--272},
  adsurl  = {https://rdcu.be/b08Wh},
  doi     = {10.1038/s41592-019-0686-2},
}

@Article{         harris2020array,
 title         = {Array programming with {NumPy}},
 author        = {Charles R. Harris and K. Jarrod Millman and St{\'{e}}fan J.
                 van der Walt and Ralf Gommers and Pauli Virtanen and David
                 Cournapeau and Eric Wieser and Julian Taylor and Sebastian
                 Berg and Nathaniel J. Smith and Robert Kern and Matti Picus
                 and Stephan Hoyer and Marten H. van Kerkwijk and Matthew
                 Brett and Allan Haldane and Jaime Fern{\'{a}}ndez del
                 R{\'{i}}o and Mark Wiebe and Pearu Peterson and Pierre
                 G{\'{e}}rard-Marchant and Kevin Sheppard and Tyler Reddy and
                 Warren Weckesser and Hameer Abbasi and Christoph Gohlke and
                 Travis E. Oliphant},
 year          = {2020},
 month         = sep,
 journal       = {Nature},
 volume        = {585},
 number        = {7825},
 pages         = {357--362},
 doi           = {10.1038/s41586-020-2649-2},
 publisher     = {Springer Science and Business Media {LLC}},
 url           = {https://doi.org/10.1038/s41586-020-2649-2}
}

@Article{matplotlib,
  Author    = {Hunter, J. D.},
  Title     = {Matplotlib: A 2D graphics environment},
  Journal   = {Computing in Science \& Engineering},
  Volume    = {9},
  Number    = {3},
  Pages     = {90--95},
  publisher = {IEEE COMPUTER SOC},
  doi       = {10.1109/MCSE.2007.55},
  year      = 2007
}

@MISC{lkurve,
   author = {{Lightkurve Collaboration} and {Cardoso}, J.~V.~d.~M. and
             {Hedges}, C. and {Gully-Santiago}, M. and {Saunders}, N. and
             {Cody}, A.~M. and {Barclay}, T. and {Hall}, O. and
             {Sagear}, S. and {Turtelboom}, E. and {Zhang}, J. and
             {Tzanidakis}, A. and {Mighell}, K. and {Coughlin}, J. and
             {Bell}, K. and {Berta-Thompson}, Z. and {Williams}, P. and
             {Dotson}, J. and {Barentsen}, G.},
    title = "{Lightkurve: Kepler and TESS time series analysis in Python}",
howpublished = {Astrophysics Source Code Library},
     year = 2018,
    month = dec,
archivePrefix = "ascl",
   eprint = {1812.013},
   adsurl = {http://adsabs.harvard.edu/abs/2018ascl.soft12013L},
}

@InProceedings{mckinney-proc-scipy-2010,
  author    = { {W}es {M}c{K}inney },
  title     = { {D}ata {S}tructures for {S}tatistical {C}omputing in {P}ython },
  booktitle = { {P}roceedings of the 9th {P}ython in {S}cience {C}onference },
  pages     = { 56 - 61 },
  year      = { 2010 },
  editor    = { {S}t\'efan van der {W}alt and {J}arrod {M}illman },
  doi       = { 10.25080/Majora-92bf1922-00a }
}

@ARTICLE{Lattanzio2015,
       author = {{Lattanzio}, J.~C. and {Siess}, L. and {Church}, R.~P. and {Angelou}, G. and {Stancliffe}, R.~J. and {Doherty}, C.~L. and {Stephen}, T. and {Campbell}, S.~W.},
        title = "{On the numerical treatment and dependence of thermohaline mixing in red giants}",
      journal = {\mnras},
         year = 2015,
        month = jan,
       volume = {446},
       number = {3},
        pages = {2673-2688},
          doi = {10.1093/mnras/stu2238},
archivePrefix = {arXiv},
       eprint = {1410.6517},
 primaryClass = {astro-ph.SR},
       adsurl = {https://ui.adsabs.harvard.edu/abs/2015MNRAS.446.2673L}
}

@ARTICLE{Lagarde2012,
       author = {{Lagarde}, N. and {Decressin}, T. and {Charbonnel}, C. and {Eggenberger}, P. and {Ekstr{\"o}m}, S. and {Palacios}, A.},
        title = "{Thermohaline instability and rotation-induced mixing. III. Grid of stellar models and asymptotic asteroseismic quantities from the pre-main sequence up to the AGB for low- and intermediate-mass stars of various metallicities}",
      journal = {\aap},
         year = 2012,
        month = jul,
       volume = {543},
          eid = {A108},
        pages = {A108},
          doi = {10.1051/0004-6361/201118331},
archivePrefix = {arXiv},
       eprint = {1204.5193},
 primaryClass = {astro-ph.SR},
       adsurl = {https://ui.adsabs.harvard.edu/abs/2012A&A...543A.108L}
}

@ARTICLE{Bedding2011,
       author = {{Bedding}, Timothy R. and {Mosser}, Benoit and {Huber}, Daniel and {Montalb{\'a}n}, Josefina and {Beck}, Paul and {Christensen-Dalsgaard}, J{\o}rgen and {Elsworth}, Yvonne P. and {Garc{\'\i}a}, Rafael A. and {Miglio}, Andrea and {Stello}, Dennis and {White}, Timothy R. and {De Ridder}, Joris and {Hekker}, Saskia and {Aerts}, Conny and {Barban}, Caroline and {Belkacem}, Kevin and {Broomhall}, Anne-Marie and {Brown}, Timothy M. and {Buzasi}, Derek L. and {Carrier}, Fabien and {Chaplin}, William J. and {di Mauro}, Maria Pia and {Dupret}, Marc-Antoine and {Frandsen}, S{\o}ren and {Gilliland}, Ronald L. and {Goupil}, Marie-Jo and {Jenkins}, Jon M. and {Kallinger}, Thomas and {Kawaler}, Steven and {Kjeldsen}, Hans and {Mathur}, Savita and {Noels}, Arlette and {Silva Aguirre}, Victor and {Ventura}, Paolo},
        title = "{Gravity modes as a way to distinguish between hydrogen- and helium-burning red giant stars}",
      journal = {\nat},
         year = 2011,
        month = mar,
       volume = {471},
       number = {7340},
        pages = {608-611},
          doi = {10.1038/nature09935},
archivePrefix = {arXiv},
       eprint = {1103.5805},
 primaryClass = {astro-ph.SR},
       adsurl = {https://ui.adsabs.harvard.edu/abs/2011Natur.471..608B}
}

@ARTICLE{gaiadr3,
       author = {{Gaia Collaboration} and {Vallenari}, A. and {Brown}, A.~G.~A. and {Prusti}, T. and {de Bruijne}, J.~H.~J. and {Arenou}, F. and {Babusiaux}, C. and {Biermann}, M. and {Creevey}, O.~L. and {Ducourant}, C. and {Evans}, D.~W. and {Eyer}, L. and {Guerra}, R. and {Hutton}, A. and {Jordi}, C. and {Klioner}, S.~A. and {Lammers}, U.~L. and {Lindegren}, L. and {Luri}, X. and {Mignard}, F. and {Panem}, C. and {Pourbaix}, D. and {Randich}, S. and {Sartoretti}, P. and {Soubiran}, C. and {Tanga}, P. and {Walton}, N.~A. and {Bailer-Jones}, C.~A.~L. and {Bastian}, U. and {Drimmel}, R. and {Jansen}, F. and {Katz}, D. and {Lattanzi}, M.~G. and {van Leeuwen}, F. and {Bakker}, J. and {Cacciari}, C. and {Casta{\~n}eda}, J. and {De Angeli}, F. and {Fabricius}, C. and {Fouesneau}, M. and {Fr{\'e}mat}, Y. and {Galluccio}, L. and {Guerrier}, A. and {Heiter}, U. and {Masana}, E. and {Messineo}, R. and {Mowlavi}, N. and {Nicolas}, C. and {Nienartowicz}, K. and {Pailler}, F. and {Panuzzo}, P. and {Riclet}, F. and {Roux}, W. and {Seabroke}, G.~M. and {Sordo}, R. and {Th{\'e}venin}, F. and {Gracia-Abril}, G. and {Portell}, J. and {Teyssier}, D. and {Altmann}, M. and {Andrae}, R. and {Audard}, M. and {Bellas-Velidis}, I. and {Benson}, K. and {Berthier}, J. and {Blomme}, R. and {Burgess}, P.~W. and {Busonero}, D. and {Busso}, G. and {C{\'a}novas}, H. and {Carry}, B. and {Cellino}, A. and {Cheek}, N. and {Clementini}, G. and {Damerdji}, Y. and {Davidson}, M. and {de Teodoro}, P. and {Nu{\~n}ez Campos}, M. and {Delchambre}, L. and {Dell'Oro}, A. and {Esquej}, P. and {Fern{\'a}ndez-Hern{\'a}ndez}, J. and {Fraile}, E. and {Garabato}, D. and {Garc{\'\i}a-Lario}, P. and {Gosset}, E. and {Haigron}, R. and {Halbwachs}, J. -L. and {Hambly}, N.~C. and {Harrison}, D.~L. and {Hern{\'a}ndez}, J. and {Hestroffer}, D. and {Hodgkin}, S.~T. and {Holl}, B. and {Jan{\ss}en}, K. and {Jevardat de Fombelle}, G. and {Jordan}, S. and {Krone-Martins}, A. and {Lanzafame}, A.~C. and {L{\"o}ffler}, W. and {Marchal}, O. and {Marrese}, P.~M. and {Moitinho}, A. and {Muinonen}, K. and {Osborne}, P. and {Pancino}, E. and {Pauwels}, T. and {Recio-Blanco}, A. and {Reyl{\'e}}, C. and {Riello}, M. and {Rimoldini}, L. and {Roegiers}, T. and {Rybizki}, J. and {Sarro}, L.~M. and {Siopis}, C. and {Smith}, M. and {Sozzetti}, A. and {Utrilla}, E. and {van Leeuwen}, M. and {Abbas}, U. and {{\'A}brah{\'a}m}, P. and {Abreu Aramburu}, A. and {Aerts}, C. and {Aguado}, J.~J. and {Ajaj}, M. and {Aldea-Montero}, F. and {Altavilla}, G. and {{\'A}lvarez}, M.~A. and {Alves}, J. and {Anders}, F. and {Anderson}, R.~I. and {Anglada Varela}, E. and {Antoja}, T. and {Baines}, D. and {Baker}, S.~G. and {Balaguer-N{\'u}{\~n}ez}, L. and {Balbinot}, E. and {Balog}, Z. and {Barache}, C. and {Barbato}, D. and {Barros}, M. and {Barstow}, M.~A. and {Bartolom{\'e}}, S. and {Bassilana}, J. -L. and {Bauchet}, N. and {Becciani}, U. and {Bellazzini}, M. and {Berihuete}, A. and {Bernet}, M. and {Bertone}, S. and {Bianchi}, L. and {Binnenfeld}, A. and {Blanco-Cuaresma}, S. and {Blazere}, A. and {Boch}, T. and {Bombrun}, A. and {Bossini}, D. and {Bouquillon}, S. and {Bragaglia}, A. and {Bramante}, L. and {Breedt}, E. and {Bressan}, A. and {Brouillet}, N. and {Brugaletta}, E. and {Bucciarelli}, B. and {Burlacu}, A. and {Butkevich}, A.~G. and {Buzzi}, R. and {Caffau}, E. and {Cancelliere}, R. and {Cantat-Gaudin}, T. and {Carballo}, R. and {Carlucci}, T. and {Carnerero}, M.~I. and {Carrasco}, J.~M. and {Casamiquela}, L. and {Castellani}, M. and {Castro-Ginard}, A. and {Chaoul}, L. and {Charlot}, P. and {Chemin}, L. and {Chiaramida}, V. and {Chiavassa}, A. and {Chornay}, N. and {Comoretto}, G. and {Contursi}, G. and {Cooper}, W.~J. and {Cornez}, T. and {Cowell}, S. and {Crifo}, F. and {Cropper}, M. and {Crosta}, M. and {Crowley}, C. and {Dafonte}, C. and {Dapergolas}, A. and {David}, M. and {David}, P. and {de Laverny}, P. and {De Luise}, F. and {De March}, R. and {De Ridder}, J. and {de Souza}, R. and {de Torres}, A. and {del Peloso}, E.~F. and {del Pozo}, E. and {Delbo}, M. and {Delgado}, A. and {Delisle}, J. -B. and {Demouchy}, C. and {Dharmawardena}, T.~E. and {Di Matteo}, P. and {Diakite}, S. and {Diener}, C. and {Distefano}, E. and {Dolding}, C. and {Edvardsson}, B. and {Enke}, H. and {Fabre}, C. and {Fabrizio}, M. and {Faigler}, S. and {Fedorets}, G. and {Fernique}, P. and {Fienga}, A. and {Figueras}, F. and {Fournier}, Y. and {Fouron}, C. and {Fragkoudi}, F. and {Gai}, M. and {Garcia-Gutierrez}, A. and {Garcia-Reinaldos}, M. and {Garc{\'\i}a-Torres}, M. and {Garofalo}, A. and {Gavel}, A. and {Gavras}, P. and {Gerlach}, E. and {Geyer}, R. and {Giacobbe}, P. and {Gilmore}, G. and {Girona}, S. and {Giuffrida}, G. and {Gomel}, R. and {Gomez}, A. and {Gonz{\'a}lez-N{\'u}{\~n}ez}, J. and {Gonz{\'a}lez-Santamar{\'\i}a}, I. and {Gonz{\'a}lez-Vidal}, J.~J. and {Granvik}, M. and {Guillout}, P. and {Guiraud}, J. and {Guti{\'e}rrez-S{\'a}nchez}, R. and {Guy}, L.~P. and {Hatzidimitriou}, D. and {Hauser}, M. and {Haywood}, M. and {Helmer}, A. and {Helmi}, A. and {Sarmiento}, M.~H. and {Hidalgo}, S.~L. and {Hilger}, T. and {H{\l}adczuk}, N. and {Hobbs}, D. and {Holland}, G. and {Huckle}, H.~E. and {Jardine}, K. and {Jasniewicz}, G. and {Jean-Antoine Piccolo}, A. and {Jim{\'e}nez-Arranz}, {\'O}. and {Jorissen}, A. and {Juaristi Campillo}, J. and {Julbe}, F. and {Karbevska}, L. and {Kervella}, P. and {Khanna}, S. and {Kontizas}, M. and {Kordopatis}, G. and {Korn}, A.~J. and {K{\'o}sp{\'a}l}, {\'A}. and {Kostrzewa-Rutkowska}, Z. and {Kruszy{\'n}ska}, K. and {Kun}, M. and {Laizeau}, P. and {Lambert}, S. and {Lanza}, A.~F. and {Lasne}, Y. and {Le Campion}, J. -F. and {Lebreton}, Y. and {Lebzelter}, T. and {Leccia}, S. and {Leclerc}, N. and {Lecoeur-Taibi}, I. and {Liao}, S. and {Licata}, E.~L. and {Lindstr{\o}m}, H.~E.~P. and {Lister}, T.~A. and {Livanou}, E. and {Lobel}, A. and {Lorca}, A. and {Loup}, C. and {Madrero Pardo}, P. and {Magdaleno Romeo}, A. and {Managau}, S. and {Mann}, R.~G. and {Manteiga}, M. and {Marchant}, J.~M. and {Marconi}, M. and {Marcos}, J. and {Marcos Santos}, M.~M.~S. and {Mar{\'\i}n Pina}, D. and {Marinoni}, S. and {Marocco}, F. and {Marshall}, D.~J. and {Martin Polo}, L. and {Mart{\'\i}n-Fleitas}, J.~M. and {Marton}, G. and {Mary}, N. and {Masip}, A. and {Massari}, D. and {Mastrobuono-Battisti}, A. and {Mazeh}, T. and {McMillan}, P.~J. and {Messina}, S. and {Michalik}, D. and {Millar}, N.~R. and {Mints}, A. and {Molina}, D. and {Molinaro}, R. and {Moln{\'a}r}, L. and {Monari}, G. and {Mongui{\'o}}, M. and {Montegriffo}, P. and {Montero}, A. and {Mor}, R. and {Mora}, A. and {Morbidelli}, R. and {Morel}, T. and {Morris}, D. and {Muraveva}, T. and {Murphy}, C.~P. and {Musella}, I. and {Nagy}, Z. and {Noval}, L. and {Oca{\~n}a}, F. and {Ogden}, A. and {Ordenovic}, C. and {Osinde}, J.~O. and {Pagani}, C. and {Pagano}, I. and {Palaversa}, L. and {Palicio}, P.~A. and {Pallas-Quintela}, L. and {Panahi}, A. and {Payne-Wardenaar}, S. and {Pe{\~n}alosa Esteller}, X. and {Penttil{\"a}}, A. and {Pichon}, B. and {Piersimoni}, A.~M. and {Pineau}, F. -X. and {Plachy}, E. and {Plum}, G. and {Poggio}, E. and {Pr{\v{s}}a}, A. and {Pulone}, L. and {Racero}, E. and {Ragaini}, S. and {Rainer}, M. and {Raiteri}, C.~M. and {Rambaux}, N. and {Ramos}, P. and {Ramos-Lerate}, M. and {Re Fiorentin}, P. and {Regibo}, S. and {Richards}, P.~J. and {Rios Diaz}, C. and {Ripepi}, V. and {Riva}, A. and {Rix}, H. -W. and {Rixon}, G. and {Robichon}, N. and {Robin}, A.~C. and {Robin}, C. and {Roelens}, M. and {Rogues}, H.~R.~O. and {Rohrbasser}, L. and {Romero-G{\'o}mez}, M. and {Rowell}, N. and {Royer}, F. and {Ruz Mieres}, D. and {Rybicki}, K.~A. and {Sadowski}, G. and {S{\'a}ez N{\'u}{\~n}ez}, A. and {Sagrist{\`a} Sell{\'e}s}, A. and {Sahlmann}, J. and {Salguero}, E. and {Samaras}, N. and {Sanchez Gimenez}, V. and {Sanna}, N. and {Santove{\~n}a}, R. and {Sarasso}, M. and {Schultheis}, M. and {Sciacca}, E. and {Segol}, M. and {Segovia}, J.~C. and {S{\'e}gransan}, D. and {Semeux}, D. and {Shahaf}, S. and {Siddiqui}, H.~I. and {Siebert}, A. and {Siltala}, L. and {Silvelo}, A. and {Slezak}, E. and {Slezak}, I. and {Smart}, R.~L. and {Snaith}, O.~N. and {Solano}, E. and {Solitro}, F. and {Souami}, D. and {Souchay}, J. and {Spagna}, A. and {Spina}, L. and {Spoto}, F. and {Steele}, I.~A. and {Steidelm{\"u}ller}, H. and {Stephenson}, C.~A. and {S{\"u}veges}, M. and {Surdej}, J. and {Szabados}, L. and {Szegedi-Elek}, E. and {Taris}, F. and {Taylor}, M.~B. and {Teixeira}, R. and {Tolomei}, L. and {Tonello}, N. and {Torra}, F. and {Torra}, J. and {Torralba Elipe}, G. and {Trabucchi}, M. and {Tsounis}, A.~T. and {Turon}, C. and {Ulla}, A. and {Unger}, N. and {Vaillant}, M.~V. and {van Dillen}, E. and {van Reeven}, W. and {Vanel}, O. and {Vecchiato}, A. and {Viala}, Y. and {Vicente}, D. and {Voutsinas}, S. and {Weiler}, M. and {Wevers}, T. and {Wyrzykowski}, {\L}. and {Yoldas}, A. and {Yvard}, P. and {Zhao}, H. and {Zorec}, J. and {Zucker}, S. and {Zwitter}, T.},
        title = "{Gaia Data Release 3. Summary of the content and survey properties}",
      journal = {\aap},
         year = 2023,
        month = jun,
       volume = {674},
          eid = {A1},
        pages = {A1},
          doi = {10.1051/0004-6361/202243940},
archivePrefix = {arXiv},
       eprint = {2208.00211},
 primaryClass = {astro-ph.GA},
       adsurl = {https://ui.adsabs.harvard.edu/abs/2023A&A...674A...1G}
}

@ARTICLE{gaia_collab,
       author = {{Gaia Collaboration} and {Prusti}, T. and {de Bruijne}, J.~H.~J. and {Brown}, A.~G.~A. and {Vallenari}, A. and {Babusiaux}, C. and {Bailer-Jones}, C.~A.~L. and {Bastian}, U. and {Biermann}, M. and {Evans}, D.~W. and {Eyer}, L. and {Jansen}, F. and {Jordi}, C. and {Klioner}, S.~A. and {Lammers}, U. and {Lindegren}, L. and {Luri}, X. and {Mignard}, F. and {Milligan}, D.~J. and {Panem}, C. and {Poinsignon}, V. and {Pourbaix}, D. and {Randich}, S. and {Sarri}, G. and {Sartoretti}, P. and {Siddiqui}, H.~I. and {Soubiran}, C. and {Valette}, V. and {van Leeuwen}, F. and {Walton}, N.~A. and {Aerts}, C. and {Arenou}, F. and {Cropper}, M. and {Drimmel}, R. and {H{\o}g}, E. and {Katz}, D. and {Lattanzi}, M.~G. and {O'Mullane}, W. and {Grebel}, E.~K. and {Holland}, A.~D. and {Huc}, C. and {Passot}, X. and {Bramante}, L. and {Cacciari}, C. and {Casta{\~n}eda}, J. and {Chaoul}, L. and {Cheek}, N. and {De Angeli}, F. and {Fabricius}, C. and {Guerra}, R. and {Hern{\'a}ndez}, J. and {Jean-Antoine-Piccolo}, A. and {Masana}, E. and {Messineo}, R. and {Mowlavi}, N. and {Nienartowicz}, K. and {Ord{\'o}{\~n}ez-Blanco}, D. and {Panuzzo}, P. and {Portell}, J. and {Richards}, P.~J. and {Riello}, M. and {Seabroke}, G.~M. and {Tanga}, P. and {Th{\'e}venin}, F. and {Torra}, J. and {Els}, S.~G. and {Gracia-Abril}, G. and {Comoretto}, G. and {Garcia-Reinaldos}, M. and {Lock}, T. and {Mercier}, E. and {Altmann}, M. and {Andrae}, R. and {Astraatmadja}, T.~L. and {Bellas-Velidis}, I. and {Benson}, K. and {Berthier}, J. and {Blomme}, R. and {Busso}, G. and {Carry}, B. and {Cellino}, A. and {Clementini}, G. and {Cowell}, S. and {Creevey}, O. and {Cuypers}, J. and {Davidson}, M. and {De Ridder}, J. and {de Torres}, A. and {Delchambre}, L. and {Dell'Oro}, A. and {Ducourant}, C. and {Fr{\'e}mat}, Y. and {Garc{\'\i}a-Torres}, M. and {Gosset}, E. and {Halbwachs}, J. -L. and {Hambly}, N.~C. and {Harrison}, D.~L. and {Hauser}, M. and {Hestroffer}, D. and {Hodgkin}, S.~T. and {Huckle}, H.~E. and {Hutton}, A. and {Jasniewicz}, G. and {Jordan}, S. and {Kontizas}, M. and {Korn}, A.~J. and {Lanzafame}, A.~C. and {Manteiga}, M. and {Moitinho}, A. and {Muinonen}, K. and {Osinde}, J. and {Pancino}, E. and {Pauwels}, T. and {Petit}, J. -M. and {Recio-Blanco}, A. and {Robin}, A.~C. and {Sarro}, L.~M. and {Siopis}, C. and {Smith}, M. and {Smith}, K.~W. and {Sozzetti}, A. and {Thuillot}, W. and {van Reeven}, W. and {Viala}, Y. and {Abbas}, U. and {Abreu Aramburu}, A. and {Accart}, S. and {Aguado}, J.~J. and {Allan}, P.~M. and {Allasia}, W. and {Altavilla}, G. and {{\'A}lvarez}, M.~A. and {Alves}, J. and {Anderson}, R.~I. and {Andrei}, A.~H. and {Anglada Varela}, E. and {Antiche}, E. and {Antoja}, T. and {Ant{\'o}n}, S. and {Arcay}, B. and {Atzei}, A. and {Ayache}, L. and {Bach}, N. and {Baker}, S.~G. and {Balaguer-N{\'u}{\~n}ez}, L. and {Barache}, C. and {Barata}, C. and {Barbier}, A. and {Barblan}, F. and {Baroni}, M. and {Barrado y Navascu{\'e}s}, D. and {Barros}, M. and {Barstow}, M.~A. and {Becciani}, U. and {Bellazzini}, M. and {Bellei}, G. and {Bello Garc{\'\i}a}, A. and {Belokurov}, V. and {Bendjoya}, P. and {Berihuete}, A. and {Bianchi}, L. and {Bienaym{\'e}}, O. and {Billebaud}, F. and {Blagorodnova}, N. and {Blanco-Cuaresma}, S. and {Boch}, T. and {Bombrun}, A. and {Borrachero}, R. and {Bouquillon}, S. and {Bourda}, G. and {Bouy}, H. and {Bragaglia}, A. and {Breddels}, M.~A. and {Brouillet}, N. and {Br{\"u}semeister}, T. and {Bucciarelli}, B. and {Budnik}, F. and {Burgess}, P. and {Burgon}, R. and {Burlacu}, A. and {Busonero}, D. and {Buzzi}, R. and {Caffau}, E. and {Cambras}, J. and {Campbell}, H. and {Cancelliere}, R. and {Cantat-Gaudin}, T. and {Carlucci}, T. and {Carrasco}, J.~M. and {Castellani}, M. and {Charlot}, P. and {Charnas}, J. and {Charvet}, P. and {Chassat}, F. and {Chiavassa}, A. and {Clotet}, M. and {Cocozza}, G. and {Collins}, R.~S. and {Collins}, P. and {Costigan}, G. and {Crifo}, F. and {Cross}, N.~J.~G. and {Crosta}, M. and {Crowley}, C. and {Dafonte}, C. and {Damerdji}, Y. and {Dapergolas}, A. and {David}, P. and {David}, M. and {De Cat}, P. and {de Felice}, F. and {de Laverny}, P. and {De Luise}, F. and {De March}, R. and {de Martino}, D. and {de Souza}, R. and {Debosscher}, J. and {del Pozo}, E. and {Delbo}, M. and {Delgado}, A. and {Delgado}, H.~E. and {di Marco}, F. and {Di Matteo}, P. and {Diakite}, S. and {Distefano}, E. and {Dolding}, C. and {Dos Anjos}, S. and {Drazinos}, P. and {Dur{\'a}n}, J. and {Dzigan}, Y. and {Ecale}, E. and {Edvardsson}, B. and {Enke}, H. and {Erdmann}, M. and {Escolar}, D. and {Espina}, M. and {Evans}, N.~W. and {Eynard Bontemps}, G. and {Fabre}, C. and {Fabrizio}, M. and {Faigler}, S. and {Falc{\~a}o}, A.~J. and {Farr{\`a}s Casas}, M. and {Faye}, F. and {Federici}, L. and {Fedorets}, G. and {Fern{\'a}ndez-Hern{\'a}ndez}, J. and {Fernique}, P. and {Fienga}, A. and {Figueras}, F. and {Filippi}, F. and {Findeisen}, K. and {Fonti}, A. and {Fouesneau}, M. and {Fraile}, E. and {Fraser}, M. and {Fuchs}, J. and {Furnell}, R. and {Gai}, M. and {Galleti}, S. and {Galluccio}, L. and {Garabato}, D. and {Garc{\'\i}a-Sedano}, F. and {Gar{\'e}}, P. and {Garofalo}, A. and {Garralda}, N. and {Gavras}, P. and {Gerssen}, J. and {Geyer}, R. and {Gilmore}, G. and {Girona}, S. and {Giuffrida}, G. and {Gomes}, M. and {Gonz{\'a}lez-Marcos}, A. and {Gonz{\'a}lez-N{\'u}{\~n}ez}, J. and {Gonz{\'a}lez-Vidal}, J.~J. and {Granvik}, M. and {Guerrier}, A. and {Guillout}, P. and {Guiraud}, J. and {G{\'u}rpide}, A. and {Guti{\'e}rrez-S{\'a}nchez}, R. and {Guy}, L.~P. and {Haigron}, R. and {Hatzidimitriou}, D. and {Haywood}, M. and {Heiter}, U. and {Helmi}, A. and {Hobbs}, D. and {Hofmann}, W. and {Holl}, B. and {Holland}, G. and {Hunt}, J.~A.~S. and {Hypki}, A. and {Icardi}, V. and {Irwin}, M. and {Jevardat de Fombelle}, G. and {Jofr{\'e}}, P. and {Jonker}, P.~G. and {Jorissen}, A. and {Julbe}, F. and {Karampelas}, A. and {Kochoska}, A. and {Kohley}, R. and {Kolenberg}, K. and {Kontizas}, E. and {Koposov}, S.~E. and {Kordopatis}, G. and {Koubsky}, P. and {Kowalczyk}, A. and {Krone-Martins}, A. and {Kudryashova}, M. and {Kull}, I. and {Bachchan}, R.~K. and {Lacoste-Seris}, F. and {Lanza}, A.~F. and {Lavigne}, J. -B. and {Le Poncin-Lafitte}, C. and {Lebreton}, Y. and {Lebzelter}, T. and {Leccia}, S. and {Leclerc}, N. and {Lecoeur-Taibi}, I. and {Lemaitre}, V. and {Lenhardt}, H. and {Leroux}, F. and {Liao}, S. and {Licata}, E. and {Lindstr{\o}m}, H.~E.~P. and {Lister}, T.~A. and {Livanou}, E. and {Lobel}, A. and {L{\"o}ffler}, W. and {L{\'o}pez}, M. and {Lopez-Lozano}, A. and {Lorenz}, D. and {Loureiro}, T. and {MacDonald}, I. and {Magalh{\~a}es Fernandes}, T. and {Managau}, S. and {Mann}, R.~G. and {Mantelet}, G. and {Marchal}, O. and {Marchant}, J.~M. and {Marconi}, M. and {Marie}, J. and {Marinoni}, S. and {Marrese}, P.~M. and {Marschalk{\'o}}, G. and {Marshall}, D.~J. and {Mart{\'\i}n-Fleitas}, J.~M. and {Martino}, M. and {Mary}, N. and {Matijevi{\v{c}}}, G. and {Mazeh}, T. and {McMillan}, P.~J. and {Messina}, S. and {Mestre}, A. and {Michalik}, D. and {Millar}, N.~R. and {Miranda}, B.~M.~H. and {Molina}, D. and {Molinaro}, R. and {Molinaro}, M. and {Moln{\'a}r}, L. and {Moniez}, M. and {Montegriffo}, P. and {Monteiro}, D. and {Mor}, R. and {Mora}, A. and {Morbidelli}, R. and {Morel}, T. and {Morgenthaler}, S. and {Morley}, T. and {Morris}, D. and {Mulone}, A.~F. and {Muraveva}, T. and {Musella}, I. and {Narbonne}, J. and {Nelemans}, G. and {Nicastro}, L. and {Noval}, L. and {Ord{\'e}novic}, C. and {Ordieres-Mer{\'e}}, J. and {Osborne}, P. and {Pagani}, C. and {Pagano}, I. and {Pailler}, F. and {Palacin}, H. and {Palaversa}, L. and {Parsons}, P. and {Paulsen}, T. and {Pecoraro}, M. and {Pedrosa}, R. and {Pentik{\"a}inen}, H. and {Pereira}, J. and {Pichon}, B. and {Piersimoni}, A.~M. and {Pineau}, F. -X. and {Plachy}, E. and {Plum}, G. and {Poujoulet}, E. and {Pr{\v{s}}a}, A. and {Pulone}, L. and {Ragaini}, S. and {Rago}, S. and {Rambaux}, N. and {Ramos-Lerate}, M. and {Ranalli}, P. and {Rauw}, G. and {Read}, A. and {Regibo}, S. and {Renk}, F. and {Reyl{\'e}}, C. and {Ribeiro}, R.~A. and {Rimoldini}, L. and {Ripepi}, V. and {Riva}, A. and {Rixon}, G. and {Roelens}, M. and {Romero-G{\'o}mez}, M. and {Rowell}, N. and {Royer}, F. and {Rudolph}, A. and {Ruiz-Dern}, L. and {Sadowski}, G. and {Sagrist{\`a} Sell{\'e}s}, T. and {Sahlmann}, J. and {Salgado}, J. and {Salguero}, E. and {Sarasso}, M. and {Savietto}, H. and {Schnorhk}, A. and {Schultheis}, M. and {Sciacca}, E. and {Segol}, M. and {Segovia}, J.~C. and {Segransan}, D. and {Serpell}, E. and {Shih}, I. -C. and {Smareglia}, R. and {Smart}, R.~L. and {Smith}, C. and {Solano}, E. and {Solitro}, F. and {Sordo}, R. and {Soria Nieto}, S. and {Souchay}, J. and {Spagna}, A. and {Spoto}, F. and {Stampa}, U. and {Steele}, I.~A. and {Steidelm{\"u}ller}, H. and {Stephenson}, C.~A. and {Stoev}, H. and {Suess}, F.~F. and {S{\"u}veges}, M. and {Surdej}, J. and {Szabados}, L. and {Szegedi-Elek}, E. and {Tapiador}, D. and {Taris}, F. and {Tauran}, G. and {Taylor}, M.~B. and {Teixeira}, R. and {Terrett}, D. and {Tingley}, B. and {Trager}, S.~C. and {Turon}, C. and {Ulla}, A. and {Utrilla}, E. and {Valentini}, G. and {van Elteren}, A. and {Van Hemelryck}, E. and {van Leeuwen}, M. and {Varadi}, M. and {Vecchiato}, A. and {Veljanoski}, J. and {Via}, T. and {Vicente}, D. and {Vogt}, S. and {Voss}, H. and {Votruba}, V. and {Voutsinas}, S. and {Walmsley}, G. and {Weiler}, M. and {Weingrill}, K. and {Werner}, D. and {Wevers}, T. and {Whitehead}, G. and {Wyrzykowski}, {\L}. and {Yoldas}, A. and {{\v{Z}}erjal}, M. and {Zucker}, S. and {Zurbach}, C. and {Zwitter}, T. and {Alecu}, A. and {Allen}, M. and {Allende Prieto}, C. and {Amorim}, A. and {Anglada-Escud{\'e}}, G. and {Arsenijevic}, V. and {Azaz}, S. and {Balm}, P. and {Beck}, M. and {Bernstein}, H. -H. and {Bigot}, L. and {Bijaoui}, A. and {Blasco}, C. and {Bonfigli}, M. and {Bono}, G. and {Boudreault}, S. and {Bressan}, A. and {Brown}, S. and {Brunet}, P. -M. and {Bunclark}, P. and {Buonanno}, R. and {Butkevich}, A.~G. and {Carret}, C. and {Carrion}, C. and {Chemin}, L. and {Ch{\'e}reau}, F. and {Corcione}, L. and {Darmigny}, E. and {de Boer}, K.~S. and {de Teodoro}, P. and {de Zeeuw}, P.~T. and {Delle Luche}, C. and {Domingues}, C.~D. and {Dubath}, P. and {Fodor}, F. and {Fr{\'e}zouls}, B. and {Fries}, A. and {Fustes}, D. and {Fyfe}, D. and {Gallardo}, E. and {Gallegos}, J. and {Gardiol}, D. and {Gebran}, M. and {Gomboc}, A. and {G{\'o}mez}, A. and {Grux}, E. and {Gueguen}, A. and {Heyrovsky}, A. and {Hoar}, J. and {Iannicola}, G. and {Isasi Parache}, Y. and {Janotto}, A. -M. and {Joliet}, E. and {Jonckheere}, A. and {Keil}, R. and {Kim}, D. -W. and {Klagyivik}, P. and {Klar}, J. and {Knude}, J. and {Kochukhov}, O. and {Kolka}, I. and {Kos}, J. and {Kutka}, A. and {Lainey}, V. and {LeBouquin}, D. and {Liu}, C. and {Loreggia}, D. and {Makarov}, V.~V. and {Marseille}, M.~G. and {Martayan}, C. and {Martinez-Rubi}, O. and {Massart}, B. and {Meynadier}, F. and {Mignot}, S. and {Munari}, U. and {Nguyen}, A. -T. and {Nordlander}, T. and {Ocvirk}, P. and {O'Flaherty}, K.~S. and {Olias Sanz}, A. and {Ortiz}, P. and {Osorio}, J. and {Oszkiewicz}, D. and {Ouzounis}, A. and {Palmer}, M. and {Park}, P. and {Pasquato}, E. and {Peltzer}, C. and {Peralta}, J. and {P{\'e}turaud}, F. and {Pieniluoma}, T. and {Pigozzi}, E. and {Poels}, J. and {Prat}, G. and {Prod'homme}, T. and {Raison}, F. and {Rebordao}, J.~M. and {Risquez}, D. and {Rocca-Volmerange}, B. and {Rosen}, S. and {Ruiz-Fuertes}, M.~I. and {Russo}, F. and {Sembay}, S. and {Serraller Vizcaino}, I. and {Short}, A. and {Siebert}, A. and {Silva}, H. and {Sinachopoulos}, D. and {Slezak}, E. and {Soffel}, M. and {Sosnowska}, D. and {Strai{\v{z}}ys}, V. and {ter Linden}, M. and {Terrell}, D. and {Theil}, S. and {Tiede}, C. and {Troisi}, L. and {Tsalmantza}, P. and {Tur}, D. and {Vaccari}, M. and {Vachier}, F. and {Valles}, P. and {Van Hamme}, W. and {Veltz}, L. and {Virtanen}, J. and {Wallut}, J. -M. and {Wichmann}, R. and {Wilkinson}, M.~I. and {Ziaeepour}, H. and {Zschocke}, S.},
        title = "{The Gaia mission}",
      journal = {\aap},
         year = 2016,
        month = nov,
       volume = {595},
          eid = {A1},
        pages = {A1},
          doi = {10.1051/0004-6361/201629272},
archivePrefix = {arXiv},
       eprint = {1609.04153},
 primaryClass = {astro-ph.IM},
       adsurl = {https://ui.adsabs.harvard.edu/abs/2016A&A...595A...1G}
}

@ARTICLE{Zhang2021,
       author = {{Zhang}, Jinghua and {Shi}, Jian-Rong and {Yan}, Hong-Liang and {Li}, Yaguang and {Gao}, Qi and {Li}, Chun-Qian and {Zhang}, Xianfei and {Liu}, Shuai and {Bi}, Shaolan and {Zhao}, Gang and {Li}, Yan},
        title = "{Lithium Evolution of Giant Stars Observed by LAMOST and Kepler}",
      journal = {\apjl},
         year = 2021,
        month = sep,
       volume = {919},
       number = {1},
          eid = {L3},
        pages = {L3},
          doi = {10.3847/2041-8213/ac224c},
archivePrefix = {arXiv},
       eprint = {2108.13030},
 primaryClass = {astro-ph.SR},
       adsurl = {https://ui.adsabs.harvard.edu/abs/2021ApJ...919L...3Z}
}

@ARTICLE{sharma_bstep,
       author = {{Sharma}, Sanjib and {Stello}, Dennis and {Buder}, Sven and {Kos}, Janez and {Bland-Hawthorn}, Joss and {Asplund}, Martin and {Duong}, Ly and {Lin}, Jane and {Lind}, Karin and {Ness}, Melissa and {Huber}, Daniel and {Zwitter}, Tomaz and {Traven}, Gregor and {Hon}, Marc and {Kafle}, Prajwal R. and {Khanna}, Shourya and {Saddon}, Hafiz and {Anguiano}, Borja and {Casey}, Andrew R. and {Freeman}, Ken and {Martell}, Sarah and {De Silva}, Gayandhi M. and {Simpson}, Jeffrey D. and {Wittenmyer}, Rob A. and {Zucker}, Daniel B.},
        title = "{The TESS-HERMES survey data release 1: high-resolution spectroscopy of the TESS southern continuous viewing zone}",
      journal = {\mnras},
         year = 2018,
        month = jan,
       volume = {473},
       number = {2},
        pages = {2004-2019},
          doi = {10.1093/mnras/stx2582},
archivePrefix = {arXiv},
       eprint = {1707.05753},
 primaryClass = {astro-ph.SR},
       adsurl = {https://ui.adsabs.harvard.edu/abs/2018MNRAS.473.2004S}
}

@ARTICLE{Mallick_2023,
       author = {{Mallick}, Anohita and {Singh}, Raghubar and {Reddy}, Bacham E.},
        title = "{Lithium Abundances in Giants as a Function of Stellar Mass: Evidence for He Flash as the Source of Li Enhancement in Low-mass Giants}",
      journal = {\apjl},
         year = 2023,
        month = feb,
       volume = {944},
       number = {1},
          eid = {L5},
        pages = {L5},
          doi = {10.3847/2041-8213/acb5f6},
archivePrefix = {arXiv},
       eprint = {2301.10436},
 primaryClass = {astro-ph.SR},
       adsurl = {https://ui.adsabs.harvard.edu/abs/2023ApJ...944L...5M}
}

@ARTICLE{Patton_2023,
       author = {{Patton}, Rachel A. and {Pinsonneault}, Marc H. and {Cao}, Lyra and {Vrard}, Mathieu and {Mathur}, Savita and {Garc{\'\i}a}, Rafael A. and {Tayar}, Jamie and {Daher}, Christine Mazzola and {Beck}, Paul G.},
        title = "{Spectroscopic identification of rapidly rotating red giant stars in APOKASC-3 and APOGEE DR16}",
      journal = {\mnras},
         year = 2024,
        month = feb,
       volume = {528},
       number = {2},
        pages = {3232-3248},
          doi = {10.1093/mnras/stae074},
archivePrefix = {arXiv},
       eprint = {2303.08151},
 primaryClass = {astro-ph.SR},
       adsurl = {https://ui.adsabs.harvard.edu/abs/2024MNRAS.528.3232P}
}

@article{karakas2014dawes,
  title={The Dawes review 2: nucleosynthesis and stellar yields of low-and intermediate-mass single stars},
  author={Karakas, Amanda I and Lattanzio, John C},
  journal={Publications of the Astronomical Society of Australia},
  volume={31},
  pages={e030},
  year={2014},
  publisher={Cambridge University Press}
}

@ARTICLE{Luhman2013,
       author = {{Luhman}, K.~L.},
        title = "{Discovery of a Binary Brown Dwarf at 2 pc from the Sun}",
      journal = {\apjl},
         year = 2013,
        month = apr,
       volume = {767},
       number = {1},
          eid = {L1},
        pages = {L1},
          doi = {10.1088/2041-8205/767/1/L1},
archivePrefix = {arXiv},
       eprint = {1303.2401},
 primaryClass = {astro-ph.GA},
       adsurl = {https://ui.adsabs.harvard.edu/abs/2013ApJ...767L...1L}
}

@ARTICLE{Faherty2014,
       author = {{Faherty}, Jacqueline K. and {Beletsky}, Yuri and {Burgasser}, Adam J. and {Tinney}, Chris and {Osip}, David J. and {Filippazzo}, Joseph C. and {Simcoe}, Robert A.},
        title = "{Signatures of Cloud, Temperature, and Gravity from Spectra of the Closest Brown Dwarfs}",
      journal = {\apj},
         year = 2014,
        month = aug,
       volume = {790},
       number = {2},
          eid = {90},
        pages = {90},
          doi = {10.1088/0004-637X/790/2/90},
archivePrefix = {arXiv},
       eprint = {1406.1518},
 primaryClass = {astro-ph.SR},
       adsurl = {https://ui.adsabs.harvard.edu/abs/2014ApJ...790...90F}
}

@ARTICLE{Lodieu2015,
       author = {{Lodieu}, N. and {Zapatero Osorio}, M.~R. and {Rebolo}, R. and {B{\'e}jar}, V.~J.~S. and {Pavlenko}, Y. and {P{\'e}rez-Garrido}, A.},
        title = "{VLT X-Shooter spectroscopy of the nearest brown dwarf binary}",
      journal = {\aap},
         year = 2015,
        month = sep,
       volume = {581},
          eid = {A73},
        pages = {A73},
          doi = {10.1051/0004-6361/201424933},
archivePrefix = {arXiv},
       eprint = {1506.08848},
 primaryClass = {astro-ph.SR},
       adsurl = {https://ui.adsabs.harvard.edu/abs/2015A&A...581A..73L}
}

@ARTICLE{Cummings2018,
       author = {{Cummings}, Jeffrey D. and {Kalirai}, Jason S. and {Tremblay}, P. -E. and {Ramirez-Ruiz}, Enrico and {Choi}, Jieun},
        title = "{The White Dwarf Initial-Final Mass Relation for Progenitor Stars from 0.85 to 7.5 M $_{{\ensuremath{\odot}}}$}",
      journal = {\apj},
         year = 2018,
        month = oct,
       volume = {866},
       number = {1},
          eid = {21},
        pages = {21},
          doi = {10.3847/1538-4357/aadfd6},
archivePrefix = {arXiv},
       eprint = {1809.01673},
 primaryClass = {astro-ph.SR},
       adsurl = {https://ui.adsabs.harvard.edu/abs/2018ApJ...866...21C}
}

@ARTICLE{ElBadry2018,
       author = {{El-Badry}, Kareem and {Rix}, Hans-Walter and {Weisz}, Daniel R.},
        title = "{An Empirical Measurement of the Initial-Final Mass Relation with Gaia White Dwarfs}",
      journal = {\apjl},
         year = 2018,
        month = jun,
       volume = {860},
       number = {2},
          eid = {L17},
        pages = {L17},
          doi = {10.3847/2041-8213/aaca9c},
archivePrefix = {arXiv},
       eprint = {1805.05849},
 primaryClass = {astro-ph.SR},
       adsurl = {https://ui.adsabs.harvard.edu/abs/2018ApJ...860L..17E}
}

@ARTICLE{Cunningham2024,
       author = {{Cunningham}, Tim and {Tremblay}, Pier-Emmanuel and {W. O'Brien}, Mairi},
        title = "{Initial-final mass relation from white dwarfs within 40 pc}",
      journal = {\mnras},
         year = 2024,
        month = jan,
       volume = {527},
       number = {2},
        pages = {3602-3611},
          doi = {10.1093/mnras/stad3275},
archivePrefix = {arXiv},
       eprint = {2310.15410},
 primaryClass = {astro-ph.SR},
       adsurl = {https://ui.adsabs.harvard.edu/abs/2024MNRAS.527.3602C}
}

@ARTICLE{Marigo2007,
       author = {{Marigo}, P. and {Girardi}, L.},
        title = "{Evolution of asymptotic giant branch stars. I. Updated synthetic TP-AGB models and their basic calibration}",
      journal = {\aap},
         year = 2007,
        month = jul,
       volume = {469},
       number = {1},
        pages = {239-263},
          doi = {10.1051/0004-6361:20066772},
archivePrefix = {arXiv},
       eprint = {astro-ph/0703139},
 primaryClass = {astro-ph},
       adsurl = {https://ui.adsabs.harvard.edu/abs/2007A&A...469..239M}
}

@ARTICLE{Meng2008,
       author = {{Meng}, X. and {Chen}, X. and {Han}, Z.},
        title = "{Initial-final mass relationship for stars of different metallicities}",
      journal = {\aap},
         year = 2008,
        month = aug,
       volume = {487},
       number = {2},
        pages = {625-635},
          doi = {10.1051/0004-6361:20078841},
archivePrefix = {arXiv},
       eprint = {0710.2397},
 primaryClass = {astro-ph},
       adsurl = {https://ui.adsabs.harvard.edu/abs/2008A&A...487..625M}
}

@ARTICLE{Mallik1999,
       author = {{Mallik}, Sushma V.},
        title = "{Lithium abundance and mass}",
      journal = {\aap},
         year = 1999,
        month = dec,
       volume = {352},
        pages = {495-507},
       adsurl = {https://ui.adsabs.harvard.edu/abs/1999A&A...352..495M}
}
\bibliographystyle{aasjournal}

\end{document}